\documentclass[aps,pra,twocolumn,floatix,amssymb,showpacs,amsmath,superscriptaddress,revtex4]{revtex4-1}
\usepackage{graphicx}
\usepackage{braket}
\usepackage[usenames,dvipsnames]{color}
\usepackage{epsfig,color}
\usepackage{amsmath,amssymb}
\usepackage{mathptmx}
\usepackage[T1]{fontenc}
\usepackage{verbatim}
\usepackage{float}
\usepackage{units}
\usepackage{amsmath}
\usepackage{bm}
\usepackage{booktabs}
\usepackage[colorlinks = true,
            linkcolor = NavyBlue,
            urlcolor  = NavyBlue,
            citecolor = NavyBlue,
            anchorcolor = blue]{hyperref}
\usepackage{epstopdf}
\usepackage[normalem]{ulem}
\renewcommand\Im{\operatorname{Im}}
\renewcommand\Re{\operatorname{Re}}
\begin{document}
\title{Emergence of large non-adiabatic effects induced by the electron-phonon interaction \\ on the complex vibrational quasi-particle spectrum of the doped monolayer MoS$_{2}$}
\author{Peio Garcia-Goiricelaya}
\affiliation{Materia Kondentsatuaren Fisika Saila, University of the Basque Country UPV/EHU, 48080 Bilbao, Basque Country, Spain}
\affiliation{Donostia International Physics Center (DIPC), Paseo Manuel de Lardizabal 4, 20018 Donostia-San Sebasti\'{a}n, Spain}
\author{Jon Lafuente-Bartolome}
\affiliation{Materia Kondentsatuaren Fisika Saila, University of the Basque Country UPV/EHU, 48080 Bilbao, Basque Country, Spain}
\affiliation{Donostia International Physics Center (DIPC), Paseo Manuel de Lardizabal 4, 20018 Donostia-San Sebasti\'{a}n, Spain}
\author{Idoia G. Gurtubay}
\affiliation{Materia Kondentsatuaren Fisika Saila, University of the Basque Country UPV/EHU, 48080 Bilbao, Basque Country, Spain}
\affiliation{Donostia International Physics Center (DIPC), Paseo Manuel de Lardizabal 4, 20018 Donostia-San Sebasti\'{a}n, Spain}
\author{Asier Eiguren}
\affiliation{Materia Kondentsatuaren Fisika Saila, University of the Basque Country UPV/EHU, 48080 Bilbao, Basque Country, Spain}
\affiliation{Donostia International Physics Center (DIPC), Paseo Manuel de Lardizabal 4, 20018 Donostia-San Sebasti\'{a}n, Spain}
\date{\today}
\begin{abstract}
  We present a comprehensive first-principles analysis of the non-adiabatic effects due to the electron-phonon interaction on the vibrational spectrum of the electron-doped monolayer MoS$_{2}$.
  Deep changes in the Fermi surface upon doping cause the linewidth broadening of the normal modes governing the spin-conserving inter-valley electronic scattering, which become unstable with the population of all the spin-split conduction valleys.
  We find that the non-adiabatic spectral effects modify dramatically the adiabatic dispersion of the long-wavelength optical phonon modes, responsible for intra-valley scattering, as soon as inequivalent valleys get populated.
  These results are illustrated by means of a simple analytical model.
  Finally, we explain the emergence of an intricate dynamical structure for the strongly interacting out-of-plane polarized $\mathrm{A'_{1}}$ optical vibrational mode spectrum by means of a multiple-phonon quasi-particle picture defined in the full complex frequency plane, showing that this intriguing spectral structure originates from the splitting of the original adiabatic branch induced by the electron-phonon coupling.
\end{abstract}
\maketitle
\section{INTRODUCTION}
Most standard first-principles calculations of both electronic and vibrational properties of solids~\cite{DFT1,DFT2,DFPT} rely on the adiabatic approximation~\cite{BOapprox}, which assumes an instantaneous response of the carriers to the motion of ions, and therefore, that the lattice dynamics is only influenced by electrostatic fields.
The adiabatic approximation has proved to be of great success, as demonstrated by the good agreement between experimental and theoretical phonon dispersions, as well as electron-phonon coupling strengths~\cite{sabrasovepis1,sabrasovepis2,liuepis,mauriepis,bauerepis}.
The reason of this success is partially due to the fact that the corrections to the vibrational spectrum induced by the interaction with electrons beyond the adiabatic approximation, i.e.~non-adiabatic effects, are in most cases weak in metals~\cite{Migdal}.
However, both Migdal~\cite{Migdal} and Engelsberg and Schriffer~\cite{ES} already indicated the possibility of large non-adiabatic effects on optical phonon dispersions in the long-wavelength limit, which was subsequently studied by means of simple models~\cite{Ipatova,Maksimov1996}.
Since then, great efforts have been made to experimentally detect such many-body effects, mainly by Raman spectroscopy~\cite{ponosovna}, but also using inelastic neutron~\cite{maksimovphSF} and x-ray spectroscopies~\cite{NAKAgiustino}, all of them being valuable tools for reporting lattice vibrational spectra.
Thereby, phonon frequency renormalizations due to electron-phonon interactions have been found in several cases, among them standing out large phonon frequency hardenings of up to $30\%$~\cite{NAKAgiustino,NAKohnGraph,failureBOgraph,NAcarbnanotube,Saittaprl2008,calandramgb2,capeluttimgb2,calandramauriprb2010}.
In all these systems, an approximated \textit{ab initio} lattice vibrational theory including non-adiabatic effects has proved to be essential in order to interpret the experimental spectra, that otherwise can not be explained only within the adiabatic approach.

More recently, Raman experiments on the monolayer MoS$_{2}$ have revealed strong sensitivity to electron-doping of the optical phonons with a dominant out-of-plane polarization, which exhibit an increasing frequency softening at low carrier concentrations $(\rho<2\times10^{13}~\text{cm}^{-2})$~\cite{1lmos2a1soft}.
Although this effect was first understood by means of adiabatic vibrational calculations, several works focussing in the small momentum range have  pointed out the importance of non-adiabatic corrections when increasing the carrier doping concentration $(\rho\sim5\times10^{13}~\text{cm}^{-2})$~\cite{sohierprx2019,1907.04766}.
Moreover, the experimentally measured superconductivity~\cite{scmos2} and the intricate carrier photoemission spectra~\cite{Kang2018} at larger doping have been theoretically explained by a considerably large strengthening of the electron-phonon coupling promoted by abrupt changes in the topology of the Fermi surface (FS)~\cite{yizhiprb2013,piattimultivalleymos2,gurepaper}.
In this respect, large adiabatic frequency softenings of the strongly interacting phonons mediating electronic inter-valley scattering have been also obtained by \textit{ab initio} calculations~\cite{phsoftmos2,rosnermos2}.

In this paper, we analyze the non-adiabatic corrections induced by the electron-phonon coupling to the whole vibrational spectrum of the electron-doped monolayer MoS$_{2}$ from first-principles.
For the phonon modes driving effective electronic inter-valley scattering, we report large spectral broadenings as soon as the corresponding spin-conserving electron-hole channels are energetically allowed with increasing doping.
Besides, we detect that these vibrational modes develop lattice instabilities precisely at the moment when all the multiple inequivalent spin-split conduction-valleys get populated.
Interestingly, large phonon frequency hardenings and sharp dispersions emerge for the optical vibrational branches accompanied by an intricate dynamical structure and the broadening of the spectral function at finite momenta close to the Brillouin zone (BZ) center.
We explain the emergent spectrum in terms of a multiple-phonon quasi-particle picture which puts in evidence the splitting of the optical phonon branch induced by the electron-phonon interaction.

This paper is organized as follows.
In Sec.\,\ref{sec:theory}, we briefly review the theory based on many-body Green's function methods for calculating from first-principles non-adiabatic effects due to electron-phonon interactions on lattice vibrations.
Section\,\ref{sec:computational_details} summarizes the computational details.
In Sec.\,\ref{sec:A}, we shortly present the electronic and vibrational adiabatic calculations of the doped monolayer MoS$_{2}$ as a necessary step before addressing in Sec.\,\ref{sec:B} the calculated phonon spectral functions including non-adiabatic effects.
In Sec.\,\ref{sec:C}, we concisely emphasize the importance of the complex frequency plane in order to correctly describe and understand the emergent and intricate spectrum in terms of phonon quasi-particles.
Section\,\ref{sec:conclusions} summarizes our conclusions.
\section{THEORY}\label{sec:theory}
Since the adiabatic approximation neglects retardation effects on the electronic response to the ionic motion, the interatomic force constants are strictly real magnitudes within this approximation.
The adiabatic phonons are defined as the solutions of the secular eigenvalue problem~\cite{DFPT},%
\begin{equation}
\sum_{s'\alpha'}D_{ss'}^{\alpha\alpha'}(\mathbf{q})e_{\nu}^{s'\alpha'}(\mathbf{q})=\omega_{\nu}^{2}(\mathbf{q})e_{\nu}^{s\alpha}(\mathbf{q}),
\label{eq:EE_AD}
\end{equation}
where $\omega_{\nu}(\mathbf{q})$ and $e_{\nu}^{s\alpha}(\mathbf{q})$ are the frequency and the polarization vector  of the lattice vibrational normal mode with branch index $\nu$ and momentum $\mathbf{q}$, respectively.
Here and in the following, atomic units are used.
$D_{ss'}^{\alpha\alpha'}(\mathbf{q})$ is the adiabatic dynamical matrix which is related to the Fourier transform of the interatomic force constant matrix, $C_{ss'}^{\alpha\alpha'}(\mathbf{q})$, as follows: $D_{ss'}^{\alpha\alpha'}(\mathbf{q})=C_{ss'}^{\alpha\alpha'}(\mathbf{q})/\sqrt{M_{s}M_{s'}}$, with $M_{s}$ the mass of the $s$-th ion within the unit cell (u.c.).
Since $D_{ss'}^{\alpha\alpha'}(\mathbf{q})$ is  Hermitian by definition, the adiabatic phonons as defined above have real frequencies and are, therefore, infinitely long-lived excitations~\cite{DFPT}.
Physically, the interatomic force constants describe the force acting on the $s$-th ion in the $\alpha$ direction when the $s'$-th ion is displaced along the $\alpha'$ direction from its equilibrium position \cite{dyprosovol3}.
In this sense, a useful expression for $C_{ss'}^{\alpha\alpha'}(\mathbf{q})$ can be obtained by differentiating the Hellmann-Feynman forces~\cite{hft} with respect to the ionic displacements from the equilibrium geometry and using the density linear response theory~\cite{dyprosovol3}
\begin{equation}
\begin{aligned}
C_{ss'}^{\alpha\alpha'}(\mathbf{q})&=\iint\frac{\partial V_{\mathrm{scf}}(\mathbf{r})}{\partial u_{s'}^{\alpha'}(\mathbf{q})}
\Bigg(\chi^{0}_{\mathbf{q}}(\mathbf{r},\mathbf{r'},0)\frac{\partial V_{\mathrm{scf}}(\mathbf{r'})}{\partial u_{s}^{\alpha}(\mathbf{q})}\Bigg)^{*}\mathrm{d}\mathbf{r}\mathrm{d}\mathbf{r'}\\
-&\iint\Bigg(\frac{\partial n(\mathbf{r})}{\partial u_{s}^{\alpha}(\mathbf{q})}\Bigg)^{*}K(\mathbf{r},\mathbf{r'})\frac{\partial n(\mathbf{r'})}{\partial u_{s'}^{\alpha'}(\mathbf{q})}\mathrm{d}\mathbf{r'}\mathrm{d}\mathbf{r}\\
+&\int n(\mathbf{r})\frac{\partial^{2}V_{\mathrm{ext}}(\mathbf{r})}{\partial u_{s}^{*\alpha}(\mathbf{q})\partial u_{s'}^{\alpha'}(\mathbf{q})}\mathrm{d}\mathbf{r}
+\frac{\partial^{2} E_{\mathrm{ion}}}{\partial u_{s}^{*\alpha}(\mathbf{q})\partial u_{s'}^{\alpha'}(\mathbf{q})}.
\label{eq:DFPT}
\end{aligned}
\end{equation}
Here, $n(\mathbf{r})$, $V_{\mathrm{ext}}(\mathbf{r})$ and $E_{\mathrm{ion}}$ are the electron charge density, the electron-ion interaction external potential and the Coulomb interaction energy between nuclei, respectively.
$K(\mathbf{r},\mathbf{r'})$ is the Hartree and exchange-correlation kernel.
$u_{s}^{\alpha}(\mathbf{q})$ corresponds to the displacement amplitude of the $s$-th ion along the $\alpha$ direction for a lattice distortion of wave vector $\mathbf{q}$.
$\partial V_{\mathrm{scf}}(\mathbf{r})/\partial u_{s}^{\alpha}(\mathbf{q})$ and $\partial n(\mathbf{r})/\partial u_{s}^{\alpha}(\mathbf{q})$
are the self-consistent static-screened change of the external potential and charge density, respectively, with respect to the ionic displacement $u_{s}^{\alpha}(\mathbf{q})$.
$\chi^{0}_{\mathbf{q}}\left(\mathbf{r},\mathbf{r'},\omega\right)$ is the density-response function of the non-interacting electronic system, defined as~\cite{mahan}
\begin{equation}
\begin{aligned}
\chi^{0}_{\mathbf{q}}\left(\mathbf{r},\mathbf{r'},\omega\right)=&\frac{1}{N_{\mathbf{k}}}\sum_{\mathbf{k}}^{\mathrm{1BZ}}\sum_{mn}\frac{f(\varepsilon_{n}^{\mathbf{k}})-f(\varepsilon_{m}^{\mathbf{k}+\mathbf{q}})}{\varepsilon_{n}^{\mathbf{k}}-\varepsilon_{m}^{\mathbf{k}+\mathbf{q}}+\omega+i\eta}\times \\&\big(\psi_{n}^{\mathbf{k}}(\mathbf{r})\big)^{*}\psi_{m}^{\mathbf{k}+\mathbf{q}}(\mathbf{r})\big(\psi_{m}^{\mathbf{k}+\mathbf{q}}(\mathbf{r'})\big)^{*}\psi_{n}^{\mathbf{k}}(\mathbf{r'})
\label{eq:chi0},
\end{aligned}
\end{equation}
where $\varepsilon_{n}^{\mathbf{k}}$ and $\psi_{n}^{\mathbf{k}}(\mathbf{r})$ are the energy and wave function of the Kohn-Sham (KS) electron state of band index $n$ and momentum $\mathbf{k}$, respectively.
$f(\varepsilon_{n}^{\mathbf{k}})$ represents the Fermi-Dirac (FD) occupation factor of the KS state, $\eta$ is a positive real infinitesimal and $N_{\mathbf{k}}$ is the number of $\mathbf{k}$-points considered for the BZ integral.
All the magnitudes presented so far are directly accessible from state-of-the-art calculations based on the density functional theory (DFT)~\cite{DFT1,DFT2} and the density functional perturbation theory (DFPT)~\cite{DFPT}.
From the first term on the right-hand side of Eq.\,\ref{eq:DFPT}, it is clear that the adiabatic dynamical matrix in Eq.\,\ref{eq:EE_AD} is valid as long as a electronic static-screening of lattice vibrations is similar to the response function at typical phonon frequencies, i.e.~$\chi^{0}_{\mathbf{q}}\left(\mathbf{r},\mathbf{r'},\omega\right)\approx\chi^{0}_{\mathbf{q}}\left(\mathbf{r},\mathbf{r'},0\right)$.
In other words, adiabatic phonons are valid as long as the phonon-mediated electronic transition energies between occupied and empty states are much greater than the vibrational frequencies themselves, i.e.~$|\varepsilon^{\mathbf{k}}_{n}-\varepsilon^{\mathbf{k+q}}_{m}|\gg\omega_{\nu}(\mathbf{q})$ in Eq.\,\ref{eq:chi0}.

The above condition is satisfied in insulators and large-gap semiconductors, but may be compromised for several metals and narrow-gap semiconductors.
This is so because the dynamical matrix should in principle incorporate the retardation effects on the electronic response to the ionic motion with a finite frequency dependence $(\omega\neq0)$.
In this case, Eq.\,\ref{eq:EE_AD} becomes a self-consistent dynamical problem given by $\big|D_{ss'}^{\alpha\alpha'}(\mathbf{q},\omega)-\omega^2\big|=0$~\cite{dyprosovol3}.
The effects due to the dynamical response of the electron gas are equivalently taken into account by means of many-body perturbation theory based on Green's functions~\cite{mahan}.
Within this formalism, the dressed phonon Green's function for normal modes with momentum $\mathbf{q}$, $\mathcal{D}_{\nu\nu'}(\mathbf{q},\omega)$, is obtained from the bare phonon Green's function, $\mathcal{D}^{0}_{\nu\nu}(\mathbf{q},\omega)$, and the phonon self-energy, $\Pi_{\nu\nu'}(\mathbf{q},\omega)$, by solving the following Dyson's equation~\cite{Grimvall}
\begin{equation}
	{\mathcal{D}_{\nu\nu'}(\mathbf{q},\omega)}^{-1}={\mathcal{D}^{0}_{\nu\nu}(\mathbf{q},\omega)}^{-1}-\Pi_{\nu\nu'}(\mathbf{q},\omega).
 \label{eq:DE}
\end{equation}
The above equation describes the dynamical perturbation due to the electron-phonon coupling, which is encoded by the $\Pi_{\nu\nu'}(\mathbf{q},\omega)$ matrix, and in principle describes the hybridization of the bare phonon modes given by the diagonal matrix $\mathcal{D}^{0}_{\nu\nu}(\mathbf{q},\omega)$.
However, we found numerically that the diagonal approximation turns out to be excellent in MoS$_{2}$, and hereinafter, only the diagonal form of the Dyson's equation is considered for simplicity.
The formal definition of the phonon self-energy in Eq.\,\ref{eq:DE} includes an impractical infinite series of electron-phonon Feynman diagrams.
From now on we will consider only the simplest polarization diagram for the phonon self-energy~\cite{Grimvall}.
This procedure leads to the extensively used expression of the so-called retarded phonon self-energy due to the electron-phonon interaction~\footnote{Interestingly, a similar expression for the phonon self-energy can be recovered from directly adopting a frequency-dependent density-response function in the first term on the right-hand side of Eq.\,\ref{eq:DFPT}, and transforming it into the normal-mode representation (see the Note\,S1 of the Supplemental Material)}
\begin{equation}
 \Pi_{\nu}(\mathbf{q},\omega)=\frac{1}{N_{\mathbf{k}}}\sum_{\mathbf{k}}^{\mathrm{1BZ}}\sum_{mn}\big|g^{\nu}_{mn}(\mathbf{k},\mathbf{q})\big|^{2}\frac{f(\varepsilon^{\mathbf{k}}_{n})-f(\varepsilon^{\mathbf{k+q}}_{m})}{\varepsilon^{\mathbf{k}}_{n}-\varepsilon^{\mathbf{k+q}}_{m}+\omega+i\eta}.
 \label{eq:ph_SE}
\end{equation}
Here, $g^{\nu}_{mn}(\mathbf{k},\mathbf{q})$ are the static-screened electron-phonon matrix elements, defined as~\cite{Grimvall}
\begin{equation}
g_{mn}^{\nu}(\mathbf{k},\mathbf{q})=\sum_{s\alpha} \frac{e_{\nu}^{s\alpha}(\mathbf{q})}{\sqrt{2\omega_{\nu}(\mathbf{q})M_{s}}}
                                    \bigg\langle\psi^{\mathbf{k+q}}_{m}\bigg|\frac{\partial\hat{V}_{\mathrm{scf}}}{\partial u^{\alpha}_{s}(\mathbf{q})}\bigg|\psi^{\mathbf{k}}_{n}\bigg\rangle
\label{eq:epme},
\end{equation}
which physically describe the strength of the effective coupling between the KS states $\ket{\psi^{\mathbf{k}}_{n}}$ and $\ket{\psi^{\mathbf{k+q}}_{m}}$ via the normal mode of branch index $\nu$ and momentum $\mathbf{q}$.
The matrix elements in Eq.\,\ref{eq:epme} as well as the retarded phonon self-energy in Eq.\,\ref{eq:ph_SE} are directly computable magnitudes from DFT and DFPT calculations.
The phonon self-energy physically accounts for both the dynamical $(\omega\neq0)$ and static $(\omega=0)$ screening of lattice vibrations by electron-hole pair excitations induced by the phonon itself.
It therefore holds the information on the vibrational frequency renormalizations and linewidth broadenings due to the electron-phonon interaction up to first order.
In this respect, from Eq.\,\ref{eq:ph_SE}, it is clear that large corrections should be expected when the phonon-mediated electronic transition energies are close to the  vibrational frequencies, i.e.~$|\varepsilon^{\mathbf{k}}_{n}-\varepsilon^{\mathbf{k+q}}_{m}|\approx\omega_{\nu}(\mathbf{q})$.

In practice, one has to be careful when accounting for the static screening contributions to the phonon self-energy in Eq.\,\ref{eq:ph_SE}, and bear in mind that these terms are already included in the adiabatic phonon calculations.
Within this scheme, the bare propagator in Eq.\,\ref{eq:DE} for a phonon mode with branch index $\nu$ and momentum $\mathbf{q}$ is defined in terms of the adiabatic phonon frequency as
\begin{equation}
 \begin{aligned}
 \mathcal{D}^{0}_{\nu}(\mathbf{q},\omega)&=\frac{1}{\omega-\omega_{\nu}(\mathbf{q})+i\eta}-\frac{1}{\omega+\omega_{\nu}(\mathbf{q})-i\eta}\\&=\frac{2\omega_{\nu}(\mathbf{q})}{\omega^{2}-\big(\omega_{\nu}(\mathbf{q})-i\eta\big)^{2}}.
 \label{eq:GF_A}
 \end{aligned}
\end{equation}
Thus, the actual expression of the phonon self-energy that  takes into account only retardation effects on the electronic response to the ionic motion, i.e.~non-adiabatic effects due to the electron-phonon interaction, is given by~\cite{Giustinorev}
\begin{equation}
 \tilde{\Pi}_{\nu}(\mathbf{q},\omega)=\Pi_{\nu}(\mathbf{q},\omega)-\Pi_{\nu}(\mathbf{q},0).
 \label{eq:ph_SE_NA}
\end{equation}
Plugging now  Eq.\,\ref{eq:GF_A} in Eq.\,\ref{eq:DE} and substituting $\Pi_{\nu}(\mathbf{q},\omega)$ with  $\tilde{\Pi}_{\nu}(\mathbf{q},\omega)$, one finds the expression for the retarded dressed phonon Green's function including non-adiabatic effects:
\begin{equation}
 \mathcal{D}_{\nu}(\mathbf{q},\omega)=\frac{2\omega_{\nu}(\mathbf{q})}{\omega^{2}-\omega^{2}_{\nu}(\mathbf{q})-2\omega_{\nu}(\mathbf{q})\tilde{\Pi}_{\nu}(\mathbf{q},\omega)}.
 \label{eq:GF}
\end{equation}
Within the Green's function formalism, the poles of the dressed phonon propagator in Eq.\,\ref{eq:GF} conform the actual vibrational quasi-particle spectrum of the system, i.e.~dressed phonons with renormalized frequencies and finite lifetimes due to many-body effects.
Thus, the solutions of the quasi-particle equation are defined as $\omega^{2}-\omega^{2}_{\nu}(\mathbf{q})-2\omega_{\nu}(\mathbf{q})\tilde{\Pi}_{\nu}(\mathbf{q},\omega)=0$.
At this point, it is essential to realize that although $\omega$ has been so far considered as a purely real excitation frequency, the Dyson's equation should in principle be solved for the entire complex plane, where the complex poles, $z_{\nu}(\mathbf{q})$, describe the phonon quasi-particle excitation frequencies, $\Omega_{\nu}(\mathbf{q})$, and  linewidths, $\gamma_{\nu}(\mathbf{q})$, in a unified way, $z_{\nu}(\mathbf{q})=\Omega_{\nu}(\mathbf{q})-i\gamma_{\nu}(\mathbf{q})$.
This is so because the phonon self-energy is itself a complex function of complex frequency argument.
The complex extension of the Dyson equation into the whole complex plane leads to two coupled non-linear equations for the renormalized phonon frecuencies and linewidths~\cite{Giustinorev}:
\begin{equation}
\begin{aligned}
 \Omega^{2}_{\nu}(\mathbf{q})=&\omega_{\nu}^{2}(\mathbf{q})+\gamma^{2}_{\nu}(\mathbf{q})+ \\ 
  &2\omega_{\nu}(\mathbf{q})\Re\tilde{\Pi}_{\nu}\big(\mathbf{q},\Omega_{\nu}(\mathbf{q})-i\gamma_{\nu}(\mathbf{q})\big) \\
 \gamma_{\nu}(\mathbf{q})=&-\frac{\omega_{\nu}(\mathbf{q})}{\Omega_{\nu}(\mathbf{q})}\Im\tilde{\Pi}_{\nu}\big(\mathbf{q},\Omega_{\nu}(\mathbf{q})-i\gamma_{\nu}(\mathbf{q})\big)
 \label{eq:z_NA}.
\end{aligned}
\end{equation}
Above, the non-linear coupling between the real and the imaginary parts of the quasi-particle poles appears explicitly obvious.
Similar to the case of electrons~\cite{Asierprl2008,Asierprb2009}, the main drawback for solving the complex phonon Dyson's equation in Eq.\,\ref{eq:z_NA} consists on extending the phonon self-energy into the whole complex frequency plane, i.e.~to consider the latter as as a complex function of complex variable.
In this respect, following Ref.\,\cite{Asierprl2008,Asierprb2009,HEDIN19701,farid}, one can recover a mathematically meaningful Dyson's equation in the whole complex plane by first replacing $(\omega\to z)$ in the phonon self-energy for the upper half complex plane, and then analytically continuing the obtained $\tilde{\Pi}_{\nu}(\mathbf{q},z)$ from the upper to the lower half complex plane.

The role of the complex frequency plane in Eq.\,\ref{eq:z_NA} has been traditionally neglected in \textit{ab initio} calculations, assuming that $\gamma_{\nu}(\mathbf{q})\ll\Omega_{\nu}(\mathbf{q})$ and $|\Im\tilde{\Pi}_{\nu}(\mathbf{q},\omega)|\ll\Re\tilde{\Pi}_{\nu}(\mathbf{q},\omega)$.
Therefore, considering the analytical continuation of the phonon self-energy is not the standard treatment.
Following the above approximations, the phonon quasi-particle frequencies may be approximated by solving the phonon Dyson's equation (Eq.\,\ref{eq:z_NA}) only along the real frequency axis, i.e.~the so-called Brillouin-Wigner perturbation theory~\cite{mahan}, as in Ref.\,\cite{NAKAgiustino,capeluttimgb2,calandramauriprb2010}, finding $\big(\Omega^{\mathrm{BW}}_{\nu}(\mathbf{q})\big)^{2}\approx\omega^{2}_{\mathrm{\nu}}(\mathbf{q})+2\omega_{\nu}(\mathbf{q})\mathrm{Re}\tilde{\Pi}_{\nu}\big(\mathbf{q},\Omega^{\mathrm{BW}}_{\mathrm{\nu}}(\mathbf{q})\big)$.
An even more drastic approximation, but most frequently considered in literature, is to consider the non-selfconsistent version of the Dyson equation as in Ref.\,\cite{NAKohnGraph,failureBOgraph,NAcarbnanotube,Saittaprl2008,calandramgb2}, $\big(\Omega^{\mathrm{RS}}_{\nu}(\mathbf{q})\big)^{2}\approx\omega^{2}_{\mathrm{\nu}}(\mathbf{q})+2\omega_{\nu}(\mathbf{q})\mathrm{Re}\tilde{\Pi}_{\nu}\big(\mathbf{q},\omega_{\mathrm{\nu}}(\mathbf{q})\big)$, which is equivalent to the Rayleigh-Schr\"odinger perturbation theory~\cite{mahan}.

In this work, we compute the retarded dressed phonon Green's function in Eq.\,\ref{eq:GF} by means of the first-principles calculations of the retarded phonon self-energy in Eq.\,\ref{eq:ph_SE_NA}.
In Sec.\,\ref{sec:B}, we present the calculated phonon spectral functions.
In Sec.\,\ref{sec:C}, we consider an approximate numerical procedure to obtain the complex phonon quasi-particle poles in the entire complex frequency plane, and discuss the possible limitations of the above standard procedures.
\section{COMPUTATIONAL METHODS}\label{sec:computational_details}
All self-consistent first-principles calculations were performed using the noncollinear DFT~\cite{DFT1,DFT2} and DFPT~\cite{DFPT} with fully relativistic norm-conserving pseudopotentials~\cite{ncpp,relncpp} as implemented in the \textsc{Quantum Espresso} package~\cite{QE}.
The Perdew-Zunger local density approximation was adopted in order to describe the exchange-correlation potential~\cite{PZLDA}.
For electronic calculations, we used a $32\times32$ $k$-mesh in combination with a Gaussian smearing of $5~\text{mRy}$ and a plane-wave basis with a cutoff of $80~\text{Ry}$.
Lattice vibrational properties were evaluated based on the calculation of the dynamical matrices on a coarse $8\times8$ $q$-mesh~\footnote{The dynamical matrices were primarily calculated with a smearing value of $5~\text{mRy}=68~\text{meV}$, an energy comparable to remarkable changes in the topology of the FS upon doping. This high value can lead to an unreal smoothing of the FS that can mask interesting phenomena as Kohn anomalies. A calculation with a smearing value of $5~\text{meV}$ on a finer $720\times720$ $k$-mesh was hence performed for each dynamical matrix of the coarse $q$-mesh (see the supplemental Note\,S2).}.
Carrier doping effects were self-consistently taken into account by the addition of excess electronic charge into the unit cell system, compensated by a uniform positive jellium background.
Electron-phonon matrix elements were calculated considering doping-sensitive full-spinor electron states, phonon states and deformation potentials, on coarse $16\times16$ and $8\times8$ $k$ and $q$-meshes for electrons and phonons, respectively.
The computation of all electron-phonon related magnitudes was carried out through fine integrals over the 1BZ, using Wannier scheme interpolated matrix elements~\cite{Asierwannier,epw,giustinoborondopeddiamond}.
In particular, the 1BZ summations of the retarded phonon self-energy in Eq.\,\ref{eq:ph_SE} are computed using a mesh of $1800\times1800$ $k$-points with a broadening parameter of $\eta=1~\text{meV}$ and a smearing energy of $5~\text{meV}$ equivalent to a temperature of $58~\text{K}$ included via the FD occupation factors.

The monolayer MoS$_{2}$ consists of a hexagonal close-packed structure of two planes of sulfur (S) atoms with an intercalated plane of molybdenum (Mo) atoms, bounded in a trigonal prismatic arrangement.
We considered the in-plane lattice parameter equal to the experimental bulk value $a=3.16$~{\AA}~\cite{aexp}, and a height of five times $a$, which provides a large enough vacuum in order to avoid any interplay between adjacent sheets.
In all cases, the equilibrium cell geometry was determined keeping the lattice parameter fixed to $a$, and relaxing all atomic forces up to at least $10^{-6}~\text{Ry/a.u.}$.
For the undoped monolayer, the relaxed interplanar distance is equal to $1.56$~{\AA}, equivalent to a Mo-S bond length of $2.40$~{\AA}.
\begin{figure}[t]
 \centering
 \begin{center}
  \includegraphics[width=1\columnwidth,angle=0,scale=1.0]{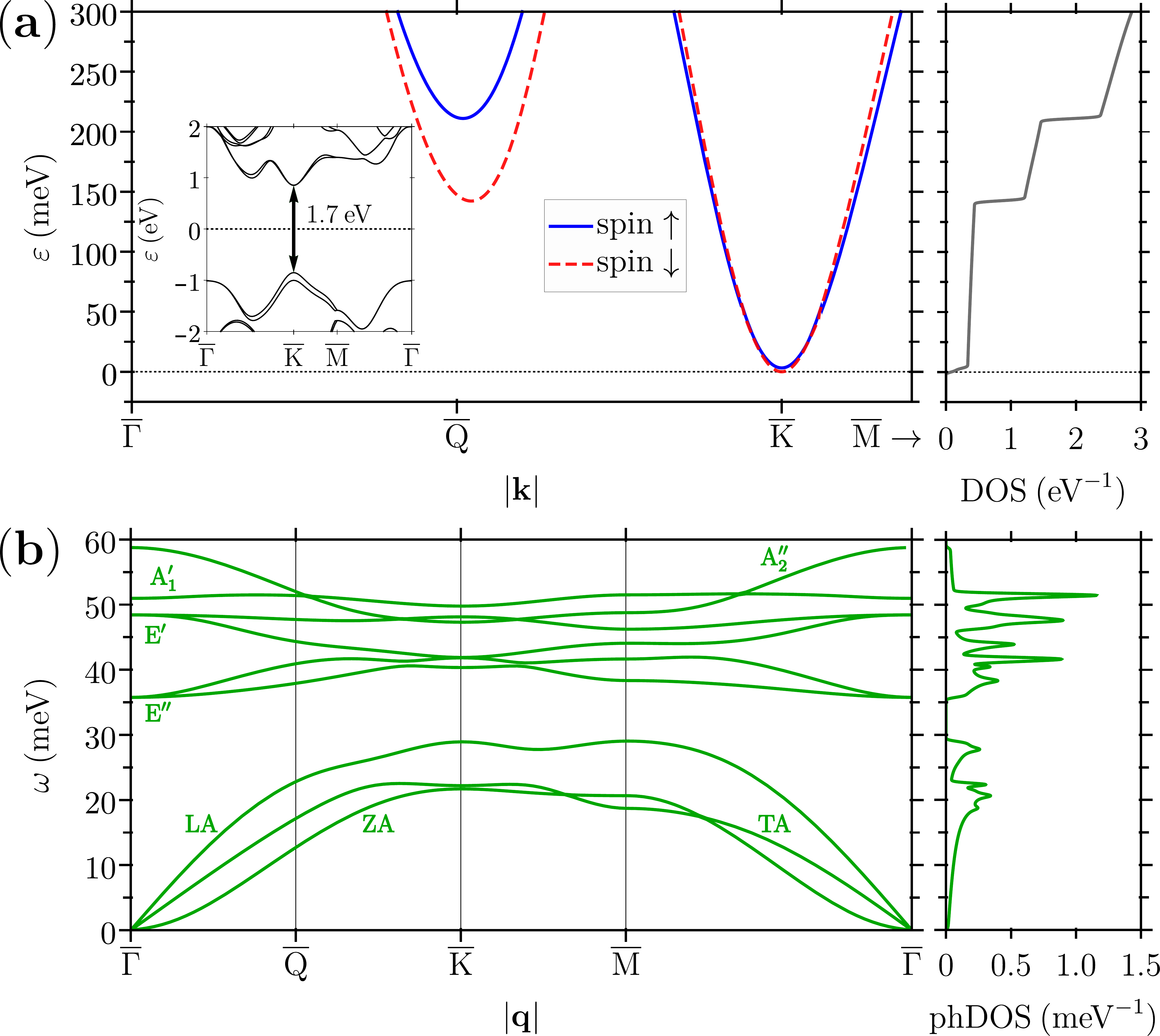}
 \end{center}
 \caption{(a) Electron conduction-band structure (left) and the corresponding DOS (right) of the undoped monolayer MoS$_{2}$ on the meV range of energy. Solid blue and dashed red lines represent opposite out-of-plane spin-polarized bands. The inset shows the band structure on the eV range centered on the semiconducting direct gap at the high-symmetry point $\overline{\mathrm{K}}(\overline{\mathrm{K'}})$, with an energy of about $1.7~\text{eV}$. (b) Phonon dispersion relation (left) and the corresponding phDOS (right) of the undoped monolayer MoS$_{2}$.}
 \label{fig:undoped}
\end{figure}
\begin{figure*}[t]
 \centering
 \begin{center}
  \includegraphics[width=1\textwidth,angle=0,scale=1.0]{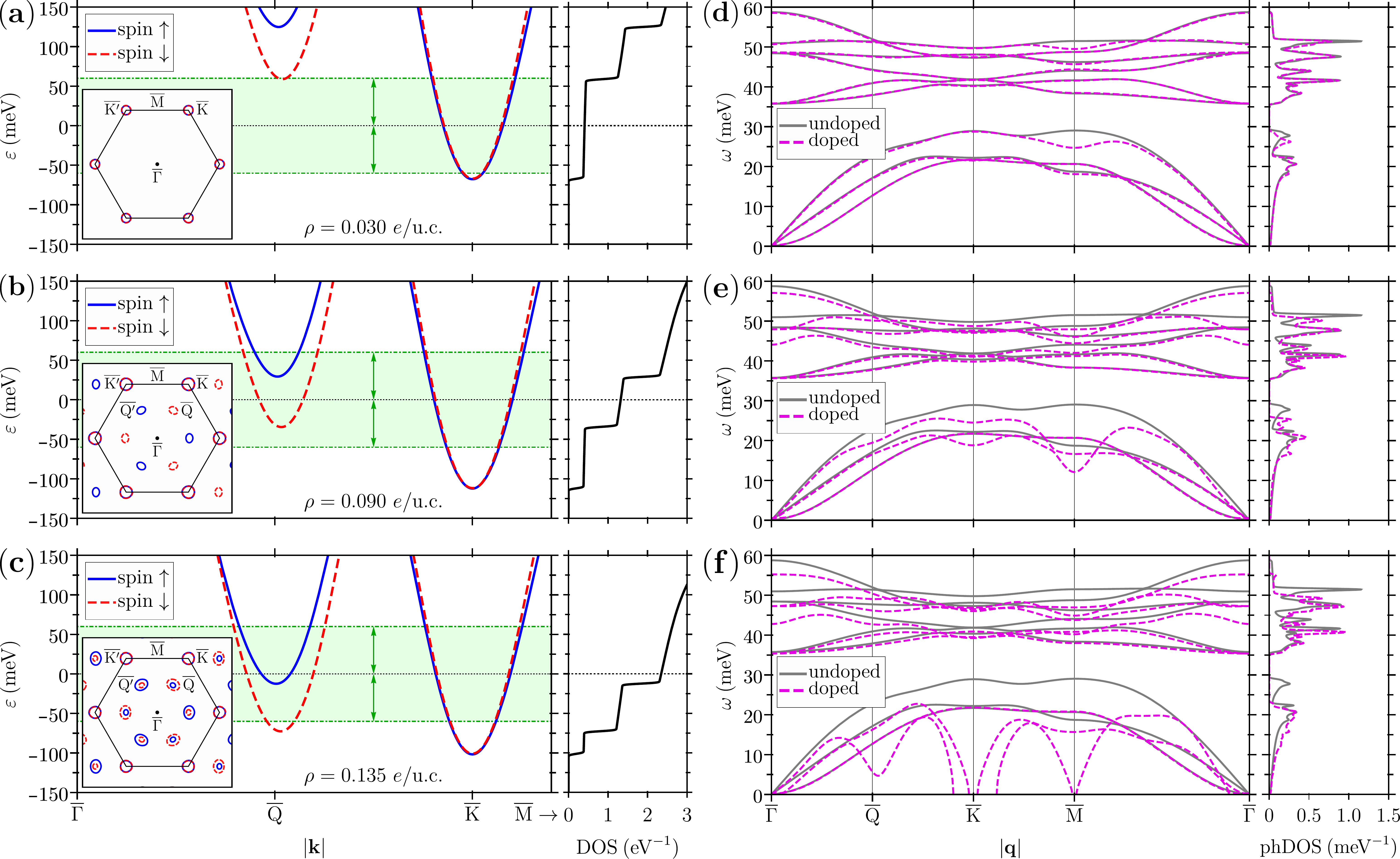}
 \end{center}
 \caption{Electron conduction-band structure (left) and its corresponding DOS (right) of the monolayer MoS$_{2}$ for some representative electron-doping concentrations $\rho=0.03$ (a), $0.09$ (b) and $0.135$~\textit{e}/u.c.~(c). The inset in left panels show the FS countour. Solid blue and dashed red lines represent opposite out-of-plane spin-polarized bands. The  Fermi level is set to zero (horizontal dotted black line). Horizontal dashed-dotted green lines delimit the energy window (shaded green area) within which an electron-hole pair can be excited (relaxed) by the decay (emission) of a phonon with maximum frequency $\omega_{\text{max}}=60~\text{meV}$. Phonon dispersion relation (left) and its corresponding phDOS (right) of the monolayer MoS$_{2}$ for the representative carrier concentrations $\rho=0.03$ (d), $0.09$ (e) and $0.135$~\textit{e}/u.c.~(f). Solid gray and dashed magenta lines represent the undoped and doping-dependent phonon branches, respectively.}
 \label{fig:bands_vs_doping}
\end{figure*}
\section{RESULTS AND DISCUSSION}\label{sec:results_discussion}
\subsection{Electronic and vibrational properties of the monolayer MoS$\mathbf{_{2}}$ within the adiabatic approximation}\label{sec:A}
We start this section by briefly presenting the electronic and vibrational structures of the undoped monolayer MoS$_{2}$ within the adiabatic approximation.
Figure\,\ref{fig:undoped}(a) shows the electron conduction-band structure (left) and its corresponding density of states (DOS) (right) on the meV range of energy, while the inset in the left panel shows the band structure on the eV range centered on the semiconducting direct band-gap.
The solid blue and dashed red lines represent the opposite out-of-plane spin-polarized conduction bands, due to the pure 2D nature of the system~\footnote{Values of the SO-splitting are strongly dependent on the orbital character of electron states. In the case of a monolayer, the SO coupling term is expected to be large for in-plane polarized states, as in the $\overline{\mathrm{Q}}(\overline{\mathrm{Q'}})$ valleys, which are a combination of Mo $d_{xy/x^{2}-y^{2}}$ and S $p$ orbitals. On the contrary, SO interaction vanishes for out-of-plane polarized states, as in the $\overline{\mathrm{K}}(\overline{\mathrm{K'}})$ valleys, which have a marked Mo $d_{z^{2}}$ orbital character.}.
An energy minimum is predicted at the $\overline{\mathrm{Q}}(\overline{\mathrm{Q'}})$ points, about $140~\text{meV}$ higher in energy than the main minimum at the $\overline{\mathrm{K}}(\overline{\mathrm{K'}})$ points.
Furthermore, while the $\overline{\mathrm{K}}(\overline{\mathrm{K'}})$ valleys are almost spin-degenerated, spin-orbit (SO) interaction induces an energy splitting of $\Delta_{\text{SO}}\sim80~\text{meV}$ at the $\overline{\mathrm{Q}}(\overline{\mathrm{Q'}})$ band edges.
Figure\,\ref{fig:undoped}(b) shows the adiabatic phonon dispersion (left) and its related density of states (phDOS) (right) of the undoped monolayer MoS$_{2}$.
The in-plane vibrating longitudinal (LA) and transverse (TA) acoustic normal modes disperse linearly near the $\overline{\Gamma}$ point, while the out-of-plane acoustic (ZA) branch exhibits a parabolic dispersion.
The $\mathrm{E''}$ and $\mathrm{E'}$ vibrational branches correspond to two pairs of in-plane longitudinal (LO) and transverse (TO) optical modes, degenerated at the $\overline{\Gamma}$ point with an energy of $36~\text{meV}$ and $48~\text{meV}$, respectively.
While $\mathrm{E''}$ modes correspond to only S atoms vibrating in counterphase, Mo and S atoms vibrate in counterphase in the $\mathrm{E'}$ branches.
The $\mathrm{A'_{1}}$ mode corresponds to the almost dispersionless optical branch at $51~\text{meV}$ at the $\overline{\Gamma}$ point, with S atoms vibrating out-of-plane in counterphase.
Finally, the $\mathrm{A''_{2}}$ mode corresponds to the highest frequency optical mode with an energy of $59~\text{meV}$ at the $\overline{\Gamma}$ point, with Mo atom and S atoms vibrating out-of-plane in counterphase.
Our calculations are in excellent agreement with previous theoretical and experimental results~\cite{1lmos2expgap,1lmos2theogap,1lmos2elbandtheo,kaasbjerg}.

Henceforth, we focus on the influence of the electron-doping on both the electronic and vibrational properties of the monolayer MoS$_{2}$, which are in fact necessary before addressing our later analysis on the electron-phonon induced non-adiabatic corrections to the phonon spectrum.
It is also a valuable examination, as it allows to do a first scan of normal modes coupled to electrons by exploring for instance the presence of frequency softenings.
Figures\,\ref{fig:bands_vs_doping}(a)-(c) present the conduction-band structure of the doped monolayer MoS$_{2}$ (left) and its related DOS (right) for three representative carrier doping concentrations $\rho=0.03$, $0.09$ and $0.135~e/\text{u.c.}$, respectively~\footnote{The conduction band and adiabatic phonon dispersions of all the doping levels considered in this work can be found in Fig.\,S1 and S2 of the supplemental Note\,S4.}.
As before, the solid blue and dashed red lines represent the opposite out-of-plane spin-polarized conduction bands.
The dashed-dotted green lines delimit the shaded green area that represents the energy window within which an electron-hole pair can be energetically excited by the emission or absorption of a phonon.
Additionaly, the inset in the left panels shows the corresponding Fermi contour for each doping level.
We can easily distinguish three different doping regimes:
the ``small'' doping regime for $\rho\leqslant0.06~e/\text{u.c.}$, where only the $\overline{\mathrm{K}}(\overline{\mathrm{K'}})$ valleys are occupied (see Fig.\,\ref{fig:bands_vs_doping}(a) and supplemental Fig.\,S1(a)-(d));
the ``intermediate'' doping regime for $0.06~e/\text{u.c}\leqslant\rho\leqslant 0.120~e/\text{u.c}$, where the lower spin-split $\overline{\mathrm{Q}}(\overline{\mathrm{Q'}})$ valleys get also populated (see Fig.\,\ref{fig:bands_vs_doping}(b) and supplemental Fig.\,S1(e)-(h));
and the ``large'' doping regime for $\rho\geqslant0.120~e/\text{u.c.}$, where the upper spin-split $\overline{\mathrm{Q}}(\overline{\mathrm{Q'}})$ states are finally occupied (see Fig.\,\ref{fig:bands_vs_doping}(c) and supplemental Fig.\,S1(i)-(j)).
\begin{figure*}[t]
 \centering
 \begin{center}
  \includegraphics[width=1\textwidth,angle=0,scale=1.0]{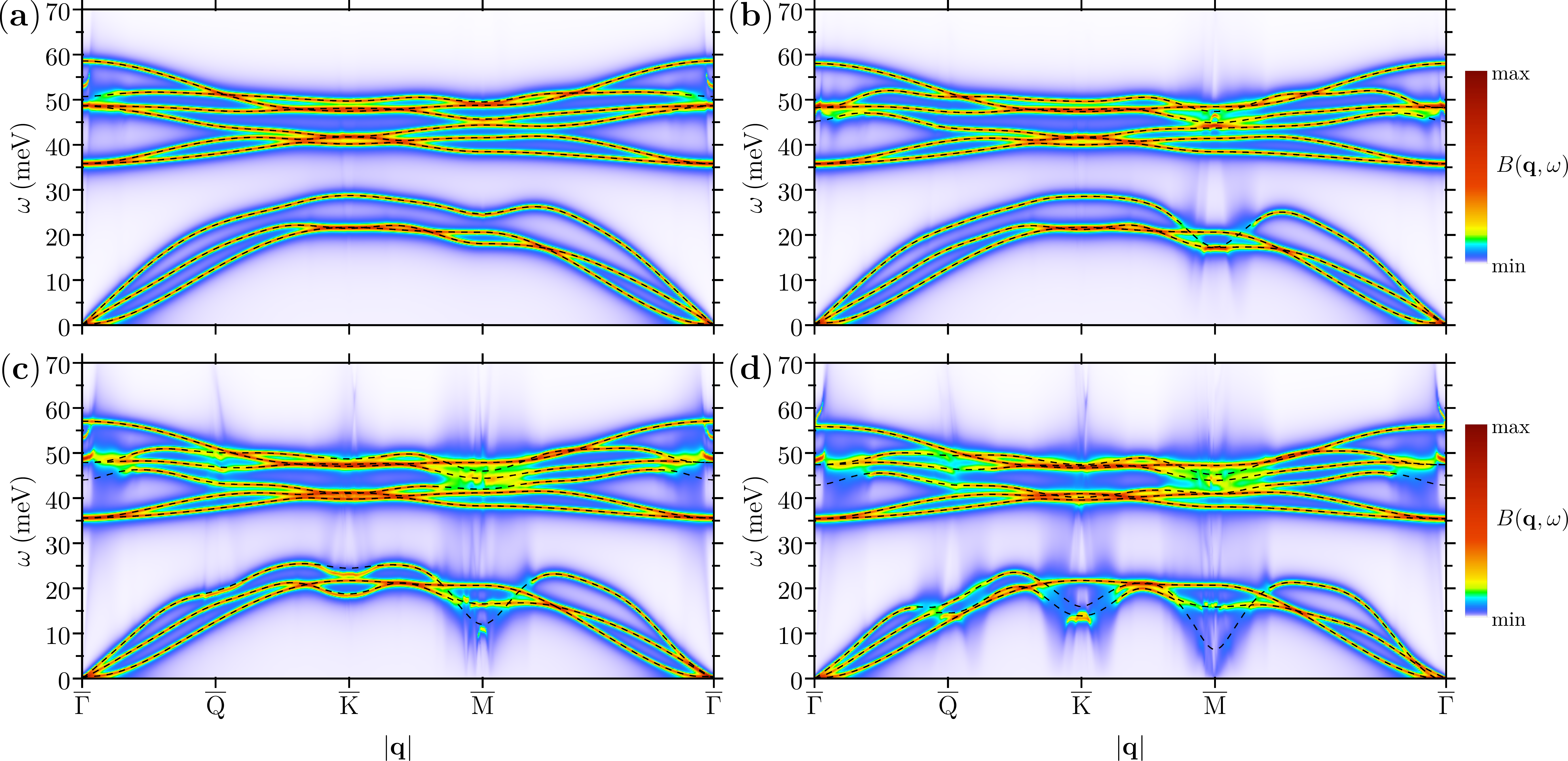}
 \end{center}
 \caption{Density plot of the phonon spectral function of the monolayer MoS$_{2}$ for some representative electron-doping concentrations $\rho=0.03$ (a), $0.06$ (b), $0.09$ (c) and $0.12$~\textit{e}/u.c.~(d). The color code scale represents the height of the spectral function. Dashed black lines represent the adiabatic dipsersions. In those areas where $\Im\tilde{\Pi}(\mathbf{q},\omega)=0$, we use a finite broadening of $\eta=0.35~\text{meV}$ in order to appreciate the different phonon peaks.}
 \label{fig:naeffects}
\end{figure*}
The high doping-sensitivity of the energy difference between the $\overline{\mathrm{K}}(\overline{\mathrm{K'}})$ and $\overline{\mathrm{Q}}(\overline{\mathrm{Q'}})$ minima highlights the importance of self-consistently incorporating the impact of doping on the vibrational calculations, where the correct band structure is necessary in order to appropiately account for the electronic static-screening (see Eq.\,\ref{eq:DFPT}).
Figure\,\ref{fig:bands_vs_doping}(d)-(f) compare the undoped (solid gray lines) and doped (dashed magenta lines) adiabatic phonon dipersion relations of the monolayer MoS$_{2}$ (left) and their corresponding phDOS (right) for the same three carrier concentrations as in Fig.\,\ref{fig:bands_vs_doping}(a)-(c).
In the small doping regime, significant frequency dips of $\sim5~\text{meV}$ are reported for the $\mathrm{A'_{1}}$ and LA branches at $\mathbf{q}=\overline{\Gamma}$ and $\overline{\mathrm{M}}$, that sink even more as doping increases (see Fig.\,\ref{fig:bands_vs_doping}(d) and supplemental Fig.\,S2(a)-(d)).
This clearly indicates that these normal modes are coupled to carrier states at $\overline{\mathrm{K}}(\overline{\mathrm{K'}})$ and $\overline{\mathrm{Q}}(\overline{\mathrm{Q'}})$~\footnote{The $\mathrm{A'_{1}}$ mode at $\mathbf{q}=\overline{\Gamma}$ and carrier states at $\mathbf{k}=\overline{\mathrm{K}}(\overline{\mathrm{K'}})$ valleys involve out-of-plane polarized large deformation potentials and orbitals at the center of the MoS$_{2}$ layer, respectively, that couple efficiently in Eq.\,\ref{eq:epme} leading to large matrix elements. Likewise, in the $\mathrm{A'_{1}}$ and LA modes at $\mathbf{q}\approx\overline{\mathrm{M}}$, the additional in-plane displacement of the Mo atoms also allows to couple with electron states of $\overline{\mathrm{Q}}(\overline{\mathrm{Q'}})$ valleys, with a marked in-plane Mo orbital character.}, and are hence effectively static-screened at this level by means of spin-conserving $\overline{\mathrm{K}}(\overline{\mathrm{K'}})$ intra-valley and $\overline{\mathrm{K}}\leftrightarrow\overline{\mathrm{Q'}}(\overline{\mathrm{K'}}\leftrightarrow\overline{\mathrm{Q}})$ inter-valley electron-hole pairs (see Fig.\,\ref{fig:bands_vs_doping}(a)).
At intermediate doping concentrations, however, frequency softenings also appear for different branches at $\mathbf{q}=\overline{\mathrm{Q}}(\overline{\mathrm{Q'}})$ and $\overline{\mathrm{K}}(\overline{\mathrm{K'}})$ (see Fig.\,\ref{fig:bands_vs_doping}(e) and supplemental Fig.\,S2(e)-(h)).
Besides, the previously observed Kohn anomalies for the LA and $\mathrm{A'_{1}}$ modes at $\mathbf{q}=\overline{\Gamma}$ and $\overline{\mathrm{M}}$ are also intensified, the former exhibiting impressively large softening values larger than $10~\text{meV}$.
This directly results from the enrichment of the FS topology, which gives rise to additional spin-conserving scattering channels connecting electron and hole states at the $\overline{\mathrm{K}}(\overline{\mathrm{K'}})$ and/or $\overline{\mathrm{Q}}(\overline{\mathrm{Q'}})$ valleys (see Supplemental Material of Ref.\cite{gurepaper}), screening the lattice vibrations themselves through the electron-phonon coupling.
Finally, within the large doping regime, the in-plane acoustic Kohn anomalies develop instabilities at the $\mathbf{q}=\overline{\mathrm{K}}(\overline{\mathrm{K'}})$ and $\overline{\mathrm{M}}$ points, and also at $\mathbf{q}=\overline{\mathrm{Q}}(\overline{\mathrm{Q'}})$ as doping increases (see Fig.\,\ref{fig:bands_vs_doping}(f) and supplemental Fig.\,S2(i)-(j)).
In this case, all the spin-polarized valleys are occupied, the number of FS-nesting channels being maximum.
Hence, the static screening induced by the electron-phonon coupling   is so large that a lattice phase transition is energetically favorable~\cite{rosnermos2}.

It is worth noting that the ZA, $\mathrm{E''}$ and $\mathrm{A''_{2}}$ normal modes are practically insensitive to doping through all BZ, as expected from symmetry arguments~\cite{kaasbjerg}.
Thus, we exclude them from our following discussion.
\subsection{Signatures of non-adiabatic corrections to the phonon spectral functions of the doped monolayer MoS$\mathbf{_{2}}$ due to the electron-phonon coupling}\label{sec:B} 
In this section, we present the first-principles phonon spectral function of the doped monolayer MoS$\mathbf{_{2}}$ including non-adiabatic renormalizations due to electron-phonon interactions.
The phonon spectral function is defined as~\cite{mahan}:
\begin{equation}
 B(\mathbf{q},\omega)=-\frac{1}{\pi}\sum_{\nu}\Im\mathcal{D}_{\nu}(\mathbf{q},\omega),
 \label{eq:ph_SF}
\end{equation}
where the retarded dressed phonon Green's function is constructed as in Eq.\,\ref{eq:GF}.
Physically, the phonon spectral function describes the probability density of phonon states in the momentum-energy $(\mathbf{q},\omega)$ space, and is experimentally accessible by means of several techniques~\cite{maksimovphSF}.

Figure\,\ref{fig:naeffects}(a)-(d) shows the calculated phonon spectral functions of the monolayer MoS$_{2}$ for the representative carrier concentrations within the small doping regime, $\rho=0.03$ and $0.06~e/\text{u.c.}$, and within the intermediate doping regime, $\rho=0.09$ and $0.12~e/\text{u.c.}$, respectively~\footnote{The phonon spectral function of all the doping levels considered in this work can be found in Fig.\,S3 of the supplemental Note\,S5.}.
The large doping regime is not shown, since the lattice becomes unstable (see Fig.\,\ref{fig:bands_vs_doping}(f)).
The dashed black lines show the corresponding adiabatic phonon dispersions.
The differences between the adiabatic branches and the main features defined by the spectral function allow to appreciate non-adiabatic corrections.

In the following, we will concentrate on studying more specifically the large and small momentum regimes, $\mathbf{q}\gg\overline{\Gamma}$ and $\mathbf{q}\approx\overline{\Gamma}$, respectively, as this will allow us to rationalize the
results with a simple model theoretical treatment.
\begin{figure}[t]
 \centering
 \begin{center}
  \includegraphics[width=1\columnwidth,angle=0,scale=1.0]{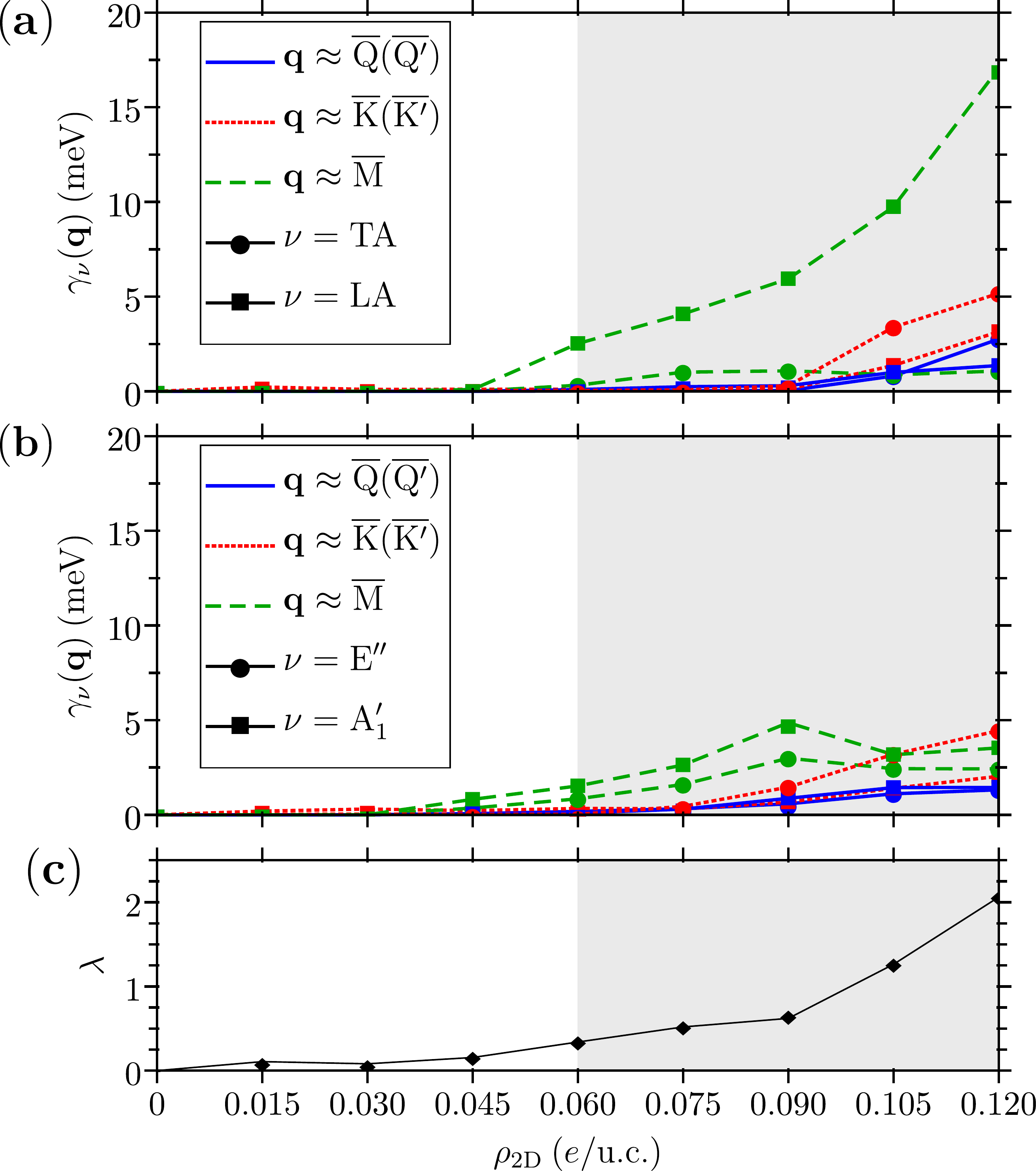}
 \end{center}
 \caption{The phonon lnewidth $\gamma_{\nu}(\mathbf{q})$ as a function of electron-doping for (a) acoustic and (b) optical vibrational modes at momenta $\mathbf{q}\approx\overline{\mathrm{Q}}(\overline{\mathrm{Q'}})$, $\overline{\mathrm{K}}(\overline{\mathrm{K'}})$ and $\overline{\mathrm{M}}$, represented by solid blue, dotted red and dashed green lines. In panel (a), circles and squares represent the acoustic TA and LA normal modes, respectively, while, in panel (b), the optical $\mathrm{E'}$ and $\mathrm{A'_{1}}$ normal modes, respectively. The white and shaded gray areas correspond to the small and intermediate doping regime, respectively. (c) The electron-phonon coupling strength, $\lambda$, as a function of doping.}
 \label{fig:imseinter}
\end{figure}
\subsubsection{Large momentum regime $(\mathbf{q}\gg\overline{\Gamma})$}
By the large momentum regime we refer to the set of wave vectors away from $\overline{\Gamma}$ that take part in the phonon-mediated inter-valley electronic scattering.
This is governed by the doping-dependent topology of the FS, as already noted in Sec.\,\ref{sec:A}, and comprises momenta near $\mathbf{q}\approx\overline{\mathrm{Q}}(\overline{\mathrm{Q'}})$,  $\overline{\mathrm{K}}(\overline{\mathrm{K'}})$ and $\overline{\mathrm{M}}$.

In Fig.\,\ref{fig:naeffects}, we show that the non-adiabatic corrections do not change the adiabatic vibrational structure within the large momentum regime, and, to a large extent, the renormalized phonon frequencies follow the adiabatic ones, i.e.~$\Omega_{\nu}(\mathbf{q})\approx\omega_{\nu}(\mathbf{q})$.
Thus, the only significant spectral feature related to the electron-phonon coupling is the broadening of the phonon spectra.
Given that in this case the phonon energy is much bigger than its lifetime broadening, we can deduce from Eq.\,\ref{eq:z_NA} that the phonon linewidth can be taken approximately equal to the imaginary part of the phonon self-energy evaluated at the adiabatic frequency itself, i.e.~$\gamma_{\nu}(\mathbf{q})\approx\big|\Im\tilde{\Pi}_{\nu}\big(\mathbf{q},\omega_{\nu}(\mathbf{q})\big)\big|$.
From the imaginary part of Eq.\,\ref{eq:ph_SE_NA}, this is given by
\begin{equation}
 \begin{aligned}
  \gamma_{\nu}(\mathbf{q})\approx&\frac{\pi}{N_\mathbf{k}}\sum_{\mathbf{k}}^{\mathrm{1BZ}}\sum_{mn}\big|g_{mn}^{\nu}(\mathbf{k,q})\big|^{2}\times \\
  &\big(f(\varepsilon^{\mathbf{k}}_{n})-f(\varepsilon^{\mathbf{k+q}}_{m})\big)\delta\big(\varepsilon^{\mathbf{k}}_{n}-\varepsilon^{\mathbf{k+q}}_{m}+\omega_{\nu}(\mathbf{q})\big),
  \label{eq:ph_SE_NA_im}
 \end{aligned}
\end{equation}
which is the same as one would obtain by Fermi's golden rule or the Rayleigh-Schr\"odinger-like scheme mentioned in Sec.\,\ref{sec:theory}.
In Fig.\,\ref{fig:imseinter}(a) and (b), we represent the calculated values of $\gamma_{\nu}(\mathbf{q})$ as a function of doping for the set of the acoustic TA and LA and the optical E$''$ and A$'_1$ phonon modes, respectively, evaluated at the momenta $\mathbf{q}\approx\overline{\mathrm{Q}}(\overline{\mathrm{Q'}})$ (solid blue lines), $\overline{\mathrm{K}}(\overline{\mathrm{K'}})$ (dotted red lines) and $\overline{\mathrm{M}}$ (dashed green lines).
At small doping concentrations (white area), only the interacting LA and $\mathrm{A'_{1}}$ normal modes (squares) with momentum $\mathbf{q}\approx\overline{\mathrm{M}}$ exhibit a weak but appreciable broadening of the linewidth.
This is so because in this regime only these specific normal modes are allowed to
decay by exciting spin-conserving $\overline{\mathrm{K}}\leftrightarrow\overline{\mathrm{Q'}}$ ($\overline{\mathrm{K'}}\leftrightarrow\overline{\mathrm{Q}}$) inter-valley electron-hole pairs, which are energetically possible via the actual phonon structure for doping concentrations $\rho\geqslant0.045~e/\text{u.c.}$ (see red dashed lines within the green shaded area in supplemental Fig.\,S1).

At intermediate doping concentrations (gray area), the additional energetically available electron-hole pair channels (see Fig.\,\ref{fig:bands_vs_doping}(b)) lead to the effective decay of normal modes also at $\mathbf{q}\approx\overline{\mathrm{Q}}(\overline{\mathrm{Q'}})$ and $\overline{\mathrm{K}}(\overline{\mathrm{K'}})$ by acquiring an appreciable finite linewidth broadening (see Fig.\,\ref{fig:naeffects}(c) and (d) and supplemental Fig.\,S3(e)-(h)).
Likewise, the broadening of the above discussed phonon peaks at $\mathbf{q}\approx\overline{\mathrm{M}}$ is also enhanced.
It is particularly interesting that the LA phonon modes exhibit broadening values as large as three times that of the other modes.
Of course, the phonon linewidth is intimately related to the electron-phonon coupling strength as~\cite{Allenprb1972}
\begin{equation}
 \lambda=\frac{2}{\pi N_{E_{\mathrm{F}}}N_{\mathbf{q}}}\sum^{\mathrm{1BZ}}_{\mathbf{q}\nu}\frac{\gamma_{\nu}(\mathbf{q})}{\omega^{2}_{\nu}(\mathbf{q})},
\end{equation}
where $N_{\mathbf{q}}$ is the number of $\mathbf{q}$-points considered for the BZ and $N_{E_{\mathrm{F}}}$ is the DOS at the Fermi energy, $E_{\mathrm{F}}$.
Therefore, the enhancement of the linewidth broadening of the LA phonon mode close to $\mathbf{q}\approx\overline{\mathrm{M}}$ is directly connected to the enhancement of the electron-phonon coupling strength in the electron-doped monolayer MoS$_{2}$ from intermediate doping concentrations $(\rho\geqslant0.06~e/\text{u.c.}\approx7\times10^{13}~e/\text{cm}^{2})$, as shown in Fig.\,\ref{fig:imseinter}(c).
This is in agreement with the experimentally measured superconducting state appearing from this doping level~\cite{scmos2}, as well as to the $\overline{\mathrm{K}}(\overline{\mathrm{K'}})$-valley intricate electron photoemission spectrum~\cite{Kang2018}, whose genuine spectral features have been recently explained in terms of three elementary many-body carrier quasi-particle states~\cite{gurepaper}.
\subsubsection{Small momentum limit $(\mathbf{q}\to\overline{\Gamma})$}
\begin{figure*}[t]
 \centering
 \begin{center}
  \includegraphics[width=1\textwidth,angle=0,scale=1.0]{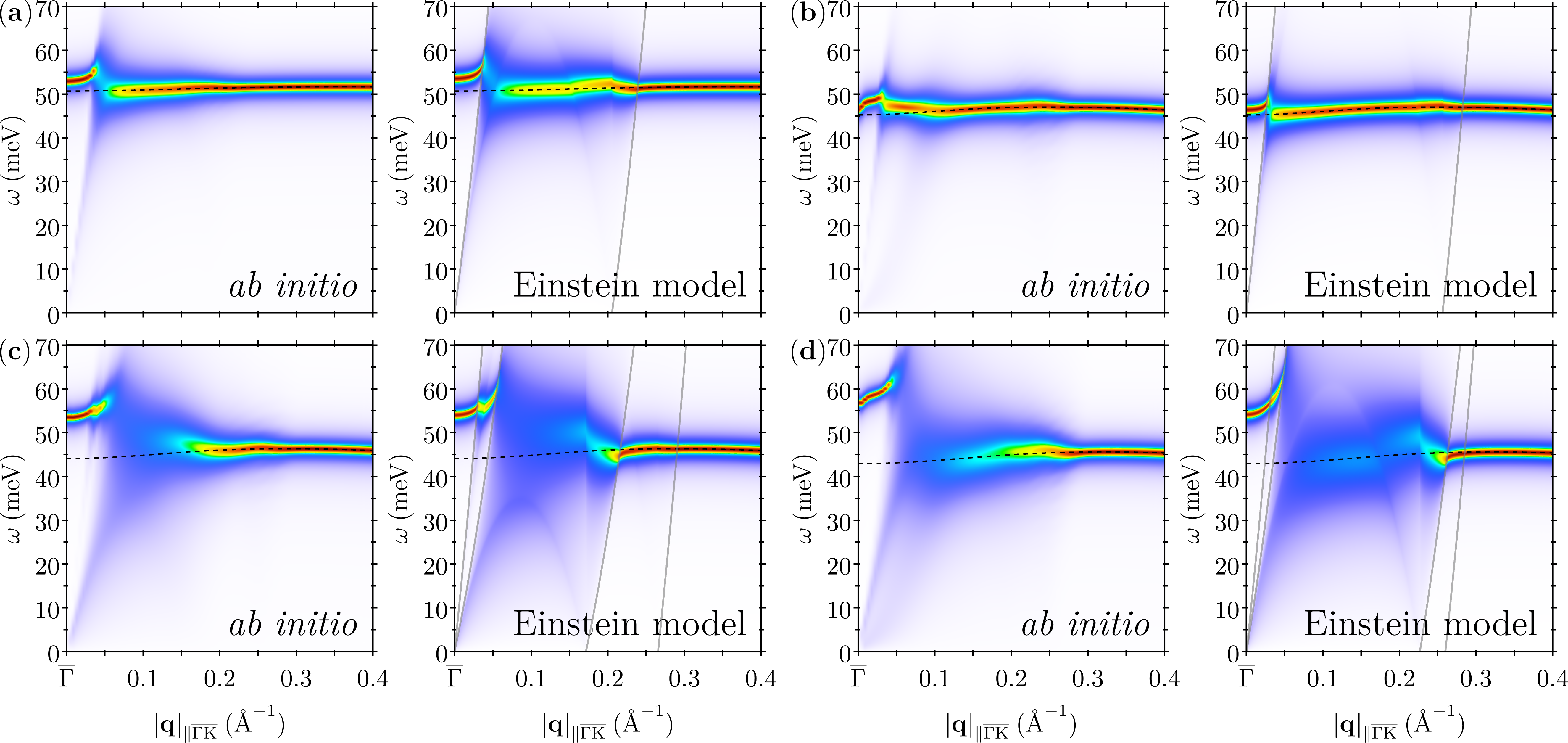}
 \end{center}
 \caption{Density plot of the spectral function of the monolayer MoS$_{2}$ for the $\mathrm{A'_{1}}$ phonon mode obtained by means of \textit{ab initio} calculations (left) and the Einstein-like model (right), for the same carrier concentrations as in Fig.\,\ref{fig:naeffects} and within the small momentum regime along the $\overline{\Gamma\mathrm{K}}$ direction. The color code scale represents the height of the phonon spectral function. Dashed black lines represent the adiabatic dispersion. Solid gray lines bound the electron-hole excitation damping continuum in the simplified model at zero temperature, delimited by: $(q^{2}+2qk_{\mathrm{F}}^{x})/2m^{*}_{x}\geqslant\omega\geqslant(q^{2}-2qk_{\mathrm{F}}^{x})/2m^{*}_{x}$, where $k_{\mathrm{F}}^{x}$ is the Fermi momentum for each $x=\overline{\mathrm{K}}$ (outer lines) and $\overline{\mathrm{Q}}$-like (inner lines) valley.}
 \label{fig:naeffectsabinitvsmodel}
\end{figure*}
In the small momentum limit, we focus on the phonon spectral function for wave vectors close to $\overline{\Gamma}$, that are the ones which take part in the phonon-mediated intra-valley electronic scattering.
We thus avoid disccussing the acoustic branches, since their interaction with electrons vanishes when $\mathbf{q}\to\overline{\Gamma}$~\cite{Grimvall}.

Unlike in the large momentum regime, in this case the non-adiabatic effects are not only limited to the broadening of the phonon linewidth, but also lead to a large hardening of vibrational frequencies at $\mathbf{q}=\overline{\Gamma}$, which is accompanied by an increasingly steeper dispersion of the branch, completely breaking the adiabatic picture~\cite{Ipatova,Maksimov1996}.
While at small doping concentrations only the out-of-plane polarized $\mathrm{A'_{1}}$ optical phonon branch exhibits significant spectral features in Fig.\,\ref{fig:naeffects}(a) and (b), at intermediate doping concentrations the in-plane polarized $\mathrm{E'}$ optical normal modes also display appreciable spectral  features in Fig.\,\ref{fig:naeffects}(c) and (d).
The latter appear rather unsensitive to doping, which is consistent with the fact that electron-phonon matrix elements do not vary with larger carrier concentrations.

Henceforth, we focus exclusively on the $\mathrm{A'_{1}}$ branch, which reveals the most interesting and complex evolution of the spectral features upon doping.
In order to get more insight about the physics of this intriguing case, we use a simple analytic model of the monolayer MoS$_{2}$ that reproduces the first-principles results (see the supplemental Note\,S3 for details).
We consider an Einstein-like model consisting on a 2D free electron gas interacting with one single optical vibrational branch with the same properties as the $\mathrm{A'_{1}}$ phonon mode.
The electron gas is composed of four equivalent $\overline{\mathrm{K}}$-like and six equivalent $\overline{\mathrm{Q}}$-like parabolic bands, one for each direction of the spin polarization and each valley (see Fig.\,\ref{fig:bands_vs_doping}(a)-(c)), with $m^{*}(\overline{\mathrm{K}})=0.60$ and $m^{*}(\overline{\mathrm{Q}})=0.80$ electron effective band masses, respectively.
Since the spin-split bands are oppositely spin-polarized, inter-band scattering can be safely neglected, and thereby, an analytical expression for the phonon self-energy of the coupled electron-phonon model is built directly from Eq.\,\ref{eq:ph_SE_NA}:
\begin{equation}
 \begin{aligned}
  \tilde{\Pi}_{\mathrm{A'_{1}}}(\mathbf{q},\omega)=4g^{2}_{\mathrm{A'_{1}}}(\overline{\mathrm{K}})\Big(\chi^{0,\overline{\mathrm{K}}}_{2\text{D}}(\mathbf{q},\omega)-\chi^{0,\overline{\mathrm{K}}}_{2\text{D}}(\mathbf{q},0)\Big)&+\\6g^{2}_{\mathrm{A'_{1}}}(\overline{\mathrm{Q}})\Big(\chi^{0,\overline{\mathrm{Q}}}_{2\text{D}}(\mathbf{q},\omega)-\chi^{0,\overline{\mathrm{Q}}}_{2\text{D}}(\mathbf{q},0)\Big)&,
 \end{aligned}
\end{equation}
where $\chi^{0,x}_{2\text{D}}(\mathbf{q},\omega)$ and $g_{\mathrm{A'_{1}}}(x)$ represent the 2D Lindhard function~\cite{lindhard2d} and the matrix element describing the coupling of the $\mathrm{A'_{1}}$ phonon mode of wave vector $\mathbf{q}=\overline{\Gamma}$ with electrons at the $x=\overline{\mathrm{K}}(\overline{\mathrm{K'}})$ and $\overline{\mathrm{Q}}(\overline{\mathrm{Q'}})$ points, respectively.
The latter are taken from first-principles calculations.
Table\,\ref{tab:abinitiovalues} gathers the values of the Fermi level as well as the matrix elements for all the considered carrier concentrations.

The color code in Fig.\,\ref{fig:naeffectsabinitvsmodel} represents the spectral function of the $\mathrm{A'_{1}}$ phonon mode obtained from \textit{ab initio} calculations (zoom in of Fig.\,\ref{fig:naeffects}) and using the Einstein-like model, both evaluated in the small momentum limit along the $\overline{\Gamma\mathrm{K}}$ direction for the same carrier concentrations as in Fig.\,\ref{fig:naeffects}~\footnote{The spectral function of the $\mathrm{A'_{1}}$ phonon mode obtained by \textit{ab initio} calculations and using the Einstein-like model, both evaluated in the small momentum limit along the $\overline{\Gamma\mathrm{K}}$ direction and for all the doping levels considered in this work can be found in Fig.\,S4 of the supplemental Note\,S5.}.
The good agreement between both vibrational spectra for all doping levels confirms that our analytic model contains the relevant physics of the non-adiabatic renormalizations due to the electron-phonon interaction.
We see that the intensity of the spectral effects decreases with larger carrier concentrations within the small doping regime, where only $\overline{\mathrm{K}}(\overline{\mathrm{K'}})$ intra-valley electron-hole scattering occurs (see Fig.\,\ref{fig:naeffectsabinitvsmodel}(a)-(b) and supplemental Fig.\,S4(a)-(d)).
This trend is explained by a screening induced reduction of the value of the matrix element $g_{\mathrm{A'_{1}}}(\overline{\mathrm{K}})$ upon charge carrier accumulation (see Table\,\ref{tab:abinitiovalues}).
Oppositely, the spectral effects are outstandingly enhanced as the doping concentration grows within the intermediate doping regime, where $\overline{\mathrm{Q}}(\overline{\mathrm{Q'}})$ intra-valley electronic transitions are also allowed (see Fig.\,\ref{fig:naeffectsabinitvsmodel}(c)-(d) and supplemental Fig.\,S4(e)-(h)).
In particular, at $\rho=0.12~e/\text{u.c.}$ in Fig.\,\ref{fig:naeffectsabinitvsmodel}(d), the renormalization of the adiabatic frequency $\omega_{\nu}(\mathbf{q})=43~\text{meV}$ results in a sharp phonon peak with maximum at frequency $\omega\approx57~\text{meV}$ at $\mathbf{q}=\overline{\Gamma}$ and $\omega\approx63~\text{meV}$ at $|\mathbf{q}|\approx0.05~\text{\AA}^{-1}$ in the $\overline{\Gamma\mathrm{K}}$ direction.
These values correspond to a frequency enhancement of $\sim33\%$ and $\sim46\%$, respectively, representing both of them the largest energy renormalization reported value so far in any material ($\sim30\%$)~\cite{Saittaprl2008}.
Indeed, at intermediate doping concentrations, $g_{\mathrm{A'_{1}}}(\overline{\mathrm{Q}})$ is larger than $g_{\mathrm{A'_{1}}}(\overline{\mathrm{K}})$ and increases with doping (see Table\,\ref{tab:abinitiovalues}), a behaviour that has been recently explained in terms of an electrostatic screening suppression caused by out-of-plane deformation potentials~\cite{sohierprx2019}.

\begin{table}[t]
\begin{center}
\caption{Ab-initio calculated parameters used in the Einstein-like model for the monolayer MoS$_{2}$. $E_{\mathrm{F}}^{\overline{\mathrm{K}}}$ and $E_{\mathrm{F}}^{\overline{\mathrm{Q}}}$ are the energies of the Fermi level with respect to the bottom of the occupied $\overline{\mathrm{K}}$ and $\overline{\mathrm{Q}}$ valleys, respectively. $g_{\mathrm{A'_{1}}}(\overline{\mathrm{K}})$ and $g_{\mathrm{A'_{1}}}(\overline{\mathrm{Q}})$ are the intra-band electron-phonon matrix elements of the $\mathrm{A'_{1}}$ phonon mode at $\mathbf{q}=\overline{\Gamma}$ interacting with electron states at strictly  $\mathbf{k}=\overline{\mathrm{K}}$ and $\mathbf{k}=\overline{\mathrm{Q}}$, respectively.}
\label{tab:abinitiovalues}
\begin{ruledtabular}
\begin{tabular}{*{5}{c}}
\multicolumn{1}{c|}{$\rho\text{ (\textit{e}/u.c.)}$} & \multicolumn{1}{c}{$E_{\mathrm{F}}^{\overline{\mathrm{K}}}\text{ (meV)}$} & \multicolumn{1}{c}{$E_{\mathrm{F}}^{\overline{\mathrm{Q}}}\text{ (meV)}$} & \multicolumn{1}{c}{$g_{\mathrm{A'_{1}}}(\overline{\mathrm{K}})\text{ (meV)}$} & \multicolumn{1}{c}{$g_{\mathrm{A'_{1}}}(\overline{\mathrm{Q}})\text{ (meV)}$} \\ \hline
\multicolumn{1}{c|}{0.000} &   - &  - & 87 & \multicolumn{1}{c}{66}   \\ 
\multicolumn{1}{c|}{0.015} &  28 &  - & 84 & \multicolumn{1}{c}{68}   \\ 
\multicolumn{1}{c|}{0.030} &  67 &  - & 82 & \multicolumn{1}{c}{69}   \\ 
\multicolumn{1}{c|}{0.045} &  91 &  - & 65 & \multicolumn{1}{c}{86}   \\ 
\multicolumn{1}{c|}{0.060} & 104 &  - & 52 & \multicolumn{1}{c}{98}   \\ 
\multicolumn{1}{c|}{0.075} & 110 & 18 & 46 & \multicolumn{1}{c}{104}  \\ 
\multicolumn{1}{c|}{0.090} & 112 & 35 & 41 & \multicolumn{1}{c}{109}  \\ 
\multicolumn{1}{c|}{0.105} & 111 & 49 & 35 & \multicolumn{1}{c}{114}  \\ 
\multicolumn{1}{c|}{0.120} & 107 & 61 & 30 & \multicolumn{1}{c}{119}  \\ 
\end{tabular}
\end{ruledtabular}
\end{center}
\end{table}
For all the considered doping levels in Fig.\,\ref{fig:naeffectsabinitvsmodel}, it is seen that the main renormalized phonon peak acquires an appreciable broadening at a given finite momentum $|\mathbf{q}|\approx0.03-0.05~\text{\AA}^{-1}$. 
By exploring the spectral functions of the Einstein-like model, we quickly notice that the broadening occurs as soon as the vibrational peak   enters the dissipative electron-hole excitation pair continua of the $\overline{\mathrm{K}}$ and $\overline{\mathrm{Q}}$-like valleys, bounded by solid gray lines in Fig.\,\ref{fig:naeffectsabinitvsmodel}.
Indeed, phonons with frequencies higher than the boundary of the Landau damping region have too much energy to decay by exciting any electron-hole pair, and are therefore well-defined quasi-particles with long lifetimes, since in this case $\Im\tilde{\Pi}_{\mathrm{A'_{1}}}(\mathbf{q},\omega)=0$.
However, from a quantum many-body point of view, these phonons are allowed to simultaneously excite and reabsorb virtual electron-hole pairs of lower-energy, even in the absence of available energy~\cite{virtualph}.
Thereby, a dressing cloud of charge carriers is produced, which oscillates with the lattice and results in an increase of the vibrational frequencies.
This gives a physical explanation to the hardening and steeper dispersions in Fig.\,\ref{fig:naeffectsabinitvsmodel}.
Note that close to the border of the damping region, the undamped renormalized phonon frequency is maximum and follows the dispersion of the dissipative electronic continuum edge.
Indeed, at this frequency, the lattice and the dressing electronic cloud vibrate in phase, and therefore, the phonon phase velocity coincides with the Fermi velocity, $\mathbf{v}_{\mathrm{F}}$, as $\Omega_{\mathrm{A'_{1}}}(\mathbf{q})/|\mathbf{q}|\approx|\mathbf{v}_{\mathrm{F}}|$~\cite{Ipatova,Maksimov1996}.
At higher momenta, phonons acquire a finite lifetime, since they are energetically allowed to decay by exciting real electron-hole pairs, which gives $\Im\tilde{\Pi}_{\mathrm{A'_{1}}}(\mathbf{q},\omega)\neq0$.
Besides, for these damped phonons, the frequency renormalizations vanish.

The above spectral signatures have been explained with a simple free electron gas model, and therefore, are not unique to the doped monolayer MoS$_{2}$.
Indeed, the occurrence of this phenomenon should be expected for any optical phonon branch within the small momentum regime in the presence of strong electron-phonon coupling.

\subsection{Vibrational quasi-particle branch splitting induced by the electron-phonon interaction in the small momentum regime}\label{sec:C}
We start this final section analyzing in more detail the dynamical structure of the spectral function for the strongly interacting $\mathrm{A'_{1}}$ optical phonon mode within the small momentum regime.
By exploring more closely Fig.\,\ref{fig:naeffectsabinitvsmodel}, we observe that along with the above illustrated main renormalized phonon peak around $\omega\sim55-65~\text{meV}$, a substantial part of the spectral weight remains in the lower frequency range of the phonon spectrum (see blue area).
This spectral feature develops inside the boundary of the dissipative electron-hole continuum, and thus, has quite a wide structure.
While inmediately close to $\mathbf{q}=\overline{\Gamma}$ its spectral weight is negligible, the low-frequency feature accumulates an increasing weight when approaching the adiabatic optical branch to the detriment of the high-frequency peak.
This weight also increases with the strength of the non-adiabatic effects, and hence with the electron-phonon coupling upon doping.

Let us try to explain the vibrational spectral details in terms of phonon quasi-particle poles.
As seen in Sec.\,\ref{sec:theory}, the quasi-particle poles of the dressed phonon propagator are properly defined in the whole complex frequency plane.
Besides, the non-linear character of the Dyson's equation in Eq.\,\ref{eq:z_NA} leads to the possibility of finding several solutions, as found for electrons~\cite{ES,Asierprb2009,Asierprl2008,gurepaper}.
Assuming that the phonon self-energy is analytic in the entire complex frequency plane, the dressed phonon Green's function is given by
\begin{equation}
 \mathcal{D}_{\nu}(\mathbf{q},z)=\frac{2\omega_{\nu}(\mathbf{q})}{z^{2}-\omega^{2}_{\nu}(\mathbf{q})-2\omega_{\nu}(\mathbf{q})\tilde{\Pi}_{\nu}(\mathbf{q},z)}.
 \label{eq:GF_z}
\end{equation}
Suppose now that this function has several poles, labelled by the index $j$, and located in the lower half complex plane at $z^{(j)}_{\nu}(\mathbf{q})=\Omega^{(j)}_{\nu}(\mathbf{q})-i\gamma^{(j)}_{\nu}(\mathbf{q})$.
In this way, one can define the first-order Laurent series expansion of $\mathcal{D}_{\nu}(\mathbf{q},z)$ (Eq.\,\ref{eq:GF_z}) around them,
\begin{equation}
 \begin{aligned}
 \mathcal{D}^{\mathrm{qp}}_{\nu}(\mathbf{q},z)&=\sum_{j}\frac{\mathbb{Z}^{(j)}_{\nu}(\mathbf{q})}{z-z^{(j)}_{\nu}(\mathbf{q})}-\frac{\mathbb{Z}^{(j)}_{\nu}(\mathbf{q})}{z+z^{(j)}_{\nu}(\mathbf{q})}\\&=\sum_{j}\mathbb{Z}^{(j)}_{\nu}(\mathbf{q})\frac{2z^{(j)}_{\nu}(\mathbf{q})}{z^{2}-\big(z^{(j)}_{\nu}(\mathbf{q})\big)^{2}},
 \label{eq:GF_z_qp}
 \end{aligned}
\end{equation}
where $\mathbb{Z}^{(j)}_{\nu}(\mathbf{q})$ is the renormalization factor of the phonon quasi-particle pole, which is mathematically defined as the complex residue of $\mathcal{D}_{\nu}(\mathbf{q},z)$ evaluated at $z^{(j)}_{\nu}(\mathbf{q})$:
\begin{equation}
 \mathbb{Z}^{(j)}_{\nu}(\mathbf{q})=\frac{1}{\big(z^{(j)}_{\nu}(\mathbf{q})/\omega_{\nu}(\mathbf{q})\big)-\tilde{\Pi}^{'}_{\nu}\big(\mathbf{q},z^{(j)}_{\nu}(\mathbf{q})\big)}.
\end{equation}
Noting that $\mathbb{Z}^{(j)}_{\nu}(\mathbf{q})$ is a complex magnitude, the spectral representation of $\mathcal{D}^{\mathrm{qp}}_{\nu}(\mathbf{q},z)$ in Eq.\,\ref{eq:GF_z_qp} can be written as follows
\begin{equation}
 \begin{aligned}
 &B^{\mathrm{qp}}_{\nu}(\mathbf{q},\omega)=-\frac{1}{\pi}\sum_{j}\mathrm{Im}\mathcal{D}^{\mathrm{qp}}_{\nu}(\mathbf{q},\omega)\\&=\sum_{j}\frac{\big(\Omega^{(j)}_{\nu}(\mathbf{q})\pm\omega\big)\mathrm{Im}\mathbb{Z}^{(j)}_{\nu}(\mathbf{q})+\gamma^{(j)}_{\nu}(\mathbf{q})\mathrm{Re}\mathbb{Z}^{(j)}_{\nu}(\mathbf{q})}{\pi\Big(\big(\Omega^{(j)}_{\nu}(\mathbf{q})\pm\omega\big)^{2}+\big(\gamma^{(j)}_{\nu}(\mathbf{q})\big)^{2}\Big)}.
 \end{aligned}
 \label{eq:B_w_qp}
\end{equation}
Again, similar to electrons, the imaginary character of $\mathbb{Z}^{(j)}_{\nu}(\mathbf{q})$ in Eq.\,\ref{eq:B_w_qp} leads to the appearence of distorted or asymmetric peaks in the vibrational spectral function~\cite{Asierprb2009}.
Besides, the total phonon spectral weight coming from the vibrational quasi-particle modes is equivalent to the sum of the real parts of $\mathbb{Z}^{(j)}_{\nu}(\mathbf{q})$ and must be smaller than or equal to unity (see the supplemental Note\,S6)~\cite{Asierprb2009}.
Still, The most standard procedure to rationalize the phonon spectral function in terms of quasi-particle solutions is to completely neglect the imaginary part of the renormalization factors.

\begin{figure}[t]
 \centering
 \begin{center}
  \includegraphics[width=1\columnwidth,angle=0,scale=1.0]{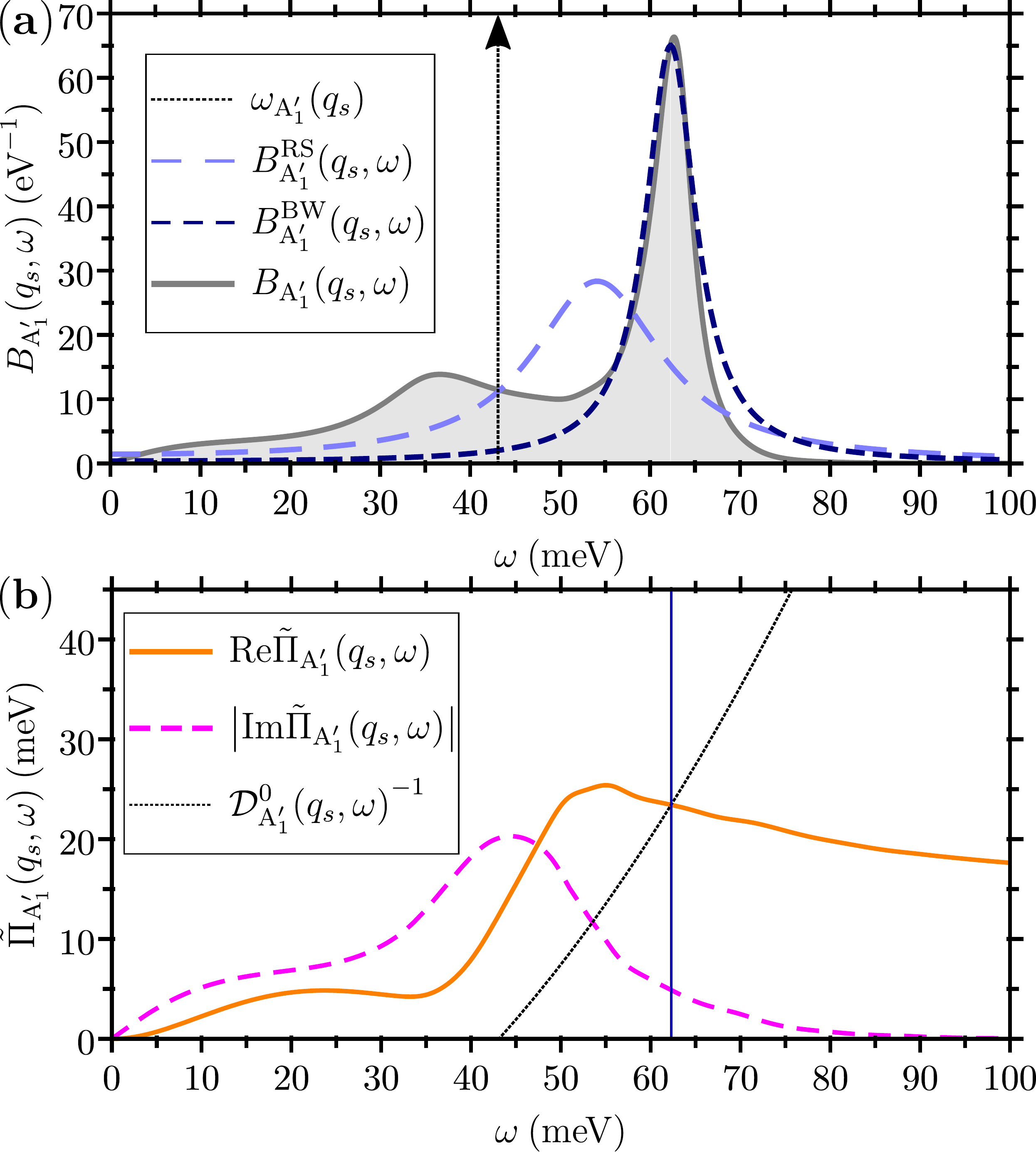}
 \end{center}
 \caption{(a) The spectral function and (b) the self-energy for the $\mathrm{A'_{1}}$ phonon mode and $\rho=0.12~e/\text{u.c.}$ evaluated at $q_{s}$. In panel (a), the gray area represents the \textit{ab initio} spectral function, $B_{\mathrm{A'_{1}}}(q_{s},\omega)$. The long-dashed light-blue and short-dashed dark-blue lines represent the quasi-particle spectral functions resulting from the Rayleigh-Schr\"odinger, $B^{\mathrm{RS}}_{\mathrm{A'_{1}}}(q_{s},\omega)$, and the Brillouin-Wigner, $B^{\mathrm{BW}}_{\mathrm{A'_{1}}}(q_{s},\omega)$, perturbation theories. The vertical dotted black arrow indicates the adiabatic spectral delta-line. In panel (b), the real and imaginary parts of $\tilde{\Pi}_{\mathrm{A'_{1}}}(q_{s},\omega)$ are represented by the solid orange and dashed magenta lines, respectively. The dotted black line represents the inverse of $\mathcal{D}^{0}_{\mathrm{A'_{1}}}(q_{s},\omega)$, whose cut with $\Re\tilde{\Pi}_{\mathrm{A'_{1}}}(q_{s},\omega)$ defines $\Omega^{\mathrm{BW}}_{\mathrm{A'_{1}}}(q_{s})$.}
 \label{fig:spectralpeaks}
\end{figure}
In Fig.\,\ref{fig:spectralpeaks}(a), we represent a cut of the dressed phonon spectral function, $B_{\mathrm{A'_{1}}}(q_{s},\omega)$ (gray area), calculated from first-principles (Eq.\,\ref{eq:ph_SF}) for the $\mathrm{A'_{1}}$ mode and $\rho=0.12~e/\text{u.c.}$, at the momentum $q_{s}=|\mathbf{q}|=0.05~\text{\AA}^{-1}$ along the $\overline{\Gamma\mathrm{K}}$ direction.
We also show the phonon quasi-particle spectral functions obtained by considering the Rayleigh-Schr\"odinger, $B^{\mathrm{RS}}_{\mathrm{A'_{1}}}(q_{s},\omega)$ (see long-dashed light-blue line), and the Brillouin-Wigner, $B^{\mathrm{BW}}_{\mathrm{A'_{1}}}(q_{s},\omega)$ (see short-dashed dark-blue line), perturbation theories.
While $B_{\mathrm{A'_{1}}}(q_{s},\omega)$ displays a two peak-like structure with maxima at $\omega\approx36$ and $62~\text{meV}$, both $B^{\mathrm{RS}}_{\mathrm{A'_{1}}}(q_{s},\omega)$ and $B^{\mathrm{BW}}_{\mathrm{A'_{1}}}(q_{s},\omega)$ exhibit a single Lorentzian peaked function.
Their phonon quasi-particle frequency, linewidth and (real) renormalization factor values are gathered in Table\,\ref{table:qpvalues}.
Note that $B^{\mathrm{RS}}_{\mathrm{A'_{1}}}(q_{s},\omega)$ completely fails describing the \textit{ab initio} spectral structure, and $B^{\mathrm{BW}}_{\mathrm{A'_{1}}}(q_{s},\omega)$ roughly approximates its high-frequency spectral feature.
This is consistent with the fact that the Brillouin-Wigner perturbation theory is only valid when $\big|\Im\tilde{\Pi}_{\mathrm{A'_{1}}}(q_{s},\omega)\big|\ll\Re\tilde{\Pi}_{\mathrm{A'_{1}}}(q_{s},\omega)$, which is satisfied in the vicinity of $\Omega^{\mathrm{BW}}_{\mathrm{A'_{1}}}(q_{s})$, as can be appreciated in Fig.\,\ref{fig:spectralpeaks}(b).
There, the real and imaginary parts of the first-principles phonon self-energy evaluated at $q_{s}$ are represented by the solid orange and dashed magenta lines, respectively.

\begin{figure}[t]
 \centering
 \begin{center}
  \includegraphics[width=1\columnwidth,angle=0,scale=1.0]{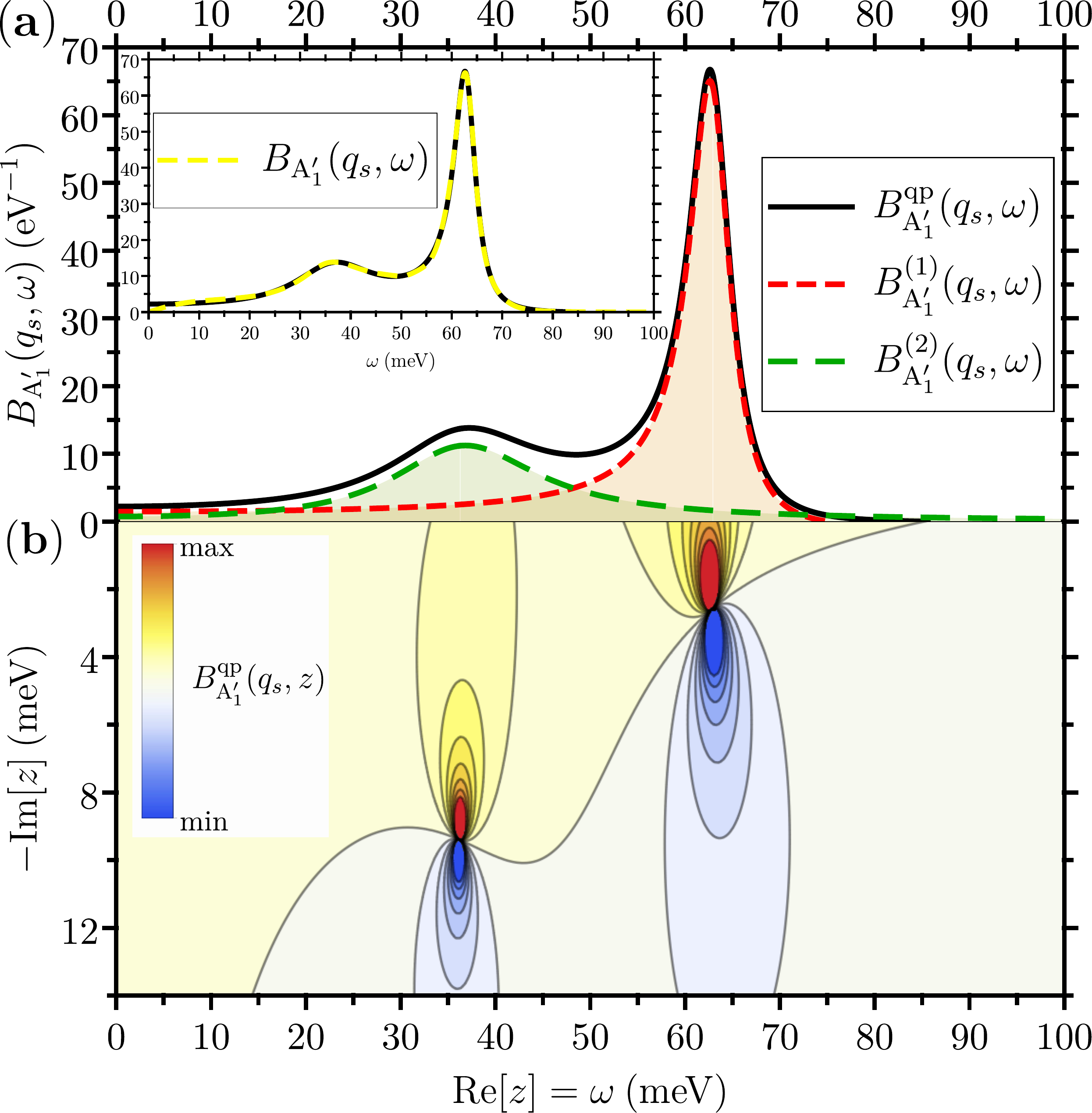}
 \end{center}
 \caption{(a) Spectral function of the double-phonon quasi-particle picture for the $\mathrm{A'_{1}}$ phonon mode and $\rho=0.12~e/\text{u.c.}$ evaluated at $q_{s}$. The spectral contributions of the high-, $B^{(1)}_{\mathrm{A'_{1}}}(q_{s},\omega)$, and low-frequency, $B^{(2)}_{\mathrm{A'_{1}}}(q_{s},\omega)$, quasi-partice poles, and their addition, $B^{\mathrm{qp}}_{\mathrm{A'_{1}}}(q_{s},\omega)$, are represented by short-dashed red, long-dashed green and solid black lines, respectively. The inset compares $B^{\mathrm{qp}}_{\mathrm{A'_{1}}}(q_{s},\omega)$ with $B_{\mathrm{A'_{1}}}(q_{s},\omega)$, represented by the dashed yellow line. Panel (b) shows the contour map of the spectral function on the complex plane $z$, where the two complex poles are well-defined at $z^{(1)}_{\mathrm{A'_{1}}}(q_{s})$ and $z^{(2)}_{\mathrm{A'_{1}}}(q_{s})$. The color code represents the height of the spectral function.}
 \label{fig:poles}
\end{figure}
\begin{table}[t]
\begin{center}
\caption{The phonon quasi-particle frequency, $\Omega^{x}_{\mathrm{A'_{1}}}(q_{s})$, linewidth, $\gamma^{x}_{\mathrm{A'_{1}}}(q_{s})$, and renormalization factor, $\mathbb{Z}^{x}_{\mathrm{A'_{1}}}(q_{s})$, within the Rayleigh-Schr\"odinger $(x=\text{RS})$ and Brillouin-Wigner $(x=\text{BW})$ perturbation theories, and within the spectral fitting procedure, $x=(1)$ and $(2)$, starting from the adiabatic frequency $\omega_{\mathrm{A'_{1}}}(q_{s})=43.1~\text{meV}$ for the $\mathrm{A'_{1}}$ optical phonon mode and $\rho=0.12~e/\text{u.c.}$ at $q_{s}$. Recall that the quasi-particle complex poles are defined as $z^{x}_{\mathrm{A'_{1}}}(q_{s})=\Omega^{x}_{\mathrm{A'_{1}}}(q_{s})-i\,\gamma^{x}_{\mathrm{A'_{1}}}(q_{s})$.}
\label{tab:qpvalues}
\begin{ruledtabular}
\begin{tabular}{*{4}{c}}
\multicolumn{1}{c|}{$x$} & \multicolumn{1}{c}{$\Omega^{x}_{\mathrm{A'_{1}}}(q_{s})$} & \multicolumn{1}{c}{$\gamma^{x}_{\mathrm{A'_{1}}}(q_{s})$} & \multicolumn{1}{c}{$\mathbb{Z}^{x}_{\mathrm{A'_{1}}}(q_{s})$} \\ \hline
\multicolumn{1}{c|}{RS} & $54.1~\text{meV}$ & $8.8~\text{meV}$ & $0.78$ \\ \hline
\multicolumn{1}{c|}{BW} & $62.3~\text{meV}$ & $3.4~\text{meV}$ & $0.69$ \\ \hline
\multicolumn{1}{c|}{$(1)$} & $62.9~\text{meV}$ & $2.6~\text{meV}$ & $0.52+i\,0.12\; (\sim65\%)$ \\ 
\multicolumn{1}{c|}{$(2)$} & $36.3~\text{meV}$ & $9.4~\text{meV}$ & $0.33+i\,0.04\; (\sim35\%)$ \\  
\end{tabular}
\end{ruledtabular}
\label{table:qpvalues}
\end{center}
\end{table}
The low-frequency \textit{ab initio} spectral feature at $\omega\approx36~\text{meV}$ develops at frequencies where the imaginary part of the self-energy is larger than the real part (see Fig.\,\ref{fig:spectralpeaks}(b)).
It is therefore reasonable to think that this spectral feature originates from an additional phonon quasi-particle pole with larger linewidth.
Using Eq.\,\ref{eq:B_w_qp}, we developed a numerical fitting procedure of $B_{\mathrm{A'_{1}}}(q_{s},\omega)$ in order to extract the complex poles and the complex renormalization factors of the phonon quasi-particles.
We find that a double complex pole picture is consistent with the structure of the \textit{ab initio} phonon spectral function, whose results are presented in Fig.\,\ref{fig:poles}.
Their phonon quasi-particle frequency, linewidth and (complex) renormalization factor values are gathered in Table\,\ref{table:qpvalues}.
Note that the spectral weight corresponding to the high- (1) and low-frequency (2) poles represent about the $65\%$ and $35\%$ of the total, respectively.
The spectral contributions from each phonon quasi-particle pole are labelled as $B^{(1)}_{\mathrm{A'_{1}}}(q_{s},\omega)$ (short-dashed green line) and $B^{(2)}_{\mathrm{A'_{1}}}(q_{s},\omega)$ (long-dashed green line), respectively (see Fig.\,\ref{fig:poles}(a)), and their simple addition, $B^{\mathrm{qp}}_{\mathrm{A'_{1}}}(q_{s},\omega)$ (solid black line), fits perfectly with the \textit{ab initio} spectral function (see yellow line in the inset panel of Fig.\,\ref{fig:poles}(a)).
Figure\,\ref{fig:poles}(b) exhibits the contour map of the phonon quasi-particle spectral function in the whole complex frequency plane, with the high- and the low-frequency poles found at $z^{(1)}_{\mathrm{A'_{1}}}(q_{s})$ and $z^{(2)}_{\mathrm{A'_{1}}}(q_{s})$, respectively.
Physically, the high-frequency phonon quasi-particle mode has its frequency within the Landau damping region of the $\overline{\mathrm{K}}(\overline{\mathrm{K'}})$ valleys and acquires a finite linewidth due to allowed intra-valley electron-hole pairs excitations.
However, it is also more energetic than the threshold of the Landau damping at $\overline{\mathrm{Q}}(\overline{\mathrm{Q'}})$ valleys.
Thereby, the excitation of virtual electron-hole processes is strongly promoted, leading to a strong frequency renormalization of this mode (see Sec.\,\ref{sec:B}).
On the other hand, the low-frequency phonon quasi-particle mode has its frequency also within the dissipative continuum of the $\overline{\mathrm{Q}}(\overline{\mathrm{Q'}})$ valleys.
Thus, this mode decays into the corresponding electron-hole pair excitations and results highly damped, with a linewidths $3.5$ times larger than that of the high-frequency mode (see Table\,\ref{tab:qpvalues}).

\begin{figure}[t]
 \centering
 \begin{center}
  \includegraphics[width=1\columnwidth,angle=0,scale=1.0]{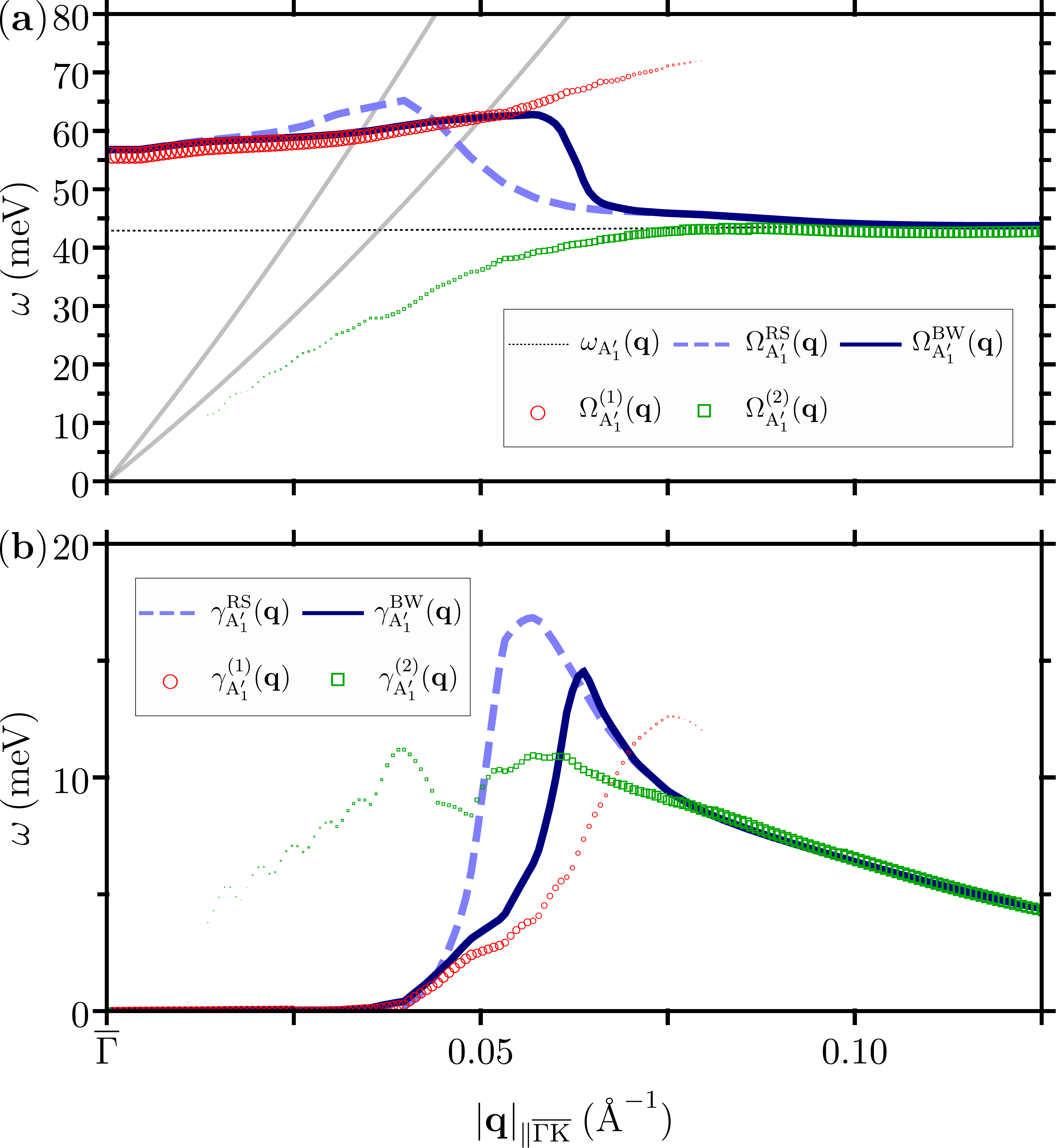}
 \end{center}
 \caption{(a) The renormalized frequency and (b) the linewidth dispersions for the $\mathrm{A'_{1}}$ phonon mode and $\rho=0.12~e/\text{u.c.}$, within the small momentum regime. The adiabatic dispersion is represented by the dotted black line. The quasi-particle frequencies and linewidths resulting from the Rayleigh-Schr\"odinger, $\Omega^{\mathrm{RS}}_{\mathrm{A'_{1}}}(\mathbf{q})$ and $\gamma^{\mathrm{RS}}_{\mathrm{A'_{1}}}(\mathbf{q})$, and the Brillouin-Wigner, $\Omega^{\mathrm{BW}}_{\mathrm{A'_{1}}}(\mathbf{q})$ and $\gamma^{\mathrm{BW}}_{\mathrm{A'_{1}}}(\mathbf{q})$, perturbation theories are represented by the long-dashed light-blue and solid dark-blue lines, respectively. The frequencies and the linewidths of the high-, $\Omega^{(1)}_{\mathrm{A'_{1}}}(\mathbf{q})$ and $\gamma^{(1)}_{\mathrm{A'_{1}}}(\mathbf{q})$, and low-frequency, $\Omega^{(2)}_{\mathrm{A'_{1}}}(\mathbf{q})$ and $\gamma^{(2)}_{\mathrm{A'_{1}}}(\mathbf{q})$, quasi-particle modes are represented by open red circles and open green squares, respectively. In panel (a), solid gray lines bound the electron-hole excitation pair continua in the simplified model at zero temperature as in Fig.\,\ref{fig:naeffectsabinitvsmodel}. The size of the markers is proportional to the spectral weight of each pole.}
 \label{fig:qpdisp}
\end{figure}
We conclude that the double-phonon quasi-particle picture puts in evidence the splitting of the adiabatic $\mathrm{A'_{1}}$ vibrational mode at $q_{s}$ into two different and well-defined non-adiabatic modes, which is an effect due to the electron-phonon coupling.
Figure\,\ref{fig:qpdisp}(a) and (b) compare the frequency and linewidth dispersions, respectively, of the phonon quasi-particle modes resulting from the standard methods and the spectral fitting procedure for the $\mathrm{A'_{1}}$ optical vibrational branch and $\rho=0.12~e/\text{u.c.}$, within the small momentum regime along the $\overline{\Gamma\mathrm{K}}$ direction.
A single-phonon quasi-particle picture (see light-dashed and dark-solid blue lines) is valid very close to $\overline{\Gamma}$ and deep within the Landau damping region, while the non-adiabatic splitting of the optical branch occurs in the vicinity of the edge of the electron-hole excitation pair continua (see gray lines).
Moving away from the $\overline{\Gamma}$ point, as the renormalized high-frequency optical branch (see red circles) approaches the Landau damping region, its spectral weight is smoothly transfered to the highly-damped low-frequency acoustic branch (see green squares).
Once the high-frequency peak overlaps with the electron-hole pair continua at momentum $|\mathbf{q}|\approx q_{s}$, its linewidth broadens rapidly, coinciding with the total transfer of its spectral weight to the emergent low-frequency branch, which follows at higher momenta a damped dispersion similar to the adiabatic one and a linewidth in the range of $5-10~\text{meV}$.
\section{CONCLUSIONS}\label{sec:conclusions}
In summary, we have studied the non-adiabatic effects due to the electron-phonon interaction on the vibrational spectrum of the electron-doped monolayer MoS$_{2}$.
We have found that the linewidth of phonon modes increases with doping as the spin-split conduction-valleys get populated.
This causes the strengthening of the electron-phonon coupling, which is at the origin of the experimentally measured superconductivity onset~\cite{scmos2}.
Likewise, we have found that as soon as all the spin-split conduction-valleys are fully occupied, the enhancement of the electron-phonon interaction is so strong that lattice instabilities are induced in the vibrational structure.
Moreover, we found that the optical branches appear very strongly renormalized when approaching the area corresponding to the Landau damping, the reason being that the virtual excitation of electron-hole pairs is heavily promoted.
In fact, the original adiabatic branch is split into two physically different quasi-particle phonon states.
Though MoS$_{2}$ represents a good example of strong electron-phonon monolayer system with a relatively simple electron valley structure, it is evident that the physics described in this article is of a general nature.
\section{ACKNOWLEDGEMENTS}
The authors acknowledge the Department of Education, Universities and Research of the Basque Government and UPV/EHU (Grant No. IT756-13), the Spanish Ministry of Economy and Competitiveness MINECO (FIS2016-75862-P) and the University of the Basque Country UPV/EHU (GIU18/138) for financial support. P.G.-G. and J.L.-B. acknowledge the University of the Basque Country UPV/EHU (Grant Nos. PIF/UPV/12/279 and PIF/UPV/16/240, respectively) and the Donostia International Physics Center (DIPC) for financial support. Computer facilities were provided by the DIPC.
\clearpage
\pagebreak
\widetext
\begin{center}
 \textbf{Supplemental Material for \\ ``Emergence of large non-adiabatic effects induced by the electron-phonon interaction on the complex vibrational quasi-particle spectrum of the doped monolayer MoS$_{2}$''}
\end{center}
\setcounter{equation}{0}
\setcounter{figure}{0}
\setcounter{table}{0}
\setcounter{page}{1}
\makeatletter
\renewcommand{\theequation}{S\arabic{equation}}
\renewcommand{\thefigure}{S\arabic{figure}}
\renewcommand{\bibnumfmt}[1]{[S#1]}
\renewcommand{\citenumfont}[1]{S#1}
\section*{S1. Phonon self-energy from the density linear response theory}
In this section, we present an alternative way for obtaining, from combining density functional perturbation theory and density linear response theory, the expression in Eq.5 of the main text for the phonon self-energy due to the electron-phonon interaction.
Let start by deducing Eq.2 of the main text.
For a relaxed crystal in equilibrium, the Fourier transform of the interatomic force constant (IFC) matrix within the adiabatic and harmonic approximations is written as follows~\cite{DFPTs}
\begin{equation}
 C_{ss'}^{\alpha\alpha'}(\mathbf{q})=\iint\Bigg(\frac{\partial n(\mathbf{r})}{\partial u_{s}^{\alpha}(\mathbf{q})}\Bigg)^{*}\frac{\partial V_{\mathrm{ext}}(\mathbf{r})}{\partial u_{s'}^{\alpha'}(\mathbf{q})}\mathrm{d}\mathbf{r}+\int n(\mathbf{r})\frac{\partial^{2}V_{\mathrm{ext}}(\mathbf{r})}{\partial u_{s}^{*\alpha}(\mathbf{q})\partial u_{s'}^{\alpha'}(\mathbf{q})}\mathrm{d}\mathbf{r}+\frac{\partial^{2} E_{\mathrm{ion}}}{\partial u_{s}^{*\alpha}(\mathbf{q})\partial u_{s'}^{\alpha'}(\mathbf{q})},
 \label{eq:DFPTs}
\end{equation}
where $n(\mathbf{r})$, $V_{\mathrm{ext}}(\mathbf{r})$ and $E_{\mathrm{ion}}$ are the electron charge density, the electron-ion interaction external potential and the Coulomb interaction energy between nuclei, respectively.
$u^{\alpha}_{s}(\mathbf{q})$ corresponds to the displacement of the $s$-th ion along the $\alpha$ direction for a lattice distortion of wave-vector $\mathbf{q}$. $\partial n(\mathbf{r})/\partial u^{\alpha}_{s}(\mathbf{q})$ is the self-consistent static-screened change of the charge density and $\partial V_{\mathrm{ext}}(\mathbf{r})/\partial u^{\alpha}_{s}(\mathbf{q})$ is the bare change of the external potential.
Within the density linear response theory, the self-consistent static-screened changes of both the charge density and the external potential are related in the following way~\cite{dyprosovol3s}
\begin{equation}
 \frac{\partial n(\mathbf{r})}{\partial u^{\alpha}_{s}(\mathbf{q})}=\int\chi^{0}_{\mathbf{q}}(\mathbf{r,r'},0)\frac{\partial V_{\mathrm{scf}}(\mathbf{r'})}{\partial u^{\alpha}_{s}(\mathbf{q})}\mathrm{d}\mathbf{r'},
 \label{eq:dn_to_chi0_dvscfs}
\end{equation}
where $\partial V_{\mathrm{scf}}(\mathbf{r})/\partial u^{\alpha}_{s}(\mathbf{q})$ is the self-consistent static-screened change of the external potential with respect to the ionic displacement $u^{\alpha}_{s}(\mathbf{q})$.
$\chi^{0}_{\mathbf{q}}(\mathbf{r,r'},0)$ is the frequency-independent ($\omega=0$) component of $\chi^{0}_{\mathbf{q}}(\mathbf{r,r'},\omega)$, which represents the density-response function of the non-interacting electronic system and is defined as
\begin{equation}
  \chi^{0}_{\mathbf{q}}(\mathbf{r},\mathbf{r'},\omega)=\lim_{\eta\to 0^{+}}\frac{1}{N_{\mathbf{k}}}\sum^{\mathrm{1BZ}}_{\mathbf{k}}\sum_{mn}\frac{f(\varepsilon^{\mathbf{k}}_{n})-f(\varepsilon^{\mathbf{k+q}}_{m})}{\varepsilon^{\mathbf{k}}_{n}-\varepsilon^{\mathbf{k+q}}_{m}+\omega+i\eta}\big(\psi^{\mathbf{k}}_{n}(\mathbf{r})\big)^{*}\psi^{\mathbf{k+q}}_{m}(\mathbf{r})\big(\psi^{\mathbf{k+q}}_{m}(\mathbf{r'})\big)^{*}\psi^{\mathbf{k}}_{n}(\mathbf{r'}),
  \label{eq:chi0s}
\end{equation}
where $\varepsilon^{\mathbf{k}}_{n}$ and $\psi^{\mathbf{k}}_{n}(\mathbf{r})$ are the energy and wave function, respectively, of the Kohn-Sham (KS) single-particle electron state of band index $n$ and momentum $\mathbf{k}$.
$f(\varepsilon^{\mathbf{k}}_{n})$ represents the Fermi-Dirac (FD) occupation factor of the KS state, $\eta$ is a positive real infinitesimal and $N_{\mathbf{k}}$ is the number of $\mathbf{k}$-points considered in the Brillouin zone (BZ).
In addition, the self-consistent static-screened change of the external potential is related to its bare change in the following way~\cite{dyprosovol3s}
\begin{equation}
 \frac{\partial V_{\mathrm{scf}}(\mathbf{r})}{\partial u^{\alpha}_{s}(\mathbf{q})}=\frac{\partial V_{\mathrm{ext}}(\mathbf{r})}{\partial u^{\alpha}_{s}(\mathbf{q})}+\int K(\mathbf{r,r'})\frac{\partial n(\mathbf{r'})}{\partial u^{\alpha}_{s}(\mathbf{q})}\mathrm{d}\mathbf{r'},
 \label{eq:dvscf_to_dvexts}
\end{equation}
which can be rewritten as
\begin{equation}
 \frac{\partial V_{\mathrm{ext}}(\mathbf{r})}{\partial u^{\alpha}_{s}(\mathbf{q})}=\frac{\partial V_{\mathrm{scf}}(\mathbf{r})}{\partial u^{\alpha}_{s}(\mathbf{q})}-\int K(\mathbf{r,r'})\frac{\partial n(\mathbf{r'})}{\partial u^{\alpha}_{s}(\mathbf{q})}\mathrm{d}\mathbf{r'},
 \label{eq:dvext_to_dvscfs}
\end{equation}
where $K(\mathbf{r,r'})$ is the Hartree and exchange-correlation kernel.
Plugging now Eqs.\,\ref{eq:dn_to_chi0_dvscfs} and \ref{eq:dvext_to_dvscfs} into Eq.\,\ref{eq:DFPTs}, one easily recovers the Eq.2 of the main text,
\begin{equation}
\begin{aligned}
C_{ss'}^{\alpha\alpha'}(\mathbf{q})=&\iint\frac{\partial V_{\mathrm{scf}}(\mathbf{r})}{\partial u_{s'}^{\alpha'}(\mathbf{q})}\big(\chi^{0}_{\mathbf{q}}(\mathbf{r},\mathbf{r'},0)\big)^{*}\Bigg(\frac{\partial V_{\mathrm{scf}}(\mathbf{r'})}{\partial u_{s}^{\alpha}(\mathbf{q})}\Bigg)^{*}\mathrm{d}\mathbf{r}\mathrm{d}\mathbf{r'}-\iint\frac{\partial n(\mathbf{r})}{\partial u_{s}^{*\alpha}(\mathbf{q})}K(\mathbf{r},\mathbf{r'})\frac{\partial n(\mathbf{r'})}{\partial u_{s'}^{\alpha'}(\mathbf{q})}\mathrm{d}\mathbf{r'}\mathrm{d}\mathbf{r}\\&+\int n(\mathbf{r})\frac{\partial^{2}V_{\mathrm{ext}}(\mathbf{r})}{\partial u_{s}^{*\alpha}(\mathbf{q})\partial u_{s'}^{\alpha'}(\mathbf{q})}\mathrm{d}\mathbf{r}+\frac{\partial^{2} E_{\mathrm{ion}}}{\partial u_{s}^{*\alpha}(\mathbf{q})\partial u_{s'}^{\alpha'}(\mathbf{q})},
\label{eq:DFPT_2s}
\end{aligned}
\end{equation}
This expression is based on the adiabatic approximation, i.e.~it does only account for the electrostatic screening of lattice vibrations, since charge carriers are assumed to respond instantaneously to the motion of ions.
In order to account for the retardation effects on the electronic response to the ionic motion, i.e.~the effects due to the electron-phonon interaction beyond the adiabatic approximation or simply non-adiabatic effects, a fully dynamic electronic screening must be incorporated in Eq.\,\ref{eq:DFPT_2s}.
This should be properly done by considering the finite-frequency ($\omega\neq0$) contributions to the non-interacting density-response function, together with self-consistent dynamic-screened changes of both the charge density and the external potential, and a frequency-dependent Hartree and exchange-correlation kernel.
Nevertheless, within the standard density functional and density functional perturbation theory (DFT and DFPT), the kernel and the self-consistent screened derivatives are only statically computed.
Therefore, in practice, the dynamic extension of the IFC matrix Fourier transform in Eq.\,\ref{eq:DFPT_2s}, i.e.~$C^{\alpha\alpha'}_{ss'}(\mathbf{q})\to C^{\alpha\alpha'}_{ss'}(\mathbf{q},\omega)$, is limited to uniquely adopting the full dynamical description of the non-interacting density-response function, i.e.~$\chi^{0}_{\mathbf{q}}(\mathbf{r,r'},0)\to\chi^{0}_{\mathbf{q}}(\mathbf{r,r'},\omega)$.
Thereby, and after some simplifications, the dynamical matrix, which is related to the Fourier transform of the IFC matrix as $D^{\alpha\alpha'}_{ss'}(\mathbf{q},\omega)=C^{\alpha\alpha'}_{ss'}(\mathbf{q},\omega)/\sqrt{M_{s}M_{s'}}$, can be written and split into two different static and dynamic contributions as~\cite{giustinorevs}
\begin{equation}
D_{ss'}^{\alpha\alpha'}(\mathbf{q},\omega)=D_{ss'}^{\alpha\alpha'}(\mathbf{q})+\tilde{\Pi}^{\alpha\alpha'}_{ss'}(\mathbf{q},\omega)
\label{eq:DM1s}
\end{equation}
where $D^{\alpha\alpha'}_{ss'}(\mathbf{q})=C^{\alpha\alpha'}_{ss'}(\mathbf{q})/\sqrt{M_{s}M_{s'}}$ is the adiabatic dynamical matrix.
The adiabatic phonon modes or lattice vibrational modes are defined as the solutions of the following eigenvalue problem~\cite{DFPTs}
\begin{equation}
 \sum_{s'\alpha'}D^{\alpha\alpha'}_{ss'}(\mathbf{q})e^{s'\alpha'}_{\nu}(\mathbf{q})=\omega^{2}_{\nu}(\mathbf{q})e^{s\alpha}_{\nu}(\mathbf{q}),
\end{equation}
where $\omega_{\nu}(\mathbf{q})$ and $e^{s\alpha}_{\nu}(\mathbf{q})$ are the frequency and polarization vector, respectively, of the phonon mode of branch index $\nu$ and momentum $\mathbf{q}$.
In Eq.\,\ref{eq:DM1s}, $\tilde{\Pi}^{\alpha\alpha'}_{ss'}(\mathbf{q},\omega)$ represents the non-adiabatic contribution to the dynamical matrix, and is written as follows
\begin{equation}
\tilde{\Pi}^{\alpha\alpha'}_{ss'}(\mathbf{q},\omega)=\lim_{\eta\to 0^{+}}\frac{1}{N_{\mathbf{k}}}\sum^{\mathrm{1BZ}}_{\mathbf{k}}\Bigg(\frac{g^{s\alpha}_{mn}(\mathbf{k,q})}{\sqrt{M_{s}}}\Bigg)^{*}\frac{g^{s'\alpha'}_{mn}(\mathbf{k,q})}{\sqrt{M_{s'}}}\Bigg(\frac{f(\varepsilon_{\mathbf{k}n})-f(\varepsilon_{\mathbf{k+q}m})}{\varepsilon_{\mathbf{k}n}-\varepsilon_{\mathbf{k+q}m}+\omega+i\eta}-\frac{f(\varepsilon_{\mathbf{k}n})-f(\varepsilon_{\mathbf{k+q}m})}{\varepsilon_{\mathbf{k}n}-\varepsilon_{\mathbf{k+q}m}}\Bigg),
\end{equation}
where the self-consistent static-screened deformation potential matrix element is defined as
\begin{equation}
 g^{s\alpha}_{mn}(\mathbf{k,q})=\int\big(\psi^{\mathbf{k+q}}_{m}(\mathbf{r})\big)^{*}\frac{\partial V_{\mathrm{scf}}(\mathbf{r})}{\partial u^{\alpha}_{s}(\mathbf{q})}\psi^{\mathbf{k}}_{n}(\mathbf{r})\mathrm{d}\mathbf{r}=\bigg\langle\psi^{\mathbf{k+q}}_{m}\bigg|\frac{\partial\hat{V}_{\mathrm{scf}}}{\partial u^{\alpha}_{s}(\mathbf{q})}\bigg|\psi^{\mathbf{k}}_{n}\bigg\rangle,
\end{equation}
which physically describes the strength or probability amplitude of the effective coupling between the KS eigenstates $|\psi^{\mathbf{k}}_{n}\rangle$ and $|\psi^{\mathbf{k+q}}_{m}\rangle$ via the self-consistent static-screened change of the external potential induced by the ionic displacement $u^{\alpha}_{s}(\mathbf{q})$.
If one rewrites now the dynamical matrix in Eq.\,\ref{eq:DM1s} in the basis of adiabatic vibrational normal modes by a unitary transformation, one finds
\begin{equation}
 \begin{aligned}
 D_{\nu\nu'}(\mathbf{q},\omega)=&\sum_{\substack{s\alpha\\s'\alpha'}}\big(e^{s\alpha}_{\nu}(\mathbf{q})\big)^{*}D^{\alpha\alpha'}_{ss'}(\mathbf{q},\omega)e^{s'\alpha'}_{\nu'}(\mathbf{q})\\=&\sum_{\substack{s\alpha\\s'\alpha'}}\big(e^{s\alpha}_{\nu}(\mathbf{q})\big)^{*}D^{\alpha\alpha'}_{ss'}(\mathbf{q})e^{s'\alpha'}_{\nu'}(\mathbf{q})+\sum_{\substack{s\alpha\\s'\alpha'}}\big(e^{s\alpha}_{\nu}(\mathbf{q})\big)^{*}\tilde{\Pi}^{\alpha\alpha'}_{ss'}(\mathbf{q},\omega)e^{s'\alpha'}_{\nu'}(\mathbf{q})\\=&\;\omega^{2}_{\nu}(\mathbf{q})\delta_{\nu\nu'}+2\sqrt{\omega_{\nu}(\mathbf{q})\omega_{\nu'}(\mathbf{q})}\tilde{\Pi}_{\nu\nu'}(\mathbf{q},\omega),
 \end{aligned}
 \label{eq:DM2s}
\end{equation}
whose eigenvalue problem is equivalent to the vibrational quasi-particle equation.
The expression for the retarded phonon self-energy that only accounts for the non-adiabatic effects due to the electron-phonon interaction is given by $\tilde{\Pi}_{\nu\nu'}(\mathbf{q},\omega)=\Pi_{\nu\nu'}(\mathbf{q},\omega)-\Pi_{\nu\nu'}(\mathbf{q},0)$, where $\Pi_{\nu\nu'}(\mathbf{q},\omega)$ is the retartded phonon self-energy that takes into account both the adiabatic and non-adiabatic effects due to the electron-phonon interaction, and is obviously defined from the above as follows~\cite{Grimvalls}
\begin{equation}
 \Pi_{\nu\nu'}(\mathbf{q},\omega)=\lim_{\eta\to0^{+}}\frac{1}{N_{\mathbf{k}}}\sum^{\mathrm{1BZ}}_{\mathbf{k}}\sum_{mn}\big(g^{\nu}_{mn}(\mathbf{k,q})\big)^{*}g^{\nu'}_{mn}(\mathbf{k,q})\frac{f(\varepsilon_{\mathbf{k}n})-f(\varepsilon_{\mathbf{k+q}m})}{\varepsilon_{\mathbf{k}n}-\varepsilon_{\mathbf{k+q}m}+\omega+i\eta}.
 \label{eq:phses}
\end{equation}
The self-consistent static-screened electron-phonon matrix element is given by~\cite{Grimvalls}
\begin{equation}
 g^{\nu}_{mn}(\mathbf{k,q})=\sum_{s\alpha}\frac{e^{s\alpha}_{\nu}(\mathbf{q})}{\sqrt{2\omega_{\nu}(\mathbf{q})M_{s}}}g^{s\alpha}_{mn}(\mathbf{k,q})=\sum_{s\alpha}\frac{e^{s\alpha}_{\nu}(\mathbf{q})}{\sqrt{2\omega_{\nu}(\mathbf{q})M_{s}}}\bigg\langle\psi^{\mathbf{k+q}}_{m}\bigg|\frac{\partial\hat{V}_{\mathrm{scf}}}{\partial u^{\alpha}_{s}(\mathbf{q})}\bigg|\psi^{\mathbf{k}}_{n}\bigg\rangle,
\end{equation}
which physically describes the strength or probability amplitude of the effective coupling between $|\psi^{\mathbf{k}}_{n}\rangle$ and $|\psi^{\mathbf{k+q}}_{m}\rangle$ via the phonon mode of branch index $\nu$ and momentum $\mathbf{q}$.
As one can easily appreciate, Eq.5 of the main text is identical to the diagonal elements ($\nu=\nu'$) of Eq.\,\ref{eq:phses}.
\clearpage
\section*{S2. Non-selfconsistent calculation of the adiabatic Dynamical matrix with smaller smearings on a finer $k$-mesh than standard DFPT calculations}\label{sec:s2}
\begin{figure}[ht!]
 \centering
 \begin{center}
  \includegraphics[width=1\columnwidth,angle=0,scale=1.0]{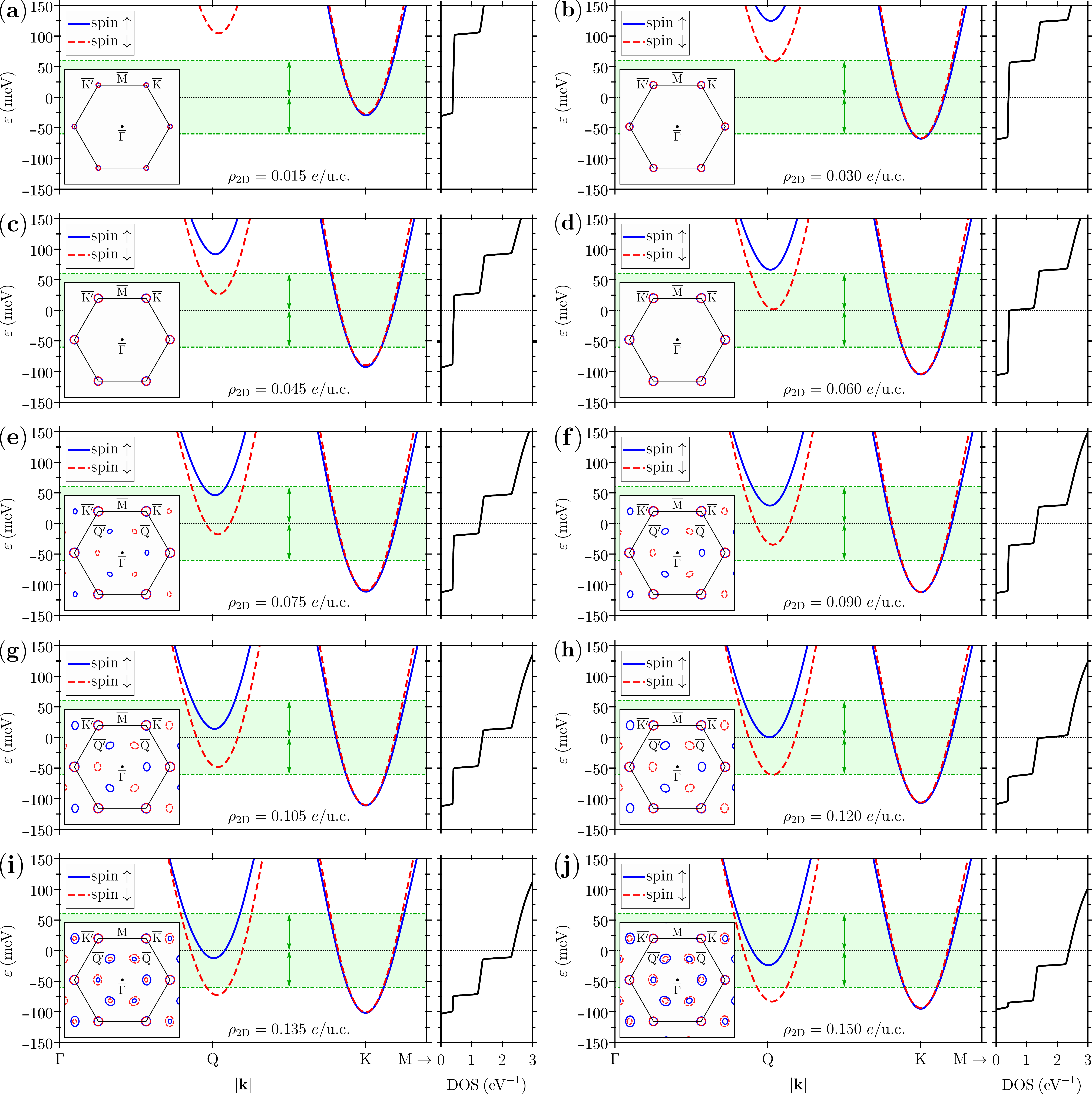}
 \end{center}
 \caption{Electron conduction-band structure (left) and its corresponding DOS (right) of the monolayer MoS$_{2}$ for the electron-doping concentrations $\rho=0.015$ (a), $0.030$ (b), $0.045$ (c), $0.060$ (d), $0.075$ (e), $0.090$ (f), $0.105$ (g), $0.120$ (h), $0.135$ (i), $0.150~e/\text{u.c.}$ (j). The insets in left panels show the FS contour. Solid blue and dashed red lines represent opposite out-of-plane spin-polarized bands. The Fermi level is set to zero (horizontal dotted black line). Horizontal dashed-dotted green lines delimit the energy window (shaded green areas) within which an electron-hole pair can be excited (relaxed) by the decay (emission) of a phonon with maximum frequency $\omega_{\mathrm{max}}=60~\text{meV}$.}
 \label{fig:figeldopall}
\end{figure}

In this section, we briefly present the simple non-selfconsistent procedure based on the Wannier interpolation of electron-phonon matrix elements for computing dynamical matrices on a finer $k$-mesh and with smaller smearing than the initial DFPT calculations.
From Eq.\,\ref{eq:DFPT_2s} and employing relations and simplications already used above, the adiabatic dynamical matrix can be rewritten in the following way
\begin{equation}
\begin{aligned}
D_{ss'}^{\alpha\alpha'}(\mathbf{q})=&\frac{1}{N_{\mathbf{k}}}\sum^{\mathrm{1BZ}}_{\mathbf{k}}\sum_{mn}\Bigg(\frac{g^{s\alpha}_{mn}(\mathbf{k,q})}{\sqrt{M_{s}}}\Bigg)^{*}\frac{g^{s'\alpha'}_{mn}(\mathbf{k,q})}{\sqrt{M_{s'}}}\frac{f(\varepsilon^{\mathbf{k}}_{n},\sigma)-f(\varepsilon^{\mathbf{k+q}}_{m},\sigma)}{\varepsilon^{\mathbf{k}}_{n}-\varepsilon^{\mathbf{k+q}}_{m}}\\&+\frac{1}{\sqrt{M_{s}M_{s'}}}\Bigg(\int n(\mathbf{r})\frac{\partial^{2}V_{\mathrm{ext}}(\mathbf{r})}{\partial u_{s}^{*\alpha}(\mathbf{q})\partial u_{s'}^{\alpha'}(\mathbf{q})}\mathrm{d}\mathbf{r}+\frac{\partial^{2} E_{\mathrm{ion}}}{\partial u_{s}^{*\alpha}(\mathbf{q})\partial u_{s'}^{\alpha'}(\mathbf{q})}-\iint\frac{\partial n(\mathbf{r})}{\partial u_{s}^{*\alpha}(\mathbf{q})}K(\mathbf{r},\mathbf{r'})\frac{\partial n(\mathbf{r'})}{\partial u_{s'}^{\alpha'}(\mathbf{q})}\mathrm{d}\mathbf{r'}\mathrm{d}\mathbf{r}\Bigg).
\label{eq:DFPT_3s}
\end{aligned}
\end{equation}
where the dependence of the FD occupations factors on the smearing $\sigma$ is now explicitly indicated, which are expressed as follows
\begin{equation}
 f(\varepsilon^{\mathbf{k}}_{n},\sigma)=\frac{1}{1+e^{(\varepsilon^{\mathbf{k}}_{n}-E_{\mathrm{F}})/\sigma}}
\end{equation}
with $E_{\mathrm{F}}$ the Fermi energy.
Our DFPT dynamical matrix calculations on a coarse $8\times8$ $q$-mesh are based on a converged DFT electronic ground-state computation that has been previously carried out for all the considered doping levels in a coarse $N^{\mathrm{c}}_{\mathbf{k}}=32\times32$ mesh in combination with a Gaussian smearing of $\sigma^{\mathrm{c}}=5~\text{mRy}$, as mentioned in Sec.III of the main text.
This smearing value is equivalent to an energy of $\sim68~\text{meV}$ and a temperature of $\sim790~\text{K}$, which, despite having been used in previous phonon calculations~\cite{1lmos2a1softs,yizhiprb2013s,phsoftmos2s,rosnermos2s}, is comparable to remarkable changes in the topology of the Fermi surface (FS) upon doping.
This differnces can be seen in Sec.IV.B of the main text and in Fig.\,\ref{fig:figeldopall}(a)-(j) where we present the conduction-band structure, corresponding density of states and FS contour of the electron-doped monolayer MoS$_{2}$ for all the considered carrier doping concentrations.
Indeed, such a high smearing leads to a smoothing of the FS that can mask pronounced fluctuations of its structure, which are at the origin of several interesting electron-phonon properties and phenomena such as Kohn anomalies (KAs).
One would like DFPT calculations to be performed with smaller smearings in order to be as reliable as possible, but this evidently demands finer $k$-meshes for achieving converged results, which makes self-consistent calculations prohibitive.

Conversely, one can take advantage from the first term on the right-hand side (r.h.s.)~of Eq.\,\ref{eq:DFPT_3s} and devise a non-selfconsistent scheme similar to that of Ref.~\cite{calandramauriprb2010s}.
In fact, the latter matrix accounts for the effective phonon-mediated electrostatic screening between occupied and empty states on the lattice vibrations, and is already accessible once KS energies as well as deformation potential matrix elements are calculated from DFT and DFPT calculations.
The idea for getting a more realistic ``approximated'' adiabatic dynamical matrix calculation is to compute the first term on the r.h.s.~of Eq.\,\ref{eq:DFPT_3s} with $N^{\mathrm{f}}_{\mathbf{k}}$ and $\sigma^{\mathrm{f}}$ finer parameters, and subsitute it with the one using $N^{\mathrm{c}}_{\mathbf{k}}$ and $\sigma^{\mathrm{c}}$ coarse parameters from the initial DFT calculation, in the following way
\begin{equation}
 \tilde{D}_{ss'}^{\alpha\alpha'}(\mathbf{q})=D_{ss'}^{\alpha\alpha'}(\mathbf{q})+\Pi^{\alpha\alpha'}_{ss'}(\mathbf{q},N^{\mathrm{f}}_{\mathbf{k}},\sigma^{\mathrm{f}})-\Pi^{\alpha\alpha'}_{ss'}(\mathbf{q},N^{\mathrm{c}}_{\mathbf{k}},\sigma^{\mathrm{c}}),
 \label{eq:Dtildes}
\end{equation}
with
\begin{equation}
 \Pi^{\alpha\alpha'}_{ss'}(\mathbf{q},N_{\mathbf{k}},\sigma)=\frac{1}{N_{\mathbf{k}}}\sum^{\mathrm{1BZ}}_{\mathbf{k}}\sum_{mn}\Bigg(\frac{g^{s\alpha}_{mn}(\mathbf{k,q})}{\sqrt{M_{s}}}\Bigg)^{*}\frac{g^{s'\alpha'}_{mn}(\mathbf{k,q})}{\sqrt{M_{s'}}}\frac{f(\varepsilon^{\mathbf{k}}_{n},\sigma)-f(\varepsilon^{\mathbf{k+q}}_{m},\sigma)}{\varepsilon^{\mathbf{k}}_{n}-\varepsilon^{\mathbf{k+q}}_{m}}.
 \label{eq:piaapssps}
\end{equation}

\begin{figure}[ht!]
 \centering
 \begin{center}
  \includegraphics[width=1\columnwidth,angle=0,scale=1.0]{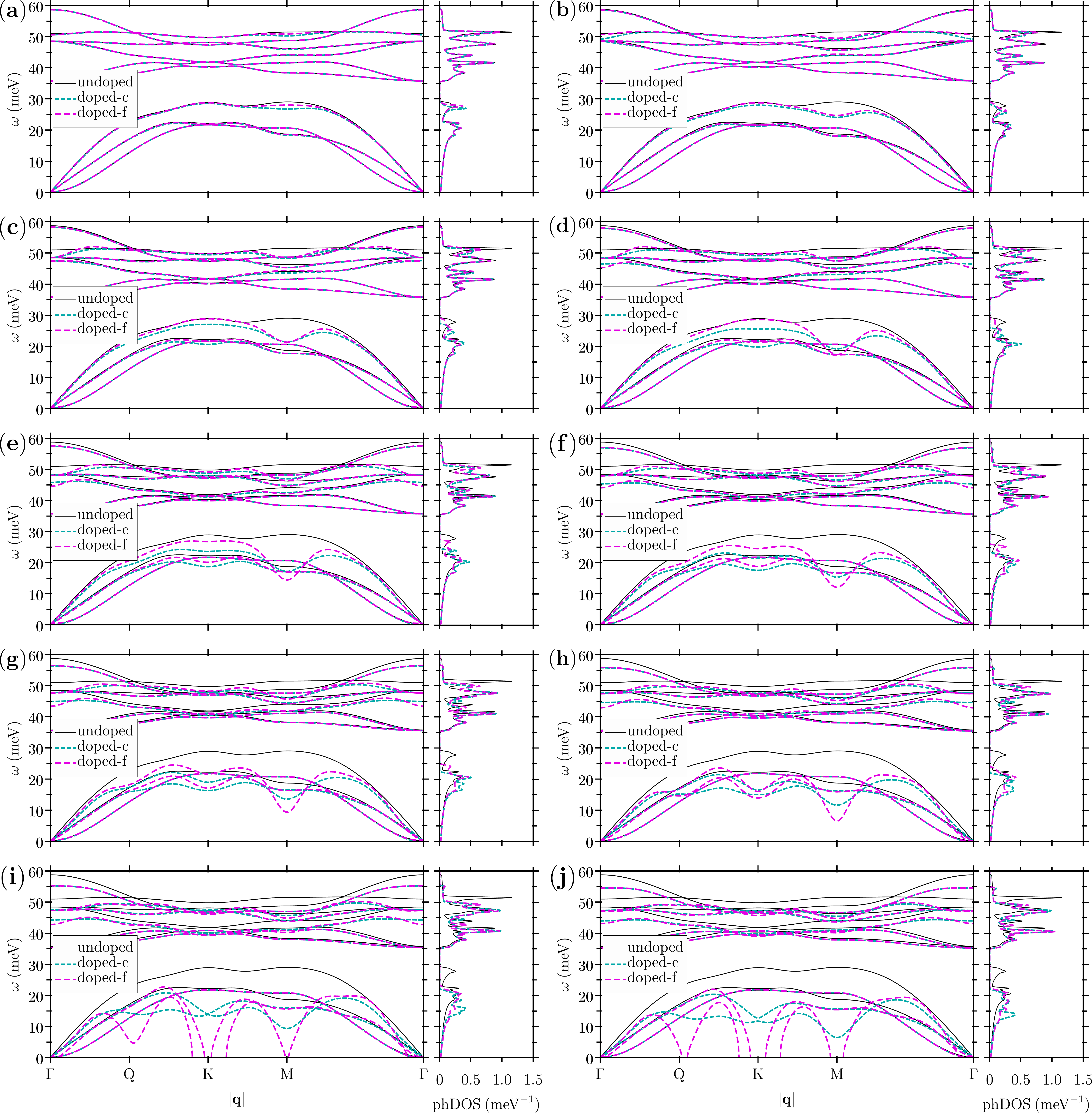}
 \end{center}
 \caption{Phonon dispersion relation (left) and its corresponding phDOS (right) of the monolayer MoS$_{2}$ for the electron-doping concentrations $\rho=0.015$ (a), $0.030$ (b), $0.045$ (c), $0.060$ (d), $0.075$ (e), $0.090$ (f), $0.105$ (g), $0.120$ (h), $0.135$ (i), $0.150~e/\text{u.c.}$ (j). Solid thin black lines represent the undoped phonon branches. Short-dashed cyan and long-dashed magenta lines represent the doping-dependent phonon branches evaluated directly from DFPT calculations and by means of the non-selfconsistent procedure described in Sec.S2, respectively}
 \label{fig:figphdopall}
\end{figure}

In the main text, the converged 1BZ summation of the finer matrix $\Pi^{\alpha\alpha'}_{ss'}(\mathbf{q},N^{\mathrm{f}}_{\mathbf{k}},\sigma^{\mathrm{f}})$ in Eq.\,\ref{eq:piaapssps} is computed using a finer $N^{\mathrm{f}}_{\mathbf{k}}=720\times720$ mesh, which allows to use a smaller smearing $\sigma^{\mathrm{f}}=5~\text{meV}$.
This is done by interpolating the KS single-electron energies, $\varepsilon_{\mathbf{k}n}$, and deformation potential matrix elements, $g^{s\alpha}_{mn}(\mathbf{k,q})$, on the corresponding finer $k$-mesh by means of Wannier functions~\cite{giustinorevs}.
The sum over the band indexes $m$ and $n$ of Eq.\,\ref{eq:piaapssps} runs over the low-energy spin-split conduction-bands.
Once this done, the dynamical matrix $\tilde{D}^{\alpha\alpha'}_{ss'}(\mathbf{q})$ is calculated by means of Eq.\,\ref{eq:Dtildes}, and the adiabatic phonons are obtained by solving the eigenvalue problem, $\mathrm{det}\big|\tilde{D}^{\alpha\alpha'}_{ss'}(\mathbf{q})-\omega^{2}\big|=0$.
Figure\,\ref{fig:figphdopall}(a)-(j) compare the undoped (see solid thin black lines) and doped adiabatic phonon dispersion relations of the electron-doped monolayer MoS$_{2}$ for all the considered carrier doping concentrations.
The doped dispersions are computed directly from DFPT calculations with the coarse $k$-mesh and smearing parameters (see short-dashed cyan lines), but also from the non-selfconsistent scheme with the finer parameters (see long-dashed magenta lines).
It is specially striking the enhancement of vibrational dips and KAs within the finer evaluation of the lattice vibrations, for which acoustic phonons become softer modes and even lattice instabilities (see Fig.\,\ref{fig:figphdopall}(i)-(j)) with the population of the higher spin-split valleys at the $\overline{\mathrm{Q}}(\overline{\mathrm{Q'}})$ points (see Fig.\,\ref{fig:figeldopall}(i)-(j)).
\clearpage
\section*{S3. Analytical expression for the phonon self-energy for an Einstein-like model of a 2D coupled electron-phonon system}
Our Einstein-like model consists on a bare free electron gas with a single and non-degenerate parabolically dispersing band, $\varepsilon^{\mathbf{k}}=|\mathbf{k}^{2}|/(2m^{*})-E_{\mathrm{F}}$, with $m^{*}$ the electron effective band mass and $E_{\mathrm{F}}$ the energy of the Fermi level.
This band interacts with a single optical phonon mode, supposed here adiabatic, of frequency $\omega_{\mathrm{o}}$ at zero temperature.
The strength of the coupling is defined by means of a constant electron-phonon matrix element $g_{\mathrm{o}}$.
From Eq.\,\ref{eq:phses}, the phonon self-energy that accounts for the non-adiabatic renormalization effects due to the electron-phonon interaction within this model is written as
\begin{equation}
 \tilde{\Pi}_{\mathrm{o}}(\mathbf{q},\omega)=\lim_{\eta\to0^{+}}g^{2}_{\mathrm{o}}\frac{1}{N_{\mathbf{k}}}\sum^{\mathrm{1BZ}}_{\mathbf{k}}\Bigg(\frac{f(\varepsilon^{\mathbf{k}})-f(\varepsilon^{\mathbf{k+q}})}{\varepsilon^{\mathbf{k}}-\varepsilon^{\mathbf{k+q}}+\omega_{\mathrm{o}}+i\eta}-\frac{f(\varepsilon^{\mathbf{k}})-f(\varepsilon^{\mathbf{k+q}})}{\varepsilon^{\mathbf{k}}-\varepsilon^{\mathbf{k+q}}}\Bigg).
 \label{eq:phsena_EMs}
\end{equation}
Intra-valley processes are exclusively considered in the present model.
Since the case we are dealing with is a monolayer, i.e.~a pure 2D material, we conveniently approach the sum in Eq.\,\ref{eq:phsena_EMs} to a 2D integral
\begin{equation}
 \lim_{N_{\mathbf{k}}\to\infty}\frac{1}{N_{\mathbf{k}}}\sum^{\mathrm{1BZ}}_{\mathbf{k}}= A\int\frac{\mathrm{d}\mathbf{k}}{(2\pi)^{2}},
\end{equation}
where $A$ is the area of the unit cell.
Performing the above substitution in Eq.\,\ref{eq:phsena_EMs}, the non-adiabatic phonon self-energy for the 2D Einstein model can be rewritten as
\begin{equation}
 \tilde{\Pi}^{\mathrm{2D}}_{\mathrm{o}}(\mathbf{q},\omega)=g^{2}_{\mathrm{o}}\Big(\chi^{0}_{\mathrm{2D}}(\mathbf{q},\omega)-\chi^{0}_{\mathrm{2D}}(\mathbf{q},0)\Big),
 \label{eq:phsena_EM_2s}
\end{equation}
where $\chi^{0}_{\mathrm{2D}}(\mathbf{q},\omega)$ is the integral expression for the 2D Lindhard function, i.e.~the polarizability or density-response function for the 2D non-interacting free electron gas, given by
\begin{equation}
 \chi^{0}_{\mathrm{2D}}(\mathbf{q},\omega)=\lim_{\eta\to0^{+}}A\int\frac{\mathrm{d}\mathbf{k}}{(2\pi)^{2}}\frac{f(\varepsilon^{\mathbf{k}})-f(\varepsilon^{\mathbf{k+q}})}{\varepsilon^{\mathbf{k}}-\varepsilon^{\mathbf{k+q}}+\omega+i\eta}.
\end{equation}
This integral has the following well-known analytical result~\cite{stern2Ds}
\begin{equation}
 \chi^{0}_{\mathrm{2D}}(\mathbf{q},\omega)=-n_{\mathrm{F}}A\Bigg(1\pm\frac{1}{2q^{2}}\frac{2m^{*}\omega\mp q^{2}}{\big|2m^{*}\omega\mp q^{2}\big|}\sqrt{\big(2m^{*}\omega\mp q^{2}\big)^{2}-\big(2k_{\mathrm{F}}q\big)^{2}}\Bigg),
 \label{eq:LF2Ds}
\end{equation}
where $n_{\mathrm{F}}=m^{*}/(2\pi)$ is the density of states (DOS) of a 2D free electron gas evaluated at the Fermi level and $k_{\mathrm{F}}$ represents the electron momentum at the Fermi level, $k_{\mathrm{F}}=\sqrt{2m^{*}E_{\mathrm{F}}}$.
The analytical expression for the 2D Lindhard function in Eq.\,\ref{eq:LF2Ds} depends explicitly on the latter magnitudes.

This model has been used for evaluating and understanding non-adiabatic renormalization effects on the $\mathrm{A'_{1}}$ optical phonon branch of the electron-doped monolayer MoS$_{2}$ in the small momentum regime, $\mathbf{q}\to\overline{\Gamma}$.
The low-energy range of conduction-bands consists of two almost spin-degenerated valleys in the $\mathbf{k}=\overline{\mathrm{K}}(\overline{\mathrm{K'}})$ points and two spin-split valleys in the $\mathbf{k}=\overline{\mathrm{Q}}(\overline{\mathrm{Q'}})$ points, which each have three equivalent points inside the BZ.
In the model, we consider $4$ equivalent $\overline{\mathrm{K}}$-like valleys as well as $6$ equivalent lower spin-split $\overline{\mathrm{Q}}$-like valleys.
Higher spin-split bands of the $\overline{\mathrm{Q}}(\overline{\mathrm{Q'}})$ points are not included within the model, since the lattice becomes unstable as soon as these valleys are populated (see Fig.2(c) of the main text and Fig.\,\ref{fig:figphdopall}(i)-(j)).
The $\overline{\mathrm{K}}$- and $\overline{\mathrm{Q}}$-like valleys energy dispersions have been characterized by means of a parabolic fit for the \textit{ab initio} calculated structures with respect to the $\overline{\Gamma\mathrm{K}}$ direction.
We obtain $m^{*}(\overline{\mathrm{K}})=0.60$ and $m^{*}(\overline{\mathrm{Q}})=0.80$, respectively.
Following all the above arguments and explanations, we arrive to the following expression for the non-adiabatic phonon self-energy of the 2D Einstein model, equivalent to the Eq.16 of the main text,
\begin{equation}
 \tilde{\Pi}_{\mathrm{A'_{1}}}(\mathbf{q},\omega)=4g^{2}_{\mathrm{A'_{1}}}(\overline{\mathrm{K}})\Big(\chi^{0,\overline{\mathrm{K}}}_{\mathrm{2D}}(\mathbf{q},\omega)-\chi^{0,\overline{\mathrm{K}}}_{\mathrm{2D}}(\mathbf{q},0)\Big)+6g^{2}_{\mathrm{A'_{1}}}(\overline{\mathrm{Q}})\Big(\chi^{0,\overline{\mathrm{Q}}}_{\mathrm{2D}}(\mathbf{q},\omega)-\chi^{0,\overline{\mathrm{Q}}}_{\mathrm{2D}}(\mathbf{q},0)\Big),
\end{equation}
where $\chi^{0,x}_{2\text{D}}(\mathbf{q},\omega)$ and $g_{\mathrm{A'_{1}}}(x)$ represent the 2D Lindhard function and the intra-band electron-phonon matrix elements for the $\mathrm{A'_{1}}$ phonon mode at $\mathbf{q}=\overline{\Gamma}$ interacting with electron states at strictly $x=\overline{\mathrm{K}}$ and $x=\overline{\mathrm{Q}}$ valleys, respectively.
The Fermi level energy, and hence the Fermi momentum, and the electron-phonon matrix elements for each equivalent valley and the $\mathrm{A'_{1}}$ optical phonon branch are taken equal from first-principles calculations for all the considered carrier concentrations and are collected in the Table\,I of the main text.
\clearpage
\section*{S4. Doping-dependent electron conduction-band structure and adiabatic phonon dispersion relation of the monolayer M$\text{o}$S$_2$}
\clearpage
\section*{S5. Doping-dependent phonon spectral function of the monolayer M$\text{o}$S$_2$}
\begin{figure}[ht!]
 \centering
 \begin{center}
  \includegraphics[width=1\columnwidth,angle=0,scale=1.0]{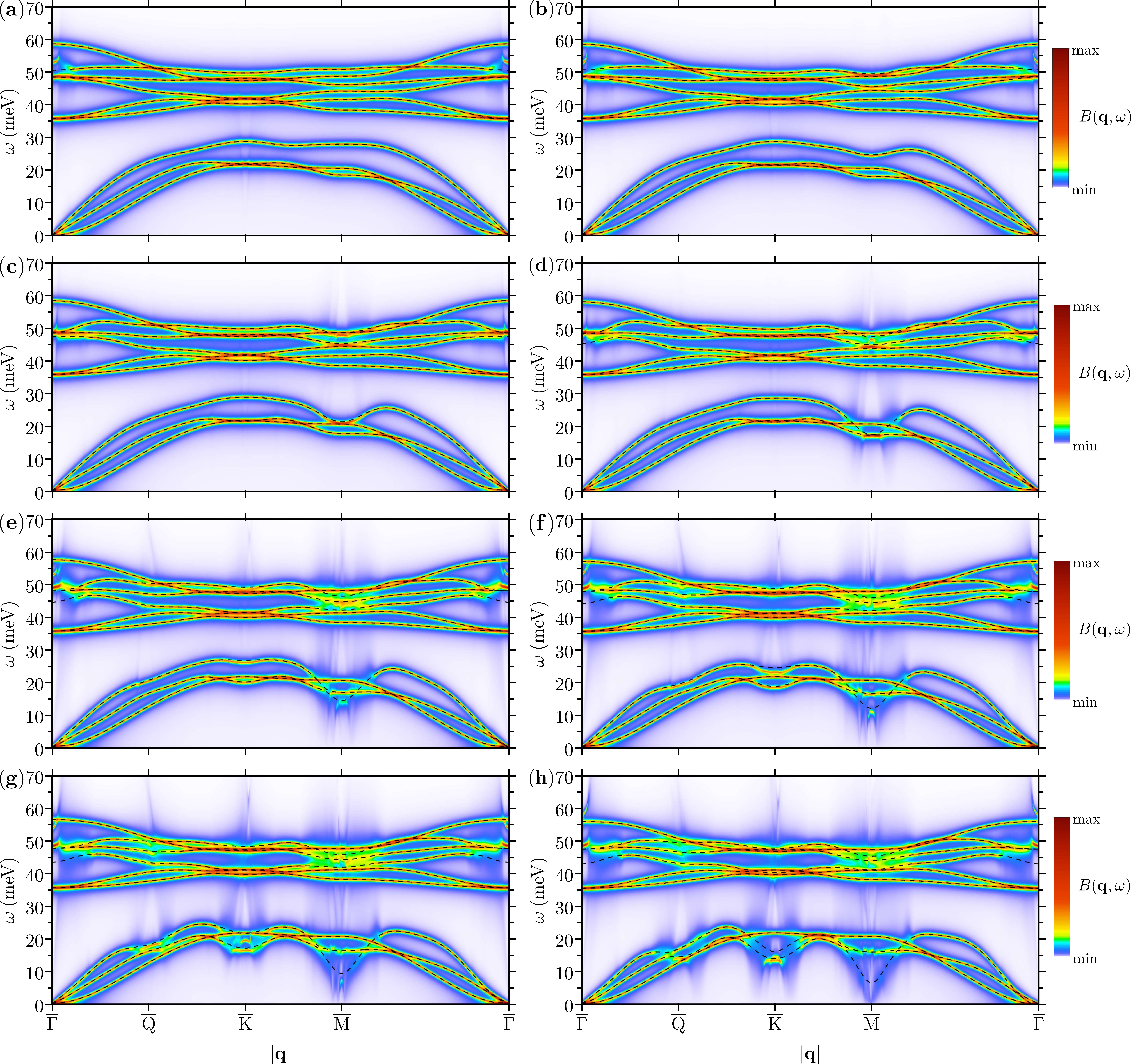}
 \end{center}
 \caption{Density plot of the phonon spectral function of the monolayer MoS$_{2}$ for the electron-doping concentrations for which the lattice is stable $\rho=0.015$ (a), $0.030$ (b), $0.045$ (c), $0.060$ (d), $0.075$ (e), $0.090$ (f), $0.105$ (g) and $0.120~e/\text{u.c.}$ (h). The color code scale represents the height of the spectral function. Dashed black lines represent the adiabatic phonon dipsersions. In those areas where $\text{Im}\tilde{\Pi}_{\nu}(\mathbf{q},\omega)=0$, we use a finite broadening of $\eta=0.35~\text{meV}$ in order to appreciate the different phonon peaks.}
 \label{fig:fignaeffects}
\end{figure}
\begin{figure}[ht!]
 \centering
 \begin{center}
  \includegraphics[width=1\columnwidth,angle=0,scale=1.0]{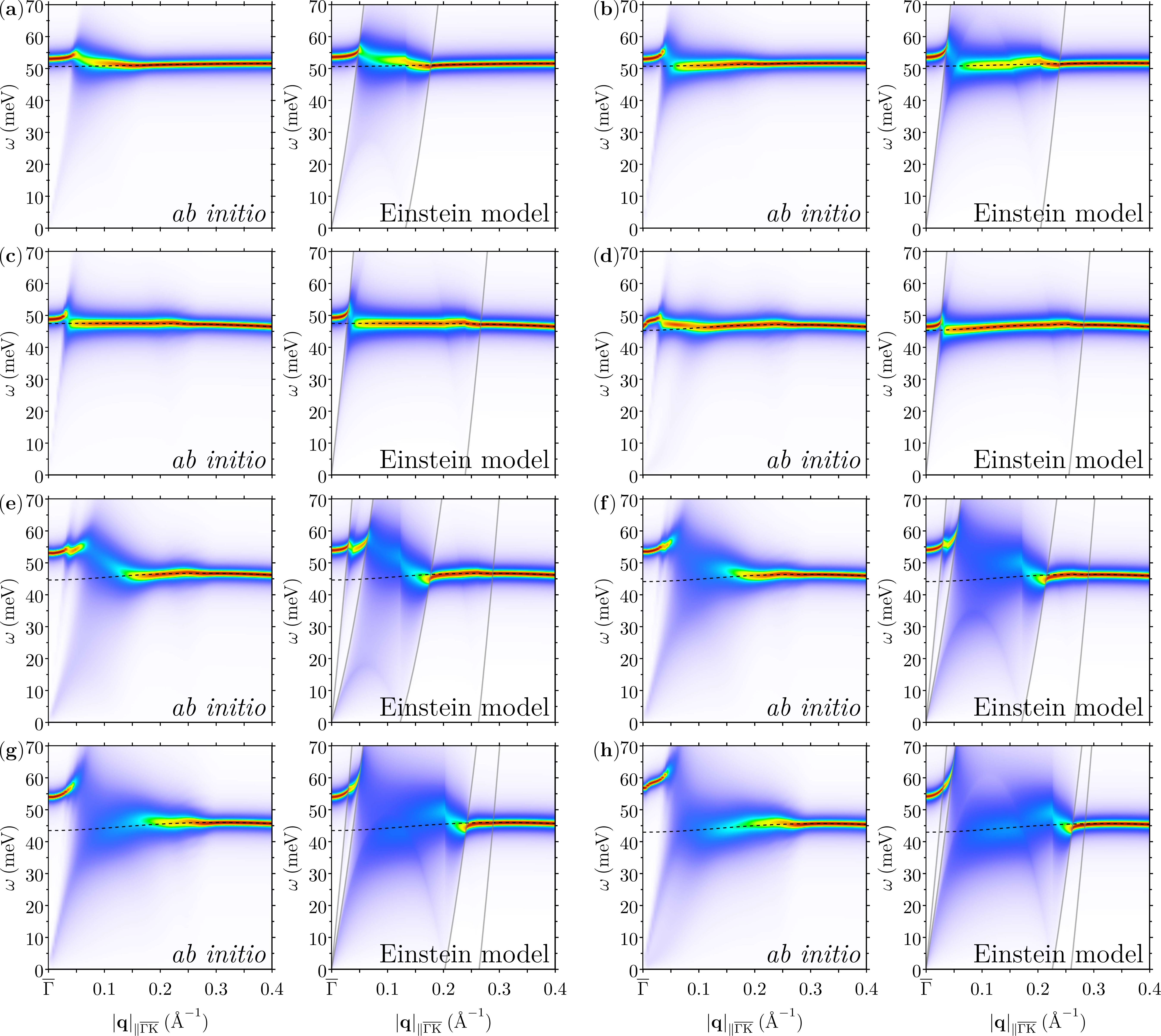}
 \end{center}
 \caption{Density plot of the spectral function of the monolayer MoS$_{2}$ for the $\mathrm{A'_{1}}$ phonon mode obtained by means of \textit{ab initio} calculations (left) and the Einstein-like model (right) for the electron-doping concentrations for which the lattice is stable and in the small momentum regime along the $\overline{\Gamma\mathrm{K}}$ direction. The color code represents the height of the phonon spectral function. Dashed black lines represent the adiabatic dispersions. Solid gray lines show the electron-hole excitation damping continuum boundaries in the simplifed model at zero temperature, $(q^{2}+2qk_{\mathrm{F}}^{x})/2m^{*}_{x}\geq\omega\geq(q^{2}-2qk_{\mathrm{F}}^{x})/2m^{*}_{x}$, where $k_{\mathrm{F}}^{x}$ is the Fermi momentum for each $x=\overline{\mathrm{K}}$ (outer lines) and $\overline{\mathrm{Q}}$-like (inner lines) valley.}
\end{figure}
\clearpage
\section*{S6. The spectral representation and weight of the phonon quasi-particle Laurent expansion}
From Eq.16 of the main text, we have that the first-order Laurent expansion of the Green's function for a dressed phonon of branch index $\nu$ and momentum $\mathbf{q}$ around the vibrational quasi-particle poles is defined as
\begin{equation}
 \mathcal{D}^{\mathrm{qp}}_{\nu}(\mathbf{q},z)=\sum_{j}\frac{\mathbb{Z}^{(j)}_{\nu}(\mathbf{q})}{z-z^{(j)}_{\nu}(\mathbf{q})}-\frac{\mathbb{Z}^{(j)}_{\nu}(\mathbf{q})}{z+z^{(j)}_{\nu}(\mathbf{q})}=\mathbb{Z}^{(j)}_{\nu}(\mathbf{q})\frac{2z^{(j)}_{\nu}(\mathbf{q})}{z^{2}-\Big(z^{(j)}_{\nu}(\mathbf{q})\Big)^{2}},
 \label{eq:Dqps}
\end{equation}
where the index $j$ labels the different solutions of the vibrational Dyson's Equation, with complex poles defined as $z^{(j)}_{\nu}(\mathbf{q})=\Omega^{(j)}_{\nu}(\mathbf{q})-i\gamma^{(j)}_{\nu}(\mathbf{q})$, where $\Omega^{(j)}_{\nu}(\mathbf{q})$ and $\gamma^{(j)}_{\nu}(\mathbf{q})$ are the frequency and linewidth of the $j$-th phonon quasi-particle mode.
$\mathbb{Z}^{(j)}_{\nu}(\mathbf{q})=|\mathbb{Z}^{(j)}_{\nu}(\mathbf{q})|e^{i\theta^{(j)}_{\nu}(\mathbf{q})}$ is the renormalization factor of the phonon quasi-particle pole, which is mathematically defined as the complex residue of $\mathcal{D}_{\nu}(\mathbf{q},z)$ evaluated at $z^{(j)}_{\nu}(\mathbf{q})$ (see Eq.15 of the main text)
The corresponding phonon quasi-particle spectral function of Eq.\,\ref{eq:Dqps} is given by
\begin{equation}
\begin{aligned}
 B^{\mathrm{qp}}_{\nu}(\mathbf{q},\omega)=&-\frac{1}{\pi}\mathrm{Im}\mathcal{D}^{\mathrm{qp}}_{\nu}(\mathbf{q},\omega)=\frac{1}{\pi}\sum_{j}\Bigg[\frac{\big(\Omega^{(j)}_{\nu}(\mathbf{q})\pm\omega\big)|\mathbb{Z}^{(j)}_{\nu}(\mathbf{q})|\sin\big(\theta^{(j)}_{\nu}(\mathbf{q})\big)}{\big(\Omega^{(j)}_{\nu}(\mathbf{q})\pm\omega\big)^{2}+\big(\gamma^{(j)}_{\nu}(\mathbf{q})\big)^{2}}+\frac{\gamma^{(j)}_{\nu}(\mathbf{q})|\mathbb{Z}^{(j)}_{\nu}(\mathbf{q})|\cos\big(\theta^{(j)}_{\nu}(\mathbf{q})\big)}{\big(\Omega^{(j)}_{\nu}(\mathbf{q})\pm\omega\big)^{2}+\big(\gamma^{(j)}_{\nu}(\mathbf{q})\big)^{2}}\Bigg]\\=&\frac{1}{\pi}\sum_{j}2|\mathbb{Z}^{(j)}_{\nu}(\mathbf{q})|\Bigg[\frac{\Big(\big(\Omega^{(j)}_{\nu}(\mathbf{q})\big)^2+(\gamma^{(j)}_{\nu}(\mathbf{q})\big)^{2}\Big)\Big(\gamma^{(j)}_{\nu}(\mathbf{q})\cos\big(\theta^{(j)}_{\nu}(\mathbf{q})\big)+\Omega^{(j)}_{\nu}(\mathbf{q})\sin\big(\theta^{(j)}_{\nu}(\mathbf{q})\big)\Big)}{\big(\Omega^{(j)}_{\nu}(\mathbf{q})\big)^{4}+2\big(\Omega^{(j)}_{\nu}(\mathbf{q})\big)^{2}\Big(\big(\gamma^{(j)}_{\nu}(\mathbf{q})\big)^{2}-\omega^{2}\Big)+\Big(\big(\gamma^{(j)}_{\nu}(\mathbf{q})\big)^{2}+\omega^{2}\Big)}+\\&\frac{\omega^{2}\Big(\gamma^{(j)}_{\nu}(\mathbf{q})\cos\big(\theta^{(j)}_{\nu}(\mathbf{q})\big)-\Omega^{(j)}_{\nu}(\mathbf{q})\sin\big(\theta^{(j)}_{\nu}(\mathbf{q})\big)\Big)}{\big(\Omega^{(j)}_{\nu}(\mathbf{q})\big)^{4}+2\big(\Omega^{(j)}_{\nu}(\mathbf{q})\big)^{2}\Big(\big(\gamma^{(j)}_{\nu}(\mathbf{q})\big)^{2}-\omega^{2}\Big)+\Big(\big(\gamma^{(j)}_{\nu}(\mathbf{q})\big)^{2}+\omega^{2}\Big)}\Bigg].
\end{aligned}
\label{eq:Bqps}
\end{equation}
As one can rapidly see from the second and third lines on the right-hand side of Eq.\,\ref{eq:Bqps}, $B^{\mathrm{qp}}_{\nu}(\mathbf{q},\omega)$ is an even function, i.e.~$B^{\mathrm{qp}}_{\nu}(\mathbf{q},\omega)=B^{\mathrm{qp}}_{\nu}(\mathbf{q},-\omega)$, and therefore the following equality is fulfilled
\begin{equation}
 \int^{\infty}_{0}B^{\mathrm{qp}}_{\nu}(\mathbf{q},\omega)\mathrm{d}\omega=\frac{1}{2}\int^{\infty}_{-\infty}B^{\mathrm{qp}}_{\nu}(\mathbf{q},\omega)\mathrm{d}\omega.
\end{equation}
In this regard, to perform the integration of the first line on the right-hand side of Eq.\,\ref{eq:Bqps} along the whole real axis is a trivial operation, since the integral of the first odd term is null, and that of the second one is a weighted Lorentzian function, whose integral is equal to unity.
Following these arguments, we obtain to the following relation
\begin{equation}
 \int^{\infty}_{0}B^{\mathrm{qp}}_{\nu}(\mathbf{q},\omega)\mathrm{d}\omega=\sum_{j}|\mathbb{Z}^{(j)}_{\nu}(\mathbf{q})|\cos\big(\theta^{(j)}_{\nu}(\mathbf{q})\big)=\sum_{j}\mathrm{Re}\mathbb{Z}^{(j)}_{\nu}(\mathbf{q})\leqslant1,
\end{equation}
which clearly shows that the total vibrational spectral weight coming from the phonon quasi-particle modes is equal to the sum of the real parts of the renormalization factors, and must be smaller than or equal to unity.

\begin{thebibliography}{64}%
\makeatletter
\providecommand \@ifxundefined [1]{%
 \@ifx{#1\undefined}
}%
\providecommand \@ifnum [1]{%
 \ifnum #1\expandafter \@firstoftwo
 \else \expandafter \@secondoftwo
 \fi
}%
\providecommand \@ifx [1]{%
 \ifx #1\expandafter \@firstoftwo
 \else \expandafter \@secondoftwo
 \fi
}%
\providecommand \natexlab [1]{#1}%
\providecommand \enquote  [1]{``#1''}%
\providecommand \bibnamefont  [1]{#1}%
\providecommand \bibfnamefont [1]{#1}%
\providecommand \citenamefont [1]{#1}%
\providecommand \href@noop [0]{\@secondoftwo}%
\providecommand \href [0]{\begingroup \@sanitize@url \@href}%
\providecommand \@href[1]{\@@startlink{#1}\@@href}%
\providecommand \@@href[1]{\endgroup#1\@@endlink}%
\providecommand \@sanitize@url [0]{\catcode `\\12\catcode `\$12\catcode
  `\&12\catcode `\#12\catcode `\^12\catcode `\_12\catcode `\%12\relax}%
\providecommand \@@startlink[1]{}%
\providecommand \@@endlink[0]{}%
\providecommand \url  [0]{\begingroup\@sanitize@url \@url }%
\providecommand \@url [1]{\endgroup\@href {#1}{\urlprefix }}%
\providecommand \urlprefix  [0]{URL }%
\providecommand \Eprint [0]{\href }%
\providecommand \doibase [0]{http://dx.doi.org/}%
\providecommand \selectlanguage [0]{\@gobble}%
\providecommand \bibinfo  [0]{\@secondoftwo}%
\providecommand \bibfield  [0]{\@secondoftwo}%
\providecommand \translation [1]{[#1]}%
\providecommand \BibitemOpen [0]{}%
\providecommand \bibitemStop [0]{}%
\providecommand \bibitemNoStop [0]{.\EOS\space}%
\providecommand \EOS [0]{\spacefactor3000\relax}%
\providecommand \BibitemShut  [1]{\csname bibitem#1\endcsname}%
\let\auto@bib@innerbib\@empty
\bibitem [{\citenamefont {Hohenberg}\ and\ \citenamefont {Kohn}(1964)}]{DFT1}%
  \BibitemOpen
  \bibfield  {author} {\bibinfo {author} {\bibfnamefont {P.}~\bibnamefont
  {Hohenberg}}\ and\ \bibinfo {author} {\bibfnamefont {W.}~\bibnamefont
  {Kohn}},\ }\href {\doibase 10.1103/PhysRev.136.B864} {\bibfield  {journal}
  {\bibinfo  {journal} {Phys. Rev.}\ }\textbf {\bibinfo {volume} {136}},\
  \bibinfo {pages} {B864} (\bibinfo {year} {1964})}\BibitemShut {NoStop}%
\bibitem [{\citenamefont {Kohn}\ and\ \citenamefont {Sham}(1965)}]{DFT2}%
  \BibitemOpen
  \bibfield  {author} {\bibinfo {author} {\bibfnamefont {W.}~\bibnamefont
  {Kohn}}\ and\ \bibinfo {author} {\bibfnamefont {L.~J.}\ \bibnamefont
  {Sham}},\ }\href {\doibase 10.1103/PhysRev.140.A1133} {\bibfield  {journal}
  {\bibinfo  {journal} {Phys. Rev.}\ }\textbf {\bibinfo {volume} {140}},\
  \bibinfo {pages} {A1133} (\bibinfo {year} {1965})}\BibitemShut {NoStop}%
\bibitem [{\citenamefont {Baroni}\ \emph {et~al.}(2001)\citenamefont {Baroni},
  \citenamefont {de~Gironcoli}, \citenamefont {Dal~Corso},\ and\ \citenamefont
  {Giannozzi}}]{DFPT}%
  \BibitemOpen
  \bibfield  {author} {\bibinfo {author} {\bibfnamefont {S.}~\bibnamefont
  {Baroni}}, \bibinfo {author} {\bibfnamefont {S.}~\bibnamefont
  {de~Gironcoli}}, \bibinfo {author} {\bibfnamefont {A.}~\bibnamefont
  {Dal~Corso}}, \ and\ \bibinfo {author} {\bibfnamefont {P.}~\bibnamefont
  {Giannozzi}},\ }\href {\doibase 10.1103/RevModPhys.73.515} {\bibfield
  {journal} {\bibinfo  {journal} {Rev. Mod. Phys.}\ }\textbf {\bibinfo {volume}
  {73}},\ \bibinfo {pages} {515} (\bibinfo {year} {2001})}\BibitemShut
  {NoStop}%
\bibitem [{\citenamefont {Born}\ and\ \citenamefont
  {Oppenheimer}(1927)}]{BOapprox}%
  \BibitemOpen
  \bibfield  {author} {\bibinfo {author} {\bibfnamefont {M.}~\bibnamefont
  {Born}}\ and\ \bibinfo {author} {\bibfnamefont {R.}~\bibnamefont
  {Oppenheimer}},\ }\href {\doibase 10.1002/andp.19273892002} {\bibfield
  {journal} {\bibinfo  {journal} {Annalen der Physik}\ }\textbf {\bibinfo
  {volume} {389}},\ \bibinfo {pages} {457} (\bibinfo {year}
  {1927})}\BibitemShut {NoStop}%
\bibitem [{\citenamefont {Savrasov}\ \emph {et~al.}(1994)\citenamefont
  {Savrasov}, \citenamefont {Savrasov},\ and\ \citenamefont
  {Andersen}}]{sabrasovepis1}%
  \BibitemOpen
  \bibfield  {author} {\bibinfo {author} {\bibfnamefont {S.~Y.}\ \bibnamefont
  {Savrasov}}, \bibinfo {author} {\bibfnamefont {D.~Y.}\ \bibnamefont
  {Savrasov}}, \ and\ \bibinfo {author} {\bibfnamefont {O.~K.}\ \bibnamefont
  {Andersen}},\ }\href {\doibase 10.1103/PhysRevLett.72.372} {\bibfield
  {journal} {\bibinfo  {journal} {Phys. Rev. Lett.}\ }\textbf {\bibinfo
  {volume} {72}},\ \bibinfo {pages} {372} (\bibinfo {year} {1994})}\BibitemShut
  {NoStop}%
\bibitem [{\citenamefont {Savrasov}\ and\ \citenamefont
  {Savrasov}(1996)}]{sabrasovepis2}%
  \BibitemOpen
  \bibfield  {author} {\bibinfo {author} {\bibfnamefont {S.~Y.}\ \bibnamefont
  {Savrasov}}\ and\ \bibinfo {author} {\bibfnamefont {D.~Y.}\ \bibnamefont
  {Savrasov}},\ }\href {\doibase 10.1103/PhysRevB.54.16487} {\bibfield
  {journal} {\bibinfo  {journal} {Phys. Rev. B}\ }\textbf {\bibinfo {volume}
  {54}},\ \bibinfo {pages} {16487} (\bibinfo {year} {1996})}\BibitemShut
  {NoStop}%
\bibitem [{\citenamefont {Liu}\ and\ \citenamefont {Quong}(1996)}]{liuepis}%
  \BibitemOpen
  \bibfield  {author} {\bibinfo {author} {\bibfnamefont {A.~Y.}\ \bibnamefont
  {Liu}}\ and\ \bibinfo {author} {\bibfnamefont {A.~A.}\ \bibnamefont
  {Quong}},\ }\href {\doibase 10.1103/PhysRevB.53.R7575} {\bibfield  {journal}
  {\bibinfo  {journal} {Phys. Rev. B}\ }\textbf {\bibinfo {volume} {53}},\
  \bibinfo {pages} {R7575} (\bibinfo {year} {1996})}\BibitemShut {NoStop}%
\bibitem [{\citenamefont {Mauri}\ \emph {et~al.}(1996)\citenamefont {Mauri},
  \citenamefont {Zakharov}, \citenamefont {de~Gironcoli}, \citenamefont
  {Louie},\ and\ \citenamefont {Cohen}}]{mauriepis}%
  \BibitemOpen
  \bibfield  {author} {\bibinfo {author} {\bibfnamefont {F.}~\bibnamefont
  {Mauri}}, \bibinfo {author} {\bibfnamefont {O.}~\bibnamefont {Zakharov}},
  \bibinfo {author} {\bibfnamefont {S.}~\bibnamefont {de~Gironcoli}}, \bibinfo
  {author} {\bibfnamefont {S.~G.}\ \bibnamefont {Louie}}, \ and\ \bibinfo
  {author} {\bibfnamefont {M.~L.}\ \bibnamefont {Cohen}},\ }\href {\doibase
  10.1103/PhysRevLett.77.1151} {\bibfield  {journal} {\bibinfo  {journal}
  {Phys. Rev. Lett.}\ }\textbf {\bibinfo {volume} {77}},\ \bibinfo {pages}
  {1151} (\bibinfo {year} {1996})}\BibitemShut {NoStop}%
\bibitem [{\citenamefont {Bauer}\ \emph {et~al.}(1998)\citenamefont {Bauer},
  \citenamefont {Schmid}, \citenamefont {Pavone},\ and\ \citenamefont
  {Strauch}}]{bauerepis}%
  \BibitemOpen
  \bibfield  {author} {\bibinfo {author} {\bibfnamefont {R.}~\bibnamefont
  {Bauer}}, \bibinfo {author} {\bibfnamefont {A.}~\bibnamefont {Schmid}},
  \bibinfo {author} {\bibfnamefont {P.}~\bibnamefont {Pavone}}, \ and\ \bibinfo
  {author} {\bibfnamefont {D.}~\bibnamefont {Strauch}},\ }\href {\doibase
  10.1103/PhysRevB.57.11276} {\bibfield  {journal} {\bibinfo  {journal} {Phys.
  Rev. B}\ }\textbf {\bibinfo {volume} {57}},\ \bibinfo {pages} {11276}
  (\bibinfo {year} {1998})}\BibitemShut {NoStop}%
\bibitem [{\citenamefont {Migdal}(1958)}]{Migdal}%
  \BibitemOpen
  \bibfield  {author} {\bibinfo {author} {\bibfnamefont {A.~B.}\ \bibnamefont
  {Migdal}},\ }\href@noop {} {\bibfield  {journal} {\bibinfo  {journal} {Sov.
  Phys. - JETP Lett.}\ }\textbf {\bibinfo {volume} {34}},\ \bibinfo {pages}
  {996} (\bibinfo {year} {1958})}\BibitemShut {NoStop}%
\bibitem [{\citenamefont {Engelsberg}\ and\ \citenamefont
  {Schrieffer}(1963)}]{ES}%
  \BibitemOpen
  \bibfield  {author} {\bibinfo {author} {\bibfnamefont {S.}~\bibnamefont
  {Engelsberg}}\ and\ \bibinfo {author} {\bibfnamefont {J.~R.}\ \bibnamefont
  {Schrieffer}},\ }\href {\doibase 10.1103/PhysRev.131.993} {\bibfield
  {journal} {\bibinfo  {journal} {Phys. Rev.}\ }\textbf {\bibinfo {volume}
  {131}},\ \bibinfo {pages} {993} (\bibinfo {year} {1963})}\BibitemShut
  {NoStop}%
\bibitem [{\citenamefont {Ipatova}\ and\ \citenamefont
  {Subashiev}(1974)}]{Ipatova}%
  \BibitemOpen
  \bibfield  {author} {\bibinfo {author} {\bibfnamefont {I.~P.}\ \bibnamefont
  {Ipatova}}\ and\ \bibinfo {author} {\bibfnamefont {A.~V.}\ \bibnamefont
  {Subashiev}},\ }\href@noop {} {\bibfield  {journal} {\bibinfo  {journal}
  {Sov. Phys. - JETP Lett.}\ }\textbf {\bibinfo {volume} {39}},\ \bibinfo
  {pages} {349} (\bibinfo {year} {1974})}\BibitemShut {NoStop}%
\bibitem [{\citenamefont {Maksimov}\ and\ \citenamefont
  {Shulga}(1996)}]{Maksimov1996}%
  \BibitemOpen
  \bibfield  {author} {\bibinfo {author} {\bibfnamefont {E.}~\bibnamefont
  {Maksimov}}\ and\ \bibinfo {author} {\bibfnamefont {S.}~\bibnamefont
  {Shulga}},\ }\href {\doibase https://doi.org/10.1016/0038-1098(95)00745-8}
  {\bibfield  {journal} {\bibinfo  {journal} {Solid State Communications}\
  }\textbf {\bibinfo {volume} {97}},\ \bibinfo {pages} {553 } (\bibinfo {year}
  {1996})}\BibitemShut {NoStop}%
\bibitem [{\citenamefont {Ponosov}\ and\ \citenamefont
  {Streltsov}(2016)}]{ponosovna}%
  \BibitemOpen
  \bibfield  {author} {\bibinfo {author} {\bibfnamefont {Y.~S.}\ \bibnamefont
  {Ponosov}}\ and\ \bibinfo {author} {\bibfnamefont {S.~V.}\ \bibnamefont
  {Streltsov}},\ }\href {\doibase 10.1103/PhysRevB.94.214302} {\bibfield
  {journal} {\bibinfo  {journal} {Phys. Rev. B}\ }\textbf {\bibinfo {volume}
  {94}},\ \bibinfo {pages} {214302} (\bibinfo {year} {2016})}\BibitemShut
  {NoStop}%
\bibitem [{\citenamefont {Maksimov}\ \emph {et~al.}(2010)\citenamefont
  {Maksimov}, \citenamefont {Kuli\'c},\ and\ \citenamefont
  {Dolgov}}]{maksimovphSF}%
  \BibitemOpen
  \bibfield  {author} {\bibinfo {author} {\bibfnamefont {E.~G.}\ \bibnamefont
  {Maksimov}}, \bibinfo {author} {\bibfnamefont {M.~L.}\ \bibnamefont
  {Kuli\'c}}, \ and\ \bibinfo {author} {\bibfnamefont {O.}~\bibnamefont
  {Dolgov}},\ }\href {\doibase 10.1155/2010/423725} {\bibfield  {journal}
  {\bibinfo  {journal} {Advances in Condensed Matter Physics}\ }\textbf
  {\bibinfo {volume} {2010}},\ \bibinfo {pages} {423725} (\bibinfo {year}
  {2010})}\BibitemShut {NoStop}%
\bibitem [{\citenamefont {Caruso}\ \emph {et~al.}(2017)\citenamefont {Caruso},
  \citenamefont {Hoesch}, \citenamefont {Achatz}, \citenamefont {Serrano},
  \citenamefont {Krisch}, \citenamefont {Bustarret},\ and\ \citenamefont
  {Giustino}}]{NAKAgiustino}%
  \BibitemOpen
  \bibfield  {author} {\bibinfo {author} {\bibfnamefont {F.}~\bibnamefont
  {Caruso}}, \bibinfo {author} {\bibfnamefont {M.}~\bibnamefont {Hoesch}},
  \bibinfo {author} {\bibfnamefont {P.}~\bibnamefont {Achatz}}, \bibinfo
  {author} {\bibfnamefont {J.}~\bibnamefont {Serrano}}, \bibinfo {author}
  {\bibfnamefont {M.}~\bibnamefont {Krisch}}, \bibinfo {author} {\bibfnamefont
  {E.}~\bibnamefont {Bustarret}}, \ and\ \bibinfo {author} {\bibfnamefont
  {F.}~\bibnamefont {Giustino}},\ }\href {\doibase
  10.1103/PhysRevLett.119.017001} {\bibfield  {journal} {\bibinfo  {journal}
  {Phys. Rev. Lett.}\ }\textbf {\bibinfo {volume} {119}},\ \bibinfo {pages}
  {017001} (\bibinfo {year} {2017})}\BibitemShut {NoStop}%
\bibitem [{\citenamefont {Lazzeri}\ and\ \citenamefont
  {Mauri}(2006)}]{NAKohnGraph}%
  \BibitemOpen
  \bibfield  {author} {\bibinfo {author} {\bibfnamefont {M.}~\bibnamefont
  {Lazzeri}}\ and\ \bibinfo {author} {\bibfnamefont {F.}~\bibnamefont
  {Mauri}},\ }\href {\doibase 10.1103/PhysRevLett.97.266407} {\bibfield
  {journal} {\bibinfo  {journal} {Phys. Rev. Lett.}\ }\textbf {\bibinfo
  {volume} {97}},\ \bibinfo {pages} {266407} (\bibinfo {year}
  {2006})}\BibitemShut {NoStop}%
\bibitem [{\citenamefont {Pisana}\ \emph {et~al.}(2007)\citenamefont {Pisana},
  \citenamefont {Lazzeri}, \citenamefont {Casiraghi}, \citenamefont
  {Novoselov}, \citenamefont {Geim}, \citenamefont {Ferrari},\ and\
  \citenamefont {Mauri}}]{failureBOgraph}%
  \BibitemOpen
  \bibfield  {author} {\bibinfo {author} {\bibfnamefont {S.}~\bibnamefont
  {Pisana}}, \bibinfo {author} {\bibfnamefont {M.}~\bibnamefont {Lazzeri}},
  \bibinfo {author} {\bibfnamefont {C.}~\bibnamefont {Casiraghi}}, \bibinfo
  {author} {\bibfnamefont {K.~S.}\ \bibnamefont {Novoselov}}, \bibinfo {author}
  {\bibfnamefont {A.~K.}\ \bibnamefont {Geim}}, \bibinfo {author}
  {\bibfnamefont {A.~C.}\ \bibnamefont {Ferrari}}, \ and\ \bibinfo {author}
  {\bibfnamefont {F.}~\bibnamefont {Mauri}},\ }\href
  {https://doi.org/10.1038/nmat1846} {\bibfield  {journal} {\bibinfo  {journal}
  {Nature Materials}\ }\textbf {\bibinfo {volume} {6}},\ \bibinfo {pages} {198}
  (\bibinfo {year} {2007})}\BibitemShut {NoStop}%
\bibitem [{\citenamefont {Piscanec}\ \emph {et~al.}(2007)\citenamefont
  {Piscanec}, \citenamefont {Lazzeri}, \citenamefont {Robertson}, \citenamefont
  {Ferrari},\ and\ \citenamefont {Mauri}}]{NAcarbnanotube}%
  \BibitemOpen
  \bibfield  {author} {\bibinfo {author} {\bibfnamefont {S.}~\bibnamefont
  {Piscanec}}, \bibinfo {author} {\bibfnamefont {M.}~\bibnamefont {Lazzeri}},
  \bibinfo {author} {\bibfnamefont {J.}~\bibnamefont {Robertson}}, \bibinfo
  {author} {\bibfnamefont {A.~C.}\ \bibnamefont {Ferrari}}, \ and\ \bibinfo
  {author} {\bibfnamefont {F.}~\bibnamefont {Mauri}},\ }\href {\doibase
  10.1103/PhysRevB.75.035427} {\bibfield  {journal} {\bibinfo  {journal} {Phys.
  Rev. B}\ }\textbf {\bibinfo {volume} {75}},\ \bibinfo {pages} {035427}
  (\bibinfo {year} {2007})}\BibitemShut {NoStop}%
\bibitem [{\citenamefont {Saitta}\ \emph {et~al.}(2008)\citenamefont {Saitta},
  \citenamefont {Lazzeri}, \citenamefont {Calandra},\ and\ \citenamefont
  {Mauri}}]{Saittaprl2008}%
  \BibitemOpen
  \bibfield  {author} {\bibinfo {author} {\bibfnamefont {A.~M.}\ \bibnamefont
  {Saitta}}, \bibinfo {author} {\bibfnamefont {M.}~\bibnamefont {Lazzeri}},
  \bibinfo {author} {\bibfnamefont {M.}~\bibnamefont {Calandra}}, \ and\
  \bibinfo {author} {\bibfnamefont {F.}~\bibnamefont {Mauri}},\ }\href
  {\doibase 10.1103/PhysRevLett.100.226401} {\bibfield  {journal} {\bibinfo
  {journal} {Phys. Rev. Lett.}\ }\textbf {\bibinfo {volume} {100}},\ \bibinfo
  {pages} {226401} (\bibinfo {year} {2008})}\BibitemShut {NoStop}%
\bibitem [{\citenamefont {Calandra}\ and\ \citenamefont
  {Mauri}(2005)}]{calandramgb2}%
  \BibitemOpen
  \bibfield  {author} {\bibinfo {author} {\bibfnamefont {M.}~\bibnamefont
  {Calandra}}\ and\ \bibinfo {author} {\bibfnamefont {F.}~\bibnamefont
  {Mauri}},\ }\href {\doibase 10.1103/PhysRevB.71.064501} {\bibfield  {journal}
  {\bibinfo  {journal} {Phys. Rev. B}\ }\textbf {\bibinfo {volume} {71}},\
  \bibinfo {pages} {064501} (\bibinfo {year} {2005})}\BibitemShut {NoStop}%
\bibitem [{\citenamefont {Cappelluti}(2006)}]{capeluttimgb2}%
  \BibitemOpen
  \bibfield  {author} {\bibinfo {author} {\bibfnamefont {E.}~\bibnamefont
  {Cappelluti}},\ }\href {\doibase 10.1103/PhysRevB.73.140505} {\bibfield
  {journal} {\bibinfo  {journal} {Phys. Rev. B}\ }\textbf {\bibinfo {volume}
  {73}},\ \bibinfo {pages} {140505(R)} (\bibinfo {year} {2006})}\BibitemShut
  {NoStop}%
\bibitem [{\citenamefont {Calandra}\ \emph {et~al.}(2010)\citenamefont
  {Calandra}, \citenamefont {Profeta},\ and\ \citenamefont
  {Mauri}}]{calandramauriprb2010}%
  \BibitemOpen
  \bibfield  {author} {\bibinfo {author} {\bibfnamefont {M.}~\bibnamefont
  {Calandra}}, \bibinfo {author} {\bibfnamefont {G.}~\bibnamefont {Profeta}}, \
  and\ \bibinfo {author} {\bibfnamefont {F.}~\bibnamefont {Mauri}},\ }\href
  {\doibase 10.1103/PhysRevB.82.165111} {\bibfield  {journal} {\bibinfo
  {journal} {Phys. Rev. B}\ }\textbf {\bibinfo {volume} {82}},\ \bibinfo
  {pages} {165111} (\bibinfo {year} {2010})}\BibitemShut {NoStop}%
\bibitem [{\citenamefont {Chakraborty}\ \emph {et~al.}(2012)\citenamefont
  {Chakraborty}, \citenamefont {Bera}, \citenamefont {Muthu}, \citenamefont
  {Bhowmick}, \citenamefont {Waghmare},\ and\ \citenamefont
  {Sood}}]{1lmos2a1soft}%
  \BibitemOpen
  \bibfield  {author} {\bibinfo {author} {\bibfnamefont {B.}~\bibnamefont
  {Chakraborty}}, \bibinfo {author} {\bibfnamefont {A.}~\bibnamefont {Bera}},
  \bibinfo {author} {\bibfnamefont {D.~V.~S.}\ \bibnamefont {Muthu}}, \bibinfo
  {author} {\bibfnamefont {S.}~\bibnamefont {Bhowmick}}, \bibinfo {author}
  {\bibfnamefont {U.~V.}\ \bibnamefont {Waghmare}}, \ and\ \bibinfo {author}
  {\bibfnamefont {A.~K.}\ \bibnamefont {Sood}},\ }\href {\doibase
  10.1103/PhysRevB.85.161403} {\bibfield  {journal} {\bibinfo  {journal} {Phys.
  Rev. B}\ }\textbf {\bibinfo {volume} {85}},\ \bibinfo {pages} {161403(R)}
  (\bibinfo {year} {2012})}\BibitemShut {NoStop}%
\bibitem [{\citenamefont {Sohier}\ \emph {et~al.}(2019)\citenamefont {Sohier},
  \citenamefont {Ponomarev}, \citenamefont {Gibertini}, \citenamefont {Berger},
  \citenamefont {Marzari}, \citenamefont {Ubrig},\ and\ \citenamefont
  {Morpurgo}}]{sohierprx2019}%
  \BibitemOpen
  \bibfield  {author} {\bibinfo {author} {\bibfnamefont {T.}~\bibnamefont
  {Sohier}}, \bibinfo {author} {\bibfnamefont {E.}~\bibnamefont {Ponomarev}},
  \bibinfo {author} {\bibfnamefont {M.}~\bibnamefont {Gibertini}}, \bibinfo
  {author} {\bibfnamefont {H.}~\bibnamefont {Berger}}, \bibinfo {author}
  {\bibfnamefont {N.}~\bibnamefont {Marzari}}, \bibinfo {author} {\bibfnamefont
  {N.}~\bibnamefont {Ubrig}}, \ and\ \bibinfo {author} {\bibfnamefont {A.~F.}\
  \bibnamefont {Morpurgo}},\ }\href {\doibase 10.1103/PhysRevX.9.031019}
  {\bibfield  {journal} {\bibinfo  {journal} {Phys. Rev. X}\ }\textbf {\bibinfo
  {volume} {9}},\ \bibinfo {pages} {031019} (\bibinfo {year}
  {2019})}\BibitemShut {NoStop}%
\bibitem [{\citenamefont {Novko}(2019)}]{1907.04766}%
  \BibitemOpen
  \bibfield  {author} {\bibinfo {author} {\bibfnamefont {D.}~\bibnamefont
  {Novko}},\ }\href@noop {} {\enquote {\bibinfo {title} {Broken adiabaticity
  induced by lifshitz transition in {M}o{S}${}_{2}$ and {W}{S}${}_{2}$ single
  layers},}\ } (\bibinfo {year} {2019}),\ \Eprint
  {http://arxiv.org/abs/arXiv:1907.04766} {arXiv:1907.04766} \BibitemShut
  {NoStop}%
\bibitem [{\citenamefont {Ye}\ \emph {et~al.}(2012)\citenamefont {Ye},
  \citenamefont {Zhang}, \citenamefont {Akashi}, \citenamefont {Bahramy},
  \citenamefont {Arita},\ and\ \citenamefont {Iwasa}}]{scmos2}%
  \BibitemOpen
  \bibfield  {author} {\bibinfo {author} {\bibfnamefont {J.~T.}\ \bibnamefont
  {Ye}}, \bibinfo {author} {\bibfnamefont {Y.~J.}\ \bibnamefont {Zhang}},
  \bibinfo {author} {\bibfnamefont {R.}~\bibnamefont {Akashi}}, \bibinfo
  {author} {\bibfnamefont {M.~S.}\ \bibnamefont {Bahramy}}, \bibinfo {author}
  {\bibfnamefont {R.}~\bibnamefont {Arita}}, \ and\ \bibinfo {author}
  {\bibfnamefont {Y.}~\bibnamefont {Iwasa}},\ }\href {\doibase
  10.1126/science.1228006} {\bibfield  {journal} {\bibinfo  {journal}
  {Science}\ }\textbf {\bibinfo {volume} {338}},\ \bibinfo {pages} {1193}
  (\bibinfo {year} {2012})}\BibitemShut {NoStop}%
\bibitem [{\citenamefont {Kang}\ \emph {et~al.}(2018)\citenamefont {Kang},
  \citenamefont {Jung}, \citenamefont {Shin}, \citenamefont {Sohn},
  \citenamefont {Ryu}, \citenamefont {Kim}, \citenamefont {Hoesch},\ and\
  \citenamefont {Kim}}]{Kang2018}%
  \BibitemOpen
  \bibfield  {author} {\bibinfo {author} {\bibfnamefont {M.}~\bibnamefont
  {Kang}}, \bibinfo {author} {\bibfnamefont {S.~W.}\ \bibnamefont {Jung}},
  \bibinfo {author} {\bibfnamefont {W.~J.}\ \bibnamefont {Shin}}, \bibinfo
  {author} {\bibfnamefont {Y.}~\bibnamefont {Sohn}}, \bibinfo {author}
  {\bibfnamefont {S.~H.}\ \bibnamefont {Ryu}}, \bibinfo {author} {\bibfnamefont
  {T.~K.}\ \bibnamefont {Kim}}, \bibinfo {author} {\bibfnamefont
  {M.}~\bibnamefont {Hoesch}}, \ and\ \bibinfo {author} {\bibfnamefont {K.~S.}\
  \bibnamefont {Kim}},\ }\href {\doibase 10.1038/s41563-018-0092-7} {\bibfield
  {journal} {\bibinfo  {journal} {Nature Materials}\ }\textbf {\bibinfo
  {volume} {17}},\ \bibinfo {pages} {676} (\bibinfo {year} {2018})}\BibitemShut
  {NoStop}%
\bibitem [{\citenamefont {Ge}\ and\ \citenamefont {Liu}(2013)}]{yizhiprb2013}%
  \BibitemOpen
  \bibfield  {author} {\bibinfo {author} {\bibfnamefont {Y.}~\bibnamefont
  {Ge}}\ and\ \bibinfo {author} {\bibfnamefont {A.~Y.}\ \bibnamefont {Liu}},\
  }\href {\doibase 10.1103/PhysRevB.87.241408} {\bibfield  {journal} {\bibinfo
  {journal} {Phys. Rev. B}\ }\textbf {\bibinfo {volume} {87}},\ \bibinfo
  {pages} {241408(R)} (\bibinfo {year} {2013})}\BibitemShut {NoStop}%
\bibitem [{\citenamefont {Piatti}\ \emph {et~al.}(2018)\citenamefont {Piatti},
  \citenamefont {De~Fazio}, \citenamefont {Daghero}, \citenamefont
  {Tamalampudi}, \citenamefont {Yoon}, \citenamefont {Ferrari},\ and\
  \citenamefont {Gonnelli}}]{piattimultivalleymos2}%
  \BibitemOpen
  \bibfield  {author} {\bibinfo {author} {\bibfnamefont {E.}~\bibnamefont
  {Piatti}}, \bibinfo {author} {\bibfnamefont {D.}~\bibnamefont {De~Fazio}},
  \bibinfo {author} {\bibfnamefont {D.}~\bibnamefont {Daghero}}, \bibinfo
  {author} {\bibfnamefont {S.~R.}\ \bibnamefont {Tamalampudi}}, \bibinfo
  {author} {\bibfnamefont {D.}~\bibnamefont {Yoon}}, \bibinfo {author}
  {\bibfnamefont {A.~C.}\ \bibnamefont {Ferrari}}, \ and\ \bibinfo {author}
  {\bibfnamefont {R.~S.}\ \bibnamefont {Gonnelli}},\ }\href {\doibase
  10.1021/acs.nanolett.8b01390} {\bibfield  {journal} {\bibinfo  {journal}
  {Nano Letters}\ }\textbf {\bibinfo {volume} {18}},\ \bibinfo {pages} {4821}
  (\bibinfo {year} {2018})},\ \bibinfo {note} {pMID: 29949374},\ \Eprint
  {http://arxiv.org/abs/https://doi.org/10.1021/acs.nanolett.8b01390}
  {https://doi.org/10.1021/acs.nanolett.8b01390} \BibitemShut {NoStop}%
\bibitem [{\citenamefont {Garcia-Goiricelaya}\ \emph
  {et~al.}(2019)\citenamefont {Garcia-Goiricelaya}, \citenamefont
  {Lafuente-Bartolome}, \citenamefont {Gurtubay},\ and\ \citenamefont
  {Eiguren}}]{gurepaper}%
  \BibitemOpen
  \bibfield  {author} {\bibinfo {author} {\bibfnamefont {P.}~\bibnamefont
  {Garcia-Goiricelaya}}, \bibinfo {author} {\bibfnamefont {J.}~\bibnamefont
  {Lafuente-Bartolome}}, \bibinfo {author} {\bibfnamefont {I.~G.}\ \bibnamefont
  {Gurtubay}}, \ and\ \bibinfo {author} {\bibfnamefont {A.}~\bibnamefont
  {Eiguren}},\ }\href {\doibase 10.1038/s42005-019-0182-0} {\bibfield
  {journal} {\bibinfo  {journal} {Communications Physics}\ }\textbf {\bibinfo
  {volume} {2}},\ \bibinfo {pages} {81} (\bibinfo {year} {2019})}\BibitemShut
  {NoStop}%
\bibitem [{\citenamefont {Fu}\ \emph {et~al.}(2017)\citenamefont {Fu},
  \citenamefont {Liu}, \citenamefont {Yuan}, \citenamefont {Tang},
  \citenamefont {Lian}, \citenamefont {Xu}, \citenamefont {Zeng}, \citenamefont
  {Chen}, \citenamefont {Wang}, \citenamefont {Zhou}, \citenamefont {Xu},
  \citenamefont {Gao}, \citenamefont {Pan}, \citenamefont {Wang}, \citenamefont
  {Wang}, \citenamefont {Zhang}, \citenamefont {Cui}, \citenamefont {Hwang},\
  and\ \citenamefont {Miao}}]{phsoftmos2}%
  \BibitemOpen
  \bibfield  {author} {\bibinfo {author} {\bibfnamefont {Y.}~\bibnamefont
  {Fu}}, \bibinfo {author} {\bibfnamefont {E.}~\bibnamefont {Liu}}, \bibinfo
  {author} {\bibfnamefont {H.}~\bibnamefont {Yuan}}, \bibinfo {author}
  {\bibfnamefont {P.}~\bibnamefont {Tang}}, \bibinfo {author} {\bibfnamefont
  {B.}~\bibnamefont {Lian}}, \bibinfo {author} {\bibfnamefont {G.}~\bibnamefont
  {Xu}}, \bibinfo {author} {\bibfnamefont {J.}~\bibnamefont {Zeng}}, \bibinfo
  {author} {\bibfnamefont {Z.}~\bibnamefont {Chen}}, \bibinfo {author}
  {\bibfnamefont {Y.}~\bibnamefont {Wang}}, \bibinfo {author} {\bibfnamefont
  {W.}~\bibnamefont {Zhou}}, \bibinfo {author} {\bibfnamefont {K.}~\bibnamefont
  {Xu}}, \bibinfo {author} {\bibfnamefont {A.}~\bibnamefont {Gao}}, \bibinfo
  {author} {\bibfnamefont {C.}~\bibnamefont {Pan}}, \bibinfo {author}
  {\bibfnamefont {M.}~\bibnamefont {Wang}}, \bibinfo {author} {\bibfnamefont
  {B.}~\bibnamefont {Wang}}, \bibinfo {author} {\bibfnamefont {S.-C.}\
  \bibnamefont {Zhang}}, \bibinfo {author} {\bibfnamefont {Y.}~\bibnamefont
  {Cui}}, \bibinfo {author} {\bibfnamefont {H.~Y.}\ \bibnamefont {Hwang}}, \
  and\ \bibinfo {author} {\bibfnamefont {F.}~\bibnamefont {Miao}},\ }\href
  {\doibase 10.1038/s41535-017-0056-1} {\bibfield  {journal} {\bibinfo
  {journal} {npj Quantum Materials}\ }\textbf {\bibinfo {volume} {2}},\
  \bibinfo {pages} {52} (\bibinfo {year} {2017})}\BibitemShut {NoStop}%
\bibitem [{\citenamefont {R\"osner}\ \emph {et~al.}(2014)\citenamefont
  {R\"osner}, \citenamefont {Haas},\ and\ \citenamefont
  {Wehling}}]{rosnermos2}%
  \BibitemOpen
  \bibfield  {author} {\bibinfo {author} {\bibfnamefont {M.}~\bibnamefont
  {R\"osner}}, \bibinfo {author} {\bibfnamefont {S.}~\bibnamefont {Haas}}, \
  and\ \bibinfo {author} {\bibfnamefont {T.~O.}\ \bibnamefont {Wehling}},\
  }\href {\doibase 10.1103/PhysRevB.90.245105} {\bibfield  {journal} {\bibinfo
  {journal} {Phys. Rev. B}\ }\textbf {\bibinfo {volume} {90}},\ \bibinfo
  {pages} {245105} (\bibinfo {year} {2014})}\BibitemShut {NoStop}%
\bibitem [{\citenamefont {Allen}(1980)}]{dyprosovol3}%
  \BibitemOpen
  \bibfield  {author} {\bibinfo {author} {\bibfnamefont {P.~B.}\ \bibnamefont
  {Allen}},\ }\href@noop {} {\emph {\bibinfo {title} {Dynamical Properties of
  Solids}}},\ edited by\ \bibinfo {editor} {\bibfnamefont {G.~K.}\ \bibnamefont
  {Horton}}\ and\ \bibinfo {editor} {\bibfnamefont {A.~A.}\ \bibnamefont
  {Maradudin}},\ Vol.~\bibinfo {volume} {3}\ (\bibinfo  {publisher}
  {North-Holland, New York},\ \bibinfo {year} {1980})\ Chap.~\bibinfo {chapter}
  {2}, pp.\ \bibinfo {pages} {95--196}\BibitemShut {NoStop}%
\bibitem [{\citenamefont {Feynman}(1939)}]{hft}%
  \BibitemOpen
  \bibfield  {author} {\bibinfo {author} {\bibfnamefont {R.~P.}\ \bibnamefont
  {Feynman}},\ }\href {\doibase 10.1103/PhysRev.56.340} {\bibfield  {journal}
  {\bibinfo  {journal} {Phys. Rev.}\ }\textbf {\bibinfo {volume} {56}},\
  \bibinfo {pages} {340} (\bibinfo {year} {1939})}\BibitemShut {NoStop}%
\bibitem [{\citenamefont {Mahan}(2000)}]{mahan}%
  \BibitemOpen
  \bibfield  {author} {\bibinfo {author} {\bibfnamefont {G.~D.}\ \bibnamefont
  {Mahan}},\ }\href@noop {} {\emph {\bibinfo {title} {Many-Particle
  Physics}}},\ \bibinfo {edition} {3rd}\ ed.\ (\bibinfo  {publisher} {Springer
  Science+Business Media, New York},\ \bibinfo {year} {2000})\BibitemShut
  {NoStop}%
\bibitem [{\citenamefont {Grimvall}(1981)}]{Grimvall}%
  \BibitemOpen
  \bibfield  {author} {\bibinfo {author} {\bibfnamefont {G.}~\bibnamefont
  {Grimvall}},\ }\href@noop {} {\emph {\bibinfo {title} {The Electron-Phonon
  Interaction in Metals, Selected Topics in Solid State Physics}}}\ (\bibinfo
  {publisher} {North-Holland, New York},\ \bibinfo {year} {1981})\BibitemShut
  {NoStop}%
\bibitem [{Note1()}]{Note1}%
  \BibitemOpen
  \bibinfo {note} {Interestingly, a similar expression for the phonon
  self-energy can be recovered from directly adopting a frequency-dependent
  density-response function in the first term on the right-hand side of
  Eq.\protect \tmspace +\thinmuskip {.1667em}\ref {eq:DFPT}, and transforming
  it into the normal-mode representation (see the Note\protect \tmspace
  +\thinmuskip {.1667em}S1 of the Supplemental Material)}\BibitemShut {NoStop}%
\bibitem [{\citenamefont {Giustino}(2017)}]{Giustinorev}%
  \BibitemOpen
  \bibfield  {author} {\bibinfo {author} {\bibfnamefont {F.}~\bibnamefont
  {Giustino}},\ }\href {\doibase 10.1103/RevModPhys.89.015003} {\bibfield
  {journal} {\bibinfo  {journal} {Rev. Mod. Phys.}\ }\textbf {\bibinfo {volume}
  {89}},\ \bibinfo {pages} {015003} (\bibinfo {year} {2017})}\BibitemShut
  {NoStop}%
\bibitem [{\citenamefont {Eiguren}\ and\ \citenamefont
  {Ambrosch-Draxl}(2008{\natexlab{a}})}]{Asierprl2008}%
  \BibitemOpen
  \bibfield  {author} {\bibinfo {author} {\bibfnamefont {A.}~\bibnamefont
  {Eiguren}}\ and\ \bibinfo {author} {\bibfnamefont {C.}~\bibnamefont
  {Ambrosch-Draxl}},\ }\href {\doibase 10.1103/PhysRevLett.101.036402}
  {\bibfield  {journal} {\bibinfo  {journal} {Phys. Rev. Lett.}\ }\textbf
  {\bibinfo {volume} {101}},\ \bibinfo {pages} {036402} (\bibinfo {year}
  {2008}{\natexlab{a}})}\BibitemShut {NoStop}%
\bibitem [{\citenamefont {Eiguren}\ \emph {et~al.}(2009)\citenamefont
  {Eiguren}, \citenamefont {Ambrosch-Draxl},\ and\ \citenamefont
  {Echenique}}]{Asierprb2009}%
  \BibitemOpen
  \bibfield  {author} {\bibinfo {author} {\bibfnamefont {A.}~\bibnamefont
  {Eiguren}}, \bibinfo {author} {\bibfnamefont {C.}~\bibnamefont
  {Ambrosch-Draxl}}, \ and\ \bibinfo {author} {\bibfnamefont {P.~M.}\
  \bibnamefont {Echenique}},\ }\href {\doibase 10.1103/PhysRevB.79.245103}
  {\bibfield  {journal} {\bibinfo  {journal} {Phys. Rev. B}\ }\textbf {\bibinfo
  {volume} {79}},\ \bibinfo {pages} {245103} (\bibinfo {year}
  {2009})}\BibitemShut {NoStop}%
\bibitem [{\citenamefont {Hedin}\ and\ \citenamefont
  {Lundqvist}(1970)}]{HEDIN19701}%
  \BibitemOpen
  \bibfield  {author} {\bibinfo {author} {\bibfnamefont {L.}~\bibnamefont
  {Hedin}}\ and\ \bibinfo {author} {\bibfnamefont {S.}~\bibnamefont
  {Lundqvist}}\ }(\bibinfo  {publisher} {Academic Press},\ \bibinfo {year}
  {1970})\ pp.\ \bibinfo {pages} {1 -- 181}\BibitemShut {NoStop}%
\bibitem [{\citenamefont {Farid}(1999)}]{farid}%
  \BibitemOpen
  \bibfield  {author} {\bibinfo {author} {\bibfnamefont {B.}~\bibnamefont
  {Farid}},\ }\enquote {\bibinfo {title} {Ground and low-lying excited states
  of interacting electron systems. a survey and some critical analyses},}\ \
  (\bibinfo {year} {1999})\ pp.\ \bibinfo {pages} {103--261}\BibitemShut
  {NoStop}%
\bibitem [{\citenamefont {Hamann}(1989)}]{ncpp}%
  \BibitemOpen
  \bibfield  {author} {\bibinfo {author} {\bibfnamefont {D.~R.}\ \bibnamefont
  {Hamann}},\ }\href {\doibase 10.1103/PhysRevB.40.2980} {\bibfield  {journal}
  {\bibinfo  {journal} {Phys. Rev. B}\ }\textbf {\bibinfo {volume} {40}},\
  \bibinfo {pages} {2980} (\bibinfo {year} {1989})}\BibitemShut {NoStop}%
\bibitem [{\citenamefont {Kleinman}(1980)}]{relncpp}%
  \BibitemOpen
  \bibfield  {author} {\bibinfo {author} {\bibfnamefont {L.}~\bibnamefont
  {Kleinman}},\ }\href {\doibase 10.1103/PhysRevB.21.2630} {\bibfield
  {journal} {\bibinfo  {journal} {Phys. Rev. B}\ }\textbf {\bibinfo {volume}
  {21}},\ \bibinfo {pages} {2630} (\bibinfo {year} {1980})}\BibitemShut
  {NoStop}%
\bibitem [{\citenamefont {Giannozzi}\ \emph {et~al.}(2009)\citenamefont
  {Giannozzi}, \citenamefont {Baroni}, \citenamefont {Bonini}, \citenamefont
  {Calandra}, \citenamefont {Car}, \citenamefont {Cavazzoni}, \citenamefont
  {Ceresoli}, \citenamefont {Chiarotti}, \citenamefont {Cococcioni},
  \citenamefont {Dabo}, \citenamefont {Corso}, \citenamefont {de~Gironcoli},
  \citenamefont {Fabris}, \citenamefont {Fratesi}, \citenamefont {Gebauer},
  \citenamefont {Gerstmann}, \citenamefont {Gougoussis}, \citenamefont
  {Kokalj}, \citenamefont {Lazzeri}, \citenamefont {Martin-Samos},
  \citenamefont {Marzari}, \citenamefont {Mauri}, \citenamefont {Mazzarello},
  \citenamefont {Paolini}, \citenamefont {Pasquarello}, \citenamefont
  {Paulatto}, \citenamefont {Sbraccia}, \citenamefont {Scandolo}, \citenamefont
  {Sclauzero}, \citenamefont {Seitsonen}, \citenamefont {Smogunov},
  \citenamefont {Umari},\ and\ \citenamefont {Wentzcovitch}}]{QE}%
  \BibitemOpen
  \bibfield  {author} {\bibinfo {author} {\bibfnamefont {P.}~\bibnamefont
  {Giannozzi}}, \bibinfo {author} {\bibfnamefont {S.}~\bibnamefont {Baroni}},
  \bibinfo {author} {\bibfnamefont {N.}~\bibnamefont {Bonini}}, \bibinfo
  {author} {\bibfnamefont {M.}~\bibnamefont {Calandra}}, \bibinfo {author}
  {\bibfnamefont {R.}~\bibnamefont {Car}}, \bibinfo {author} {\bibfnamefont
  {C.}~\bibnamefont {Cavazzoni}}, \bibinfo {author} {\bibfnamefont
  {D.}~\bibnamefont {Ceresoli}}, \bibinfo {author} {\bibfnamefont {G.~L.}\
  \bibnamefont {Chiarotti}}, \bibinfo {author} {\bibfnamefont {M.}~\bibnamefont
  {Cococcioni}}, \bibinfo {author} {\bibfnamefont {I.}~\bibnamefont {Dabo}},
  \bibinfo {author} {\bibfnamefont {A.~D.}\ \bibnamefont {Corso}}, \bibinfo
  {author} {\bibfnamefont {S.}~\bibnamefont {de~Gironcoli}}, \bibinfo {author}
  {\bibfnamefont {S.}~\bibnamefont {Fabris}}, \bibinfo {author} {\bibfnamefont
  {G.}~\bibnamefont {Fratesi}}, \bibinfo {author} {\bibfnamefont
  {R.}~\bibnamefont {Gebauer}}, \bibinfo {author} {\bibfnamefont
  {U.}~\bibnamefont {Gerstmann}}, \bibinfo {author} {\bibfnamefont
  {C.}~\bibnamefont {Gougoussis}}, \bibinfo {author} {\bibfnamefont
  {A.}~\bibnamefont {Kokalj}}, \bibinfo {author} {\bibfnamefont
  {M.}~\bibnamefont {Lazzeri}}, \bibinfo {author} {\bibfnamefont
  {L.}~\bibnamefont {Martin-Samos}}, \bibinfo {author} {\bibfnamefont
  {N.}~\bibnamefont {Marzari}}, \bibinfo {author} {\bibfnamefont
  {F.}~\bibnamefont {Mauri}}, \bibinfo {author} {\bibfnamefont
  {R.}~\bibnamefont {Mazzarello}}, \bibinfo {author} {\bibfnamefont
  {S.}~\bibnamefont {Paolini}}, \bibinfo {author} {\bibfnamefont
  {A.}~\bibnamefont {Pasquarello}}, \bibinfo {author} {\bibfnamefont
  {L.}~\bibnamefont {Paulatto}}, \bibinfo {author} {\bibfnamefont
  {C.}~\bibnamefont {Sbraccia}}, \bibinfo {author} {\bibfnamefont
  {S.}~\bibnamefont {Scandolo}}, \bibinfo {author} {\bibfnamefont
  {G.}~\bibnamefont {Sclauzero}}, \bibinfo {author} {\bibfnamefont {A.~P.}\
  \bibnamefont {Seitsonen}}, \bibinfo {author} {\bibfnamefont {A.}~\bibnamefont
  {Smogunov}}, \bibinfo {author} {\bibfnamefont {P.}~\bibnamefont {Umari}}, \
  and\ \bibinfo {author} {\bibfnamefont {R.~M.}\ \bibnamefont {Wentzcovitch}},\
  }\href {http://stacks.iop.org/0953-8984/21/i=39/a=395502} {\bibfield
  {journal} {\bibinfo  {journal} {Journal of Physics: Condensed Matter}\
  }\textbf {\bibinfo {volume} {21}},\ \bibinfo {pages} {395502} (\bibinfo
  {year} {2009})}\BibitemShut {NoStop}%
\bibitem [{\citenamefont {Perdew}\ and\ \citenamefont {Zunger}(1981)}]{PZLDA}%
  \BibitemOpen
  \bibfield  {author} {\bibinfo {author} {\bibfnamefont {J.~P.}\ \bibnamefont
  {Perdew}}\ and\ \bibinfo {author} {\bibfnamefont {A.}~\bibnamefont
  {Zunger}},\ }\href {\doibase 10.1103/PhysRevB.23.5048} {\bibfield  {journal}
  {\bibinfo  {journal} {Phys. Rev. B}\ }\textbf {\bibinfo {volume} {23}},\
  \bibinfo {pages} {5048} (\bibinfo {year} {1981})}\BibitemShut {NoStop}%
\bibitem [{Note2()}]{Note2}%
  \BibitemOpen
  \bibinfo {note} {The dynamical matrices were primarily calculated with a
  smearing value of $5~\protect \text {mRy}=68~\protect \text {meV}$, an energy
  comparable to remarkable changes in the topology of the FS upon doping. This
  high value can lead to an unreal smoothing of the FS that can mask
  interesting phenomena as Kohn anomalies. A calculation with a smearing value
  of $5~\protect \text {meV}$ on a finer $720\times 720$ $k$-mesh was hence
  performed for each dynamical matrix of the coarse $q$-mesh (see the
  supplemental Note\protect \tmspace +\thinmuskip {.1667em}S2).}\BibitemShut
  {Stop}%
\bibitem [{\citenamefont {Eiguren}\ and\ \citenamefont
  {Ambrosch-Draxl}(2008{\natexlab{b}})}]{Asierwannier}%
  \BibitemOpen
  \bibfield  {author} {\bibinfo {author} {\bibfnamefont {A.}~\bibnamefont
  {Eiguren}}\ and\ \bibinfo {author} {\bibfnamefont {C.}~\bibnamefont
  {Ambrosch-Draxl}},\ }\href {\doibase 10.1103/PhysRevB.78.045124} {\bibfield
  {journal} {\bibinfo  {journal} {Phys. Rev. B}\ }\textbf {\bibinfo {volume}
  {78}},\ \bibinfo {pages} {045124} (\bibinfo {year}
  {2008}{\natexlab{b}})}\BibitemShut {NoStop}%
\bibitem [{\citenamefont {Giustino}\ \emph
  {et~al.}(2007{\natexlab{a}})\citenamefont {Giustino}, \citenamefont {Cohen},\
  and\ \citenamefont {Louie}}]{epw}%
  \BibitemOpen
  \bibfield  {author} {\bibinfo {author} {\bibfnamefont {F.}~\bibnamefont
  {Giustino}}, \bibinfo {author} {\bibfnamefont {M.~L.}\ \bibnamefont {Cohen}},
  \ and\ \bibinfo {author} {\bibfnamefont {S.~G.}\ \bibnamefont {Louie}},\
  }\href {\doibase 10.1103/PhysRevB.76.165108} {\bibfield  {journal} {\bibinfo
  {journal} {Phys. Rev. B}\ }\textbf {\bibinfo {volume} {76}},\ \bibinfo
  {pages} {165108} (\bibinfo {year} {2007}{\natexlab{a}})}\BibitemShut
  {NoStop}%
\bibitem [{\citenamefont {Giustino}\ \emph
  {et~al.}(2007{\natexlab{b}})\citenamefont {Giustino}, \citenamefont {Yates},
  \citenamefont {Souza}, \citenamefont {Cohen},\ and\ \citenamefont
  {Louie}}]{giustinoborondopeddiamond}%
  \BibitemOpen
  \bibfield  {author} {\bibinfo {author} {\bibfnamefont {F.}~\bibnamefont
  {Giustino}}, \bibinfo {author} {\bibfnamefont {J.}~\bibnamefont {Yates}},
  \bibinfo {author} {\bibfnamefont {I.}~\bibnamefont {Souza}}, \bibinfo
  {author} {\bibfnamefont {M.~L.}\ \bibnamefont {Cohen}}, \ and\ \bibinfo
  {author} {\bibfnamefont {S.~G.}\ \bibnamefont {Louie}},\ }\href {\doibase
  10.1103/PhysRevLett.98.047005} {\bibfield  {journal} {\bibinfo  {journal}
  {Phys. Rev. Lett.}\ }\textbf {\bibinfo {volume} {98}},\ \bibinfo {pages}
  {047005} (\bibinfo {year} {2007}{\natexlab{b}})}\BibitemShut {NoStop}%
\bibitem [{\citenamefont {Young}(1968)}]{aexp}%
  \BibitemOpen
  \bibfield  {author} {\bibinfo {author} {\bibfnamefont {P.~A.}\ \bibnamefont
  {Young}},\ }\href {http://stacks.iop.org/0022-3727/1/i=7/a=416} {\bibfield
  {journal} {\bibinfo  {journal} {Journal of Physics D: Applied Physics}\
  }\textbf {\bibinfo {volume} {1}},\ \bibinfo {pages} {936} (\bibinfo {year}
  {1968})}\BibitemShut {NoStop}%
\bibitem [{Note3()}]{Note3}%
  \BibitemOpen
  \bibinfo {note} {Values of the SO-splitting are strongly dependent on the
  orbital character of electron states. In the case of a monolayer, the SO
  coupling term is expected to be large for in-plane polarized states, as in
  the $\protect \overline {\protect \mathrm {Q}}(\protect \overline {\protect
  \mathrm {Q'}})$ valleys, which are a combination of Mo $d_{xy/x^{2}-y^{2}}$
  and S $p$ orbitals. On the contrary, SO interaction vanishes for out-of-plane
  polarized states, as in the $\protect \overline {\protect \mathrm
  {K}}(\protect \overline {\protect \mathrm {K'}})$ valleys, which have a
  marked Mo $d_{z^{2}}$ orbital character.}\BibitemShut {Stop}%
\bibitem [{\citenamefont {Mak}\ \emph {et~al.}(2010)\citenamefont {Mak},
  \citenamefont {Lee}, \citenamefont {Hone}, \citenamefont {Shan},\ and\
  \citenamefont {Heinz}}]{1lmos2expgap}%
  \BibitemOpen
  \bibfield  {author} {\bibinfo {author} {\bibfnamefont {K.~F.}\ \bibnamefont
  {Mak}}, \bibinfo {author} {\bibfnamefont {C.}~\bibnamefont {Lee}}, \bibinfo
  {author} {\bibfnamefont {J.}~\bibnamefont {Hone}}, \bibinfo {author}
  {\bibfnamefont {J.}~\bibnamefont {Shan}}, \ and\ \bibinfo {author}
  {\bibfnamefont {T.~F.}\ \bibnamefont {Heinz}},\ }\href {\doibase
  10.1103/PhysRevLett.105.136805} {\bibfield  {journal} {\bibinfo  {journal}
  {Phys. Rev. Lett.}\ }\textbf {\bibinfo {volume} {105}},\ \bibinfo {pages}
  {136805} (\bibinfo {year} {2010})}\BibitemShut {NoStop}%
\bibitem [{\citenamefont {Kuc}\ \emph {et~al.}(2011)\citenamefont {Kuc},
  \citenamefont {Zibouche},\ and\ \citenamefont {Heine}}]{1lmos2theogap}%
  \BibitemOpen
  \bibfield  {author} {\bibinfo {author} {\bibfnamefont {A.}~\bibnamefont
  {Kuc}}, \bibinfo {author} {\bibfnamefont {N.}~\bibnamefont {Zibouche}}, \
  and\ \bibinfo {author} {\bibfnamefont {T.}~\bibnamefont {Heine}},\ }\href
  {\doibase 10.1103/PhysRevB.83.245213} {\bibfield  {journal} {\bibinfo
  {journal} {Phys. Rev. B}\ }\textbf {\bibinfo {volume} {83}},\ \bibinfo
  {pages} {245213} (\bibinfo {year} {2011})}\BibitemShut {NoStop}%
\bibitem [{\citenamefont {Rold\'an}\ \emph {et~al.}(2014)\citenamefont
  {Rold\'an}, \citenamefont {Silva-Guill\'en}, \citenamefont {L\'opez-Sancho},
  \citenamefont {Guinea}, \citenamefont {Cappelluti},\ and\ \citenamefont
  {Ordej\'on}}]{1lmos2elbandtheo}%
  \BibitemOpen
  \bibfield  {author} {\bibinfo {author} {\bibfnamefont {R.}~\bibnamefont
  {Rold\'an}}, \bibinfo {author} {\bibfnamefont {J.~A.}\ \bibnamefont
  {Silva-Guill\'en}}, \bibinfo {author} {\bibfnamefont {M.~P.}\ \bibnamefont
  {L\'opez-Sancho}}, \bibinfo {author} {\bibfnamefont {F.}~\bibnamefont
  {Guinea}}, \bibinfo {author} {\bibfnamefont {E.}~\bibnamefont {Cappelluti}},
  \ and\ \bibinfo {author} {\bibfnamefont {P.}~\bibnamefont {Ordej\'on}},\
  }\href {\doibase 10.1002/andp.201400128} {\bibfield  {journal} {\bibinfo
  {journal} {Annalen der Physik}\ }\textbf {\bibinfo {volume} {526}},\ \bibinfo
  {pages} {347} (\bibinfo {year} {2014})},\ \Eprint
  {http://arxiv.org/abs/https://onlinelibrary.wiley.com/doi/pdf/10.1002/andp.201400128}
  {https://onlinelibrary.wiley.com/doi/pdf/10.1002/andp.201400128} \BibitemShut
  {NoStop}%
\bibitem [{\citenamefont {Kaasbjerg}\ \emph {et~al.}(2012)\citenamefont
  {Kaasbjerg}, \citenamefont {Thygesen},\ and\ \citenamefont
  {Jacobsen}}]{kaasbjerg}%
  \BibitemOpen
  \bibfield  {author} {\bibinfo {author} {\bibfnamefont {K.}~\bibnamefont
  {Kaasbjerg}}, \bibinfo {author} {\bibfnamefont {K.~S.}\ \bibnamefont
  {Thygesen}}, \ and\ \bibinfo {author} {\bibfnamefont {K.~W.}\ \bibnamefont
  {Jacobsen}},\ }\href {\doibase 10.1103/PhysRevB.85.115317} {\bibfield
  {journal} {\bibinfo  {journal} {Phys. Rev. B}\ }\textbf {\bibinfo {volume}
  {85}},\ \bibinfo {pages} {115317} (\bibinfo {year} {2012})}\BibitemShut
  {NoStop}%
\bibitem [{Note4()}]{Note4}%
  \BibitemOpen
  \bibinfo {note} {The conduction band and adiabatic phonon dispersions of all
  the doping levels considered in this work can be found in Fig.\protect
  \tmspace +\thinmuskip {.1667em}S1 and S2 of the supplemental Note\protect
  \tmspace +\thinmuskip {.1667em}S4.}\BibitemShut {Stop}%
\bibitem [{Note5()}]{Note5}%
  \BibitemOpen
  \bibinfo {note} {The $\protect \mathrm {A'_{1}}$ mode at $\protect \mathbf
  {q}=\protect \overline {\Gamma }$ and carrier states at $\protect \mathbf
  {k}=\protect \overline {\protect \mathrm {K}}(\protect \overline {\protect
  \mathrm {K'}})$ valleys involve out-of-plane polarized large deformation
  potentials and orbitals at the center of the MoS$_{2}$ layer, respectively,
  that couple efficiently in Eq.\protect \tmspace +\thinmuskip {.1667em}\ref
  {eq:epme} leading to large matrix elements. Likewise, in the $\protect
  \mathrm {A'_{1}}$ and LA modes at $\protect \mathbf {q}\approx \protect
  \overline {\protect \mathrm {M}}$, the additional in-plane displacement of
  the Mo atoms also allows to couple with electron states of $\protect
  \overline {\protect \mathrm {Q}}(\protect \overline {\protect \mathrm {Q'}})$
  valleys, with a marked in-plane Mo orbital character.}\BibitemShut {Stop}%
\bibitem [{Note6()}]{Note6}%
  \BibitemOpen
  \bibinfo {note} {The phonon spectral function of all the doping levels
  considered in this work can be found in Fig.\protect \tmspace +\thinmuskip
  {.1667em}S3 of the supplemental Note\protect \tmspace +\thinmuskip
  {.1667em}S5.}\BibitemShut {Stop}%
\bibitem [{\citenamefont {Allen}(1972)}]{Allenprb1972}%
  \BibitemOpen
  \bibfield  {author} {\bibinfo {author} {\bibfnamefont {P.~B.}\ \bibnamefont
  {Allen}},\ }\href {\doibase 10.1103/PhysRevB.6.2577} {\bibfield  {journal}
  {\bibinfo  {journal} {Phys. Rev. B}\ }\textbf {\bibinfo {volume} {6}},\
  \bibinfo {pages} {2577} (\bibinfo {year} {1972})}\BibitemShut {NoStop}%
\bibitem [{\citenamefont {Stern}(1967)}]{lindhard2d}%
  \BibitemOpen
  \bibfield  {author} {\bibinfo {author} {\bibfnamefont {F.}~\bibnamefont
  {Stern}},\ }\href {\doibase 10.1103/PhysRevLett.18.546} {\bibfield  {journal}
  {\bibinfo  {journal} {Phys. Rev. Lett.}\ }\textbf {\bibinfo {volume} {18}},\
  \bibinfo {pages} {546} (\bibinfo {year} {1967})}\BibitemShut {NoStop}%
\bibitem [{Note7()}]{Note7}%
  \BibitemOpen
  \bibinfo {note} {The spectral function of the $\protect \mathrm {A'_{1}}$
  phonon mode obtained by \protect \textit {ab initio} calculations and using
  the Einstein-like model, both evaluated in the small momentum limit along the
  $\protect \overline {\Gamma \protect \mathrm {K}}$ direction and for all the
  doping levels considered in this work can be found in Fig.\protect \tmspace
  +\thinmuskip {.1667em}S4 of the supplemental Note\protect \tmspace
  +\thinmuskip {.1667em}S5.}\BibitemShut {Stop}%
\bibitem [{\citenamefont {Coleman}(2015)}]{virtualph}%
  \BibitemOpen
  \bibfield  {author} {\bibinfo {author} {\bibfnamefont {P.}~\bibnamefont
  {Coleman}},\ }\href {https://books.google.es/books?id=ESB0CwAAQBAJ} {\emph
  {\bibinfo {title} {Introduction to Many-Body Physics}}}\ (\bibinfo
  {publisher} {Cambridge University Press},\ \bibinfo {year}
  {2015})\BibitemShut {NoStop}%
\end{thebibliography}

\begin{thebibliography}{12}%
\makeatletter
\providecommand \@ifxundefined [1]{%
 \@ifx{#1\undefined}
}%
\providecommand \@ifnum [1]{%
 \ifnum #1\expandafter \@firstoftwo
 \else \expandafter \@secondoftwo
 \fi
}%
\providecommand \@ifx [1]{%
 \ifx #1\expandafter \@firstoftwo
 \else \expandafter \@secondoftwo
 \fi
}%
\providecommand \natexlab [1]{#1}%
\providecommand \enquote  [1]{``#1''}%
\providecommand \bibnamefont  [1]{#1}%
\providecommand \bibfnamefont [1]{#1}%
\providecommand \citenamefont [1]{#1}%
\providecommand \href@noop [0]{\@secondoftwo}%
\providecommand \href [0]{\begingroup \@sanitize@url \@href}%
\providecommand \@href[1]{\@@startlink{#1}\@@href}%
\providecommand \@@href[1]{\endgroup#1\@@endlink}%
\providecommand \@sanitize@url [0]{\catcode `\\12\catcode `\$12\catcode
  `\&12\catcode `\#12\catcode `\^12\catcode `\_12\catcode `\%12\relax}%
\providecommand \@@startlink[1]{}%
\providecommand \@@endlink[0]{}%
\providecommand \url  [0]{\begingroup\@sanitize@url \@url }%
\providecommand \@url [1]{\endgroup\@href {#1}{\urlprefix }}%
\providecommand \urlprefix  [0]{URL }%
\providecommand \Eprint [0]{\href }%
\providecommand \doibase [0]{http://dx.doi.org/}%
\providecommand \selectlanguage [0]{\@gobble}%
\providecommand \bibinfo  [0]{\@secondoftwo}%
\providecommand \bibfield  [0]{\@secondoftwo}%
\providecommand \translation [1]{[#1]}%
\providecommand \BibitemOpen [0]{}%
\providecommand \bibitemStop [0]{}%
\providecommand \bibitemNoStop [0]{.\EOS\space}%
\providecommand \EOS [0]{\spacefactor3000\relax}%
\providecommand \BibitemShut  [1]{\csname bibitem#1\endcsname}%
\let\auto@bib@innerbib\@empty
\bibitem [{\citenamefont {Baroni}\ \emph {et~al.}(2001)\citenamefont {Baroni},
  \citenamefont {de~Gironcoli}, \citenamefont {Dal~Corso},\ and\ \citenamefont
  {Giannozzi}}]{DFPTs}%
  \BibitemOpen
  \bibfield  {author} {\bibinfo {author} {\bibfnamefont {S.}~\bibnamefont
  {Baroni}}, \bibinfo {author} {\bibfnamefont {S.}~\bibnamefont
  {de~Gironcoli}}, \bibinfo {author} {\bibfnamefont {A.}~\bibnamefont
  {Dal~Corso}}, \ and\ \bibinfo {author} {\bibfnamefont {P.}~\bibnamefont
  {Giannozzi}},\ }\href {\doibase 10.1103/RevModPhys.73.515} {\bibfield
  {journal} {\bibinfo  {journal} {Rev. Mod. Phys.}\ }\textbf {\bibinfo {volume}
  {73}},\ \bibinfo {pages} {515} (\bibinfo {year} {2001})}\BibitemShut
  {NoStop}%
\bibitem [{\citenamefont {Allen}(1980)}]{dyprosovol3s}%
  \BibitemOpen
  \bibfield  {author} {\bibinfo {author} {\bibfnamefont {P.~B.}\ \bibnamefont
  {Allen}},\ }\href@noop {} {\emph {\bibinfo {title} {Dynamical Properties of
  Solids}}},\ edited by\ \bibinfo {editor} {\bibfnamefont {G.~K.}\ \bibnamefont
  {Horton}}\ and\ \bibinfo {editor} {\bibfnamefont {A.~A.}\ \bibnamefont
  {Maradudin}},\ Vol.~\bibinfo {volume} {3}\ (\bibinfo  {publisher}
  {North-Holland, New York},\ \bibinfo {year} {1980})\ Chap.~\bibinfo {chapter}
  {2}, pp.\ \bibinfo {pages} {95--196}\BibitemShut {NoStop}%
\bibitem [{\citenamefont {Giustino}(2017)}]{giustinorevs}%
  \BibitemOpen
  \bibfield  {author} {\bibinfo {author} {\bibfnamefont {F.}~\bibnamefont
  {Giustino}},\ }\href {\doibase 10.1103/RevModPhys.89.015003} {\bibfield
  {journal} {\bibinfo  {journal} {Rev. Mod. Phys.}\ }\textbf {\bibinfo {volume}
  {89}},\ \bibinfo {pages} {015003} (\bibinfo {year} {2017})}\BibitemShut
  {NoStop}%
\bibitem [{\citenamefont {Grimvall}(1981)}]{Grimvalls}%
  \BibitemOpen
  \bibfield  {author} {\bibinfo {author} {\bibfnamefont {G.}~\bibnamefont
  {Grimvall}},\ }\href@noop {} {\emph {\bibinfo {title} {The Electron-Phonon
  Interaction in Metals, Selected Topics in Solid State Physics}}}\ (\bibinfo
  {publisher} {North-Holland, New York},\ \bibinfo {year} {1981})\BibitemShut
  {NoStop}%
\bibitem [{\citenamefont {Chakraborty}\ \emph {et~al.}(2012)\citenamefont
  {Chakraborty}, \citenamefont {Bera}, \citenamefont {Muthu}, \citenamefont
  {Bhowmick}, \citenamefont {Waghmare},\ and\ \citenamefont
  {Sood}}]{1lmos2a1softs}%
  \BibitemOpen
  \bibfield  {author} {\bibinfo {author} {\bibfnamefont {B.}~\bibnamefont
  {Chakraborty}}, \bibinfo {author} {\bibfnamefont {A.}~\bibnamefont {Bera}},
  \bibinfo {author} {\bibfnamefont {D.~V.~S.}\ \bibnamefont {Muthu}}, \bibinfo
  {author} {\bibfnamefont {S.}~\bibnamefont {Bhowmick}}, \bibinfo {author}
  {\bibfnamefont {U.~V.}\ \bibnamefont {Waghmare}}, \ and\ \bibinfo {author}
  {\bibfnamefont {A.~K.}\ \bibnamefont {Sood}},\ }\href {\doibase
  10.1103/PhysRevB.85.161403} {\bibfield  {journal} {\bibinfo  {journal} {Phys.
  Rev. B}\ }\textbf {\bibinfo {volume} {85}},\ \bibinfo {pages} {161403}
  (\bibinfo {year} {2012})}\BibitemShut {NoStop}%
\bibitem [{\citenamefont {Sohier}\ \emph {et~al.}(2019)\citenamefont {Sohier},
  \citenamefont {Ponomarev}, \citenamefont {Gibertini}, \citenamefont {Berger},
  \citenamefont {Marzari}, \citenamefont {Ubrig},\ and\ \citenamefont
  {Morpurgo}}]{sohierprx2019s}%
  \BibitemOpen
  \bibfield  {author} {\bibinfo {author} {\bibfnamefont {T.}~\bibnamefont
  {Sohier}}, \bibinfo {author} {\bibfnamefont {E.}~\bibnamefont {Ponomarev}},
  \bibinfo {author} {\bibfnamefont {M.}~\bibnamefont {Gibertini}}, \bibinfo
  {author} {\bibfnamefont {H.}~\bibnamefont {Berger}}, \bibinfo {author}
  {\bibfnamefont {N.}~\bibnamefont {Marzari}}, \bibinfo {author} {\bibfnamefont
  {N.}~\bibnamefont {Ubrig}}, \ and\ \bibinfo {author} {\bibfnamefont {A.~F.}\
  \bibnamefont {Morpurgo}},\ }\href {\doibase 10.1103/PhysRevX.9.031019}
  {\bibfield  {journal} {\bibinfo  {journal} {Phys. Rev. X}\ }\textbf {\bibinfo
  {volume} {9}},\ \bibinfo {pages} {031019} (\bibinfo {year}
  {2019})}\BibitemShut {NoStop}%
\bibitem [{\citenamefont {Novko}(2019)}]{1907.04766s}%
  \BibitemOpen
  \bibfield  {author} {\bibinfo {author} {\bibfnamefont {D.}~\bibnamefont
  {Novko}},\ }\href@noop {} {\enquote {\bibinfo {title} {Broken adiabaticity
  induced by lifshitz transition in {M}o{S}${}_{2}$ and {W}{S}${}_{2}$ single
  layers},}\ } (\bibinfo {year} {2019}),\ \Eprint
  {http://arxiv.org/abs/arXiv:1907.04766} {arXiv:1907.04766} \BibitemShut
  {NoStop}%
\bibitem [{\citenamefont {Ge}\ and\ \citenamefont {Liu}(2013)}]{yizhiprb2013s}%
  \BibitemOpen
  \bibfield  {author} {\bibinfo {author} {\bibfnamefont {Y.}~\bibnamefont
  {Ge}}\ and\ \bibinfo {author} {\bibfnamefont {A.~Y.}\ \bibnamefont {Liu}},\
  }\href {\doibase 10.1103/PhysRevB.87.241408} {\bibfield  {journal} {\bibinfo
  {journal} {Phys. Rev. B}\ }\textbf {\bibinfo {volume} {87}},\ \bibinfo
  {pages} {241408} (\bibinfo {year} {2013})}\BibitemShut {NoStop}%
\bibitem [{\citenamefont {Fu}\ \emph {et~al.}(2017)\citenamefont {Fu},
  \citenamefont {Liu}, \citenamefont {Yuan}, \citenamefont {Tang},
  \citenamefont {Lian}, \citenamefont {Xu}, \citenamefont {Zeng}, \citenamefont
  {Chen}, \citenamefont {Wang}, \citenamefont {Zhou}, \citenamefont {Xu},
  \citenamefont {Gao}, \citenamefont {Pan}, \citenamefont {Wang}, \citenamefont
  {Wang}, \citenamefont {Zhang}, \citenamefont {Cui}, \citenamefont {Hwang},\
  and\ \citenamefont {Miao}}]{phsoftmos2s}%
  \BibitemOpen
  \bibfield  {author} {\bibinfo {author} {\bibfnamefont {Y.}~\bibnamefont
  {Fu}}, \bibinfo {author} {\bibfnamefont {E.}~\bibnamefont {Liu}}, \bibinfo
  {author} {\bibfnamefont {H.}~\bibnamefont {Yuan}}, \bibinfo {author}
  {\bibfnamefont {P.}~\bibnamefont {Tang}}, \bibinfo {author} {\bibfnamefont
  {B.}~\bibnamefont {Lian}}, \bibinfo {author} {\bibfnamefont {G.}~\bibnamefont
  {Xu}}, \bibinfo {author} {\bibfnamefont {J.}~\bibnamefont {Zeng}}, \bibinfo
  {author} {\bibfnamefont {Z.}~\bibnamefont {Chen}}, \bibinfo {author}
  {\bibfnamefont {Y.}~\bibnamefont {Wang}}, \bibinfo {author} {\bibfnamefont
  {W.}~\bibnamefont {Zhou}}, \bibinfo {author} {\bibfnamefont {K.}~\bibnamefont
  {Xu}}, \bibinfo {author} {\bibfnamefont {A.}~\bibnamefont {Gao}}, \bibinfo
  {author} {\bibfnamefont {C.}~\bibnamefont {Pan}}, \bibinfo {author}
  {\bibfnamefont {M.}~\bibnamefont {Wang}}, \bibinfo {author} {\bibfnamefont
  {B.}~\bibnamefont {Wang}}, \bibinfo {author} {\bibfnamefont {S.-C.}\
  \bibnamefont {Zhang}}, \bibinfo {author} {\bibfnamefont {Y.}~\bibnamefont
  {Cui}}, \bibinfo {author} {\bibfnamefont {H.~Y.}\ \bibnamefont {Hwang}}, \
  and\ \bibinfo {author} {\bibfnamefont {F.}~\bibnamefont {Miao}},\ }\href
  {\doibase 10.1038/s41535-017-0056-1} {\bibfield  {journal} {\bibinfo
  {journal} {npj Quantum Materials}\ }\textbf {\bibinfo {volume} {2}},\
  \bibinfo {pages} {52} (\bibinfo {year} {2017})}\BibitemShut {NoStop}%
\bibitem [{\citenamefont {R\"osner}\ \emph {et~al.}(2014)\citenamefont
  {R\"osner}, \citenamefont {Haas},\ and\ \citenamefont
  {Wehling}}]{rosnermos2s}%
  \BibitemOpen
  \bibfield  {author} {\bibinfo {author} {\bibfnamefont {M.}~\bibnamefont
  {R\"osner}}, \bibinfo {author} {\bibfnamefont {S.}~\bibnamefont {Haas}}, \
  and\ \bibinfo {author} {\bibfnamefont {T.~O.}\ \bibnamefont {Wehling}},\
  }\href {\doibase 10.1103/PhysRevB.90.245105} {\bibfield  {journal} {\bibinfo
  {journal} {Phys. Rev. B}\ }\textbf {\bibinfo {volume} {90}},\ \bibinfo
  {pages} {245105} (\bibinfo {year} {2014})}\BibitemShut {NoStop}%
\bibitem [{\citenamefont {Calandra}\ \emph {et~al.}(2010)\citenamefont
  {Calandra}, \citenamefont {Profeta},\ and\ \citenamefont
  {Mauri}}]{calandramauriprb2010s}%
  \BibitemOpen
  \bibfield  {author} {\bibinfo {author} {\bibfnamefont {M.}~\bibnamefont
  {Calandra}}, \bibinfo {author} {\bibfnamefont {G.}~\bibnamefont {Profeta}}, \
  and\ \bibinfo {author} {\bibfnamefont {F.}~\bibnamefont {Mauri}},\ }\href
  {\doibase 10.1103/PhysRevB.82.165111} {\bibfield  {journal} {\bibinfo
  {journal} {Phys. Rev. B}\ }\textbf {\bibinfo {volume} {82}},\ \bibinfo
  {pages} {165111} (\bibinfo {year} {2010})}\BibitemShut {NoStop}%
\bibitem [{\citenamefont {Stern}(1967)}]{stern2Ds}%
  \BibitemOpen
  \bibfield  {author} {\bibinfo {author} {\bibfnamefont {F.}~\bibnamefont
  {Stern}},\ }\href {\doibase 10.1103/PhysRevLett.18.546} {\bibfield  {journal}
  {\bibinfo  {journal} {Phys. Rev. Lett.}\ }\textbf {\bibinfo {volume} {18}},\
  \bibinfo {pages} {546} (\bibinfo {year} {1967})}\BibitemShut {NoStop}%
\end{thebibliography}
\end{document}